# Information Systems Dynamics: Foundations and Applications

Xu Jianfeng[1*], Liu Zhenyu[2], Wang Shuliang[3], Zheng Tao[3], Wang Yashi[2], Wang Yingfei[1], Dang Yingxu[3]

[1]The Information Technology Service Center of the People's Court
[2] China University of Political Science and Law
[3] Beijing University of Technology
[4] Chinese Academy of Electronic Sciences

[*] Corresponding author. E-mail: xjfcetc@163.com

**Abstract**
This article firstly reviews and summarizes the rapid development of information technology, characterized by the close combination of computer and network communication, which leads to a series of investigations, including the analyses of the important role of a series of technological achievements in the context of information movement and application, the interrelationship between the real-world, information space and information system, and the integrated framework of the real-world and information system, and the modifications and improvements of the Xu's previous mathematical theory on information models, properties and metrics. Based on the mathematical foundations, eleven types of information measure efficacies and their distribution across information systems are put forward, and then the dynamics configurations of information systems are comprehensively analyzed, which constitutes the basic theoretical framework of information systems dynamics with general significance. Finally, Smart Court SoSs (System of Systems) Engineering Project of China are introduced as the exemplified application of the theoretical work, which aims at providing a reference for the analysis, design, development and evaluation of large-scale complex information systems.

# 1 Introduction

Gleick's a Brief History of Information makes it clear at the outset that information is the blood, food, and life force on which our world runs [1]. With the increasing number of types of information systems, the wider the scope, the higher the degree of connectivity, and the stronger the demand for collaboration, it is more and more difficult for people to understand and grasp information systems, especially the *System of Systems* (SoSs) composed of many information systems. Due to the influence of various internal and external uncertainties, the evolution of dynamic behavior of these particularly complex information systems may deviate from their original purpose, and even appear unstable phenomenon. At the same time, in the construction and application of such large-scale information systems, the only emphasis on the order and ignore the vitality will inevitably lead to the rigidity of the systems, in contrast, the only emphasis on the vitality and ignorance of the order, will inevitably lead to the chaos of the systems [2]. Therefore, there is a large need of a set of standardized methods, similar to traditional dynamics to guide large-scale mechanical system engineering, to guide the design, development, application and evaluation of large-scale information systems.

However, although the development and application of information technology are progressing rapidly in the past days, a comprehensive, rigorous and complete basic theoretical system of information systems dynamics has not yet been established. The main reasons can be summarized as follows:

First, there is a lack of a generally accepted mathematical basis for the concept of information. In 1948, Shannon published The Mathematical Theory of Communication [3], which is regarded as the first work of information theory. It reveals that the essence of communication process is to reduce uncertainty, and the uncertainty of random events with a certain probability distribution can be expressed as entropy, so some people think that information is "negative entropy". However, because the application of information is far beyond the scope of communication, and the form of information system is far beyond the scope of communication system, it is difficult to explain information with "negative entropy" to meet the needs of the development of contemporary information science and technology, especially the practice of information system engineering. In fact, up to now, people's understanding of information is far from a consensus [4]. There are even differences on the fundamental question of whether information belongs to the objective category or the subjective category [5]. It is particularly important that people have not paid enough attention to the use of mature mathematical theory to express and deconstruct information, which makes the research lack a unified, clear, convenient and feasible mathematical basis, and naturally it is difficult to use rigorous and profound mathematical tools to establish a complete and detailed theoretical system of information system dynamics.

Second, there is a lack of a clear and comprehensive metric system for information value. The "entropy" proposed by [3] is indeed an important metric that can reflect the value of information, which is also the fundamental reason why Shannon's information theory has developed vigorously and has always been the backbone of the theoretical building of information science. A series of conceptual indexes such as cumulative residual entropy [6], cross entropy [7], relative entropy [8], conditional entropy [9], joint entropy [10] and fuzzy entropy [11] have also appeared in the follow-up study. However, "entropy" index is far from covering the practical needs of modern information system engineering, because "uncertainty" is a relative concept, and the realization of information systems needs to meet the needs of thousands of different users, any different user has its own "uncertainty", so there are few examples of direct use of "entropy" index in the design and analysis of modern information systems. On the other hand, [3] proposed that the amount of information can be expressed by binary bits, that is, bits, which has become the most popular and easily understood information metric, further reflecting the great contribution of Shannon's information theory to the development of information technology. However, bits do not reflect all the meanings of information, because the same number of bits of information can have very different values. However, apart from bits, there are hardly any other strictly defined and universally accepted metrics of information today. For example, some people summarize the 5V characteristics of big data, namely Volume, Velocity, Variety, Value and Veracity [12]. These characteristics obviously have metric attributes, but most of them lack strict connotation consensus and mathematical definition. Due to the lack of information metrics system with clear definition and rich types, it is difficult to establish the essential basic frame of reference for the study of information mechanism.

Third, there is a lack of scientific and reasonable framework for information space. Information space is the objective reality of information movement and function, and information system is the basic carrier of using information to serve human beings. Nowadays, from the Internet system covering the whole world to the smart phone owned by everyone, it is not only an independent information system, but also an integral part of another huge information system. Computer, Internet, big data, cloud computing, block chain, Internet of Things, robots, virtual reality and so on are not only the symbolic achievements of the development of information technology, but also the various forms of rich and colorful information systems. But what is the relationship between these information systems and the real-world we live in? What is the role of emerging information technologies in various information systems? Can the set of all information systems be reasonably classified? Can we build a scientific and rational framework of information space? All these problems affect people's analysis, planning and expectation of information systems. In general, input-behavior-output is a description applicable to any system process [13]. However, for the complex information space and information system, this model is obviously too simple to generally support the research and analysis of the system. If they are confused by the ever-changing information space, they may be bound by all kinds of tedious details and unable to extricate themselves. Therefore, it is urgent to classify the basic components of information system reasonably and describe the framework of the whole information space in order to support the formation of the theoretical system of information system dynamics.

Fourth, there is a lack of clear efficacy analysis of the functions of information. Functional mechanism describes how and how much one thing affects another. After the birth of Newtonian mechanics, mathematical tools such as calculus and differential equations were used to describe and analyze the mechanism of dynamic system through efficiency indicators such as speed, force, energy consumption and power, which makes it possible to establish the complete and mature dynamics theory system [14]. Such methodologies have also been applied to other fields besides mechanical systems, such as chemical kinetics [15], economic dynamics[16], etc., and become an effective tool for analyzing the operation mechanism of specific fields. For information, although people can't leave for a moment in life, there is a lack of widely accepted and universal efficacy definition and analysis. Shannon's information theory confirms that information has the effect of eliminating uncertainty, but this effect alone is far from enough to describe and analyze the complex interaction between information systems. Many other specific information systems, such as radar, sonar, computer, data center, control center, etc., have specific definitions and indicators of specific information efficacy in their respective fields, which can support the analysis and research of corresponding information action mechanism. However, the information efficacy of these respective fields is far from the same as that of speed and force, which has universal significance. Therefore, it is difficult to meet the needs of the analysis and research of integrating many information systems belonging to different fields and different functions into a large-scale system. It can be seen that only by establishing an information efficacy system with broad connotation and universal significance can we accurately analyze the influence patterns and degree between the components of large-scale complex information systems.

Due to the reasons mentioned above, there is still a lack of computable mathematical models to describe the intrinsic dynamics of information systems. In fact, the concept of *Information Dynamics* has existed for a long period of time [17][18][19], but the concepts expressed in the literature are mainly qualitative and lack of quantitative regularity [20]. Yan Bin elaborated the concepts of information, information quantity and information limit, and explained the relationship between information dynamics and factors such as information flow, transmission process, statistical state, physical concept, theoretical expression and actual metrics [20]. However, as only *entropy* or information quantity is considered as the metric or efficacy of information, it is difficult to apply the information dynamics described in [20] to the analysis and study of various information systems. The concept of *Information System Dynamics* has also appeared for a long time [21]. A. Flory and J. Kouloumdjian have studied the design of a database model in the name of Information System Dynamics [22], which is far from the traditional concept of dynamics. Ahmed Bounfour and Surinder Batra introduced an international research project on information system dynamics in the fields of information system and information technology acting on business models, human resources and social organizations [23], which did not mention the idea of applying basic mathematics to establish information metric and efficacy system and form the basis of information system dynamics research.

Because of the failure to form a mathematical basis with universal significance and supporting the analysis of many information movement mechanisms, a reasonable research paradigm of information science has not yet started [24]. Aiming at these fundamental problems, this paper reviews and summarizes the rapid development of information technology, characterized by the close combination of computer and network communication. Such retrospections of information technology in the context of information movement and applications, as well as the interrelationship between the real-world, information space and information system, leads to the analyses of the important role of a series of technological achievements, which arouses the integrated framework of the real-world and information system, and the modifications and improvements of our previous mathematical theory on information models, properties and metrics [25] [26]. Based on these mathematical foundations, eleven types of information metric efficacies and their distribution across information systems are put forward, and then the dynamics configurations of information systems are comprehensively analyzed, which constitutes the basic theoretical framework of information systems dynamics with universal significance. Finally, Smart Court SoSs Engineering Project of China are introduced as the exemplified application of the theoretical work in this paper, which aims at providing a reference for the analysis, design, development and evaluation of large-scale complex information systems.

# 2 Problem Statement

Like material and energy, information is the blood, food, and vitality upon which our world operates [27]. However, different from material and energy, people still lack a comprehensive and quantitative understanding of how information drives the development and change of things in the world. Compared with the mathematical formula of how many calories a given weight of food, how much force can make an object of a certain mass produce acceleration and so on, the consensus on measuring information is limited to the measure of bit, so it is difficult to understand the intricate mechanism of information, let alone to establish a complete, quantitatively rigorous and universally applicable theoretical system of information systems dynamics (ISD) for information flows, just as the foundation theoretical systems for material flows and energy flows.

## 2.1 Why do we need ISDs

According to L. V. Bertalofi, the founder of general systems theory, a system is an organic whole with certain functions, which is composed of several elements connected in a certain structural form[28]. For hundreds of years since the beginning of the Industrial Age, machinery, construction and power systems, such as automobiles, ships, airplanes, high-rises, electric telephones, electric trains, have become indispensable to people's normal lives and can no longer leave them. There is no doubt that the traditional system dynamics theory, which is composed of the basic mathematics such as calculus, differential equation, Newtonian mechanics, thermodynamic law, Maxwell's electromagnetic field theory and other classical physics, has played a vital role in supporting and leading the analysis, design, construction, integration and evaluation of mechanical, architectural and power systems in the industrial age. Such methods are also applied to other fields other than dynamical systems, such as chemical kinetics [29], economic kinetics [30], etc., and become effective tools for analyzing operational mechanisms in specific fields.

From the middle of the twentieth century, with the rapid development of computer and communication technology, especially the advent of the Internet, human beings into the information age. The extensive application of information technologies such as cloud computing, big data, mobile interconnection, supercomputing, blockchain and artificial intelligence has fundamentally changed the production and life style of human beings compared with industrial civilization, and all kinds of information systems are far superior to traditional electromechanical systems in terms of geographical coverage, people involved and business fields with penetration. But people's understanding of information and the mechanism of information system is far less clear and rigorous than the traditional electromechanical system. Different from the three laws of Newton and thermodynamics, which support the industrial civilization, the three laws of the network in modern times, namely, Moore's law, Metcalfe's law and Gild's law, are empirical formulas based on judgment. It can be said that information systems have developed spontaneously to such a large scale as they are today, rather than under the guidance of various classical laws of dynamics, as the traditional electromechanical systems. It is enough to see that there is only one measure of information, called the amount of information, or capacity, that can be accepted. It is difficult to accurately measure an object without adequate and effective measurement, and it is also difficult to form a quantitative action mechanism or kinetic theory system.

In the absence of the guidance of ISD theoretical system, the lack of quantitative analysis and constraints on various components in the development of large-scale information systems is common, and the result is that either the system is difficult to integrate, resulting in the failure of development, or the whole system is inefficient due to the limitations of certain links after the development is completed; moreover, in the case of large-scale and complex information systems that are being developed everywhere, such as smart cities and smart societies, due to the lack of universal structural morphological specifications and evaluation indicator systems, similar information systems in different cities and fields have adopted a large number of low-level redundant technologies, and it is difficult to learn from each other's strong points to offset their weaknesses, which generally leads to isolated system islands with inconsistent interfaces, interconnected interfaces and information interconnection, and at the same time significantly reduces the usability and easiness of the entire information system by consuming a lot of manpower and cost that could have been saved; and more importantly, due to the lack of quantitative analysis and adequate research on the effectiveness of many advanced technologies applied to information systems, there are plenty of examples of such problems as moral, ethical and legal issues arising from blind application of information systems. Whether the future of artificial intelligence technology will bring humans the fear of unpredictable disasters, in fact, the lack of ISD theoretical system may be related.

Therefore, in today's era when information systems are ubiquitous and large-scale information systems play a decisive role in the future development of human society, we must, as soon as possible, base ourselves on the achievements of the development of information technology, apply a complete range of basic mathematical theories, and construct an ISD theoretical system, so as to guide and regulate the sound, orderly and scientific development of large-scale information systems or information system systems by using mathematically-based, quantitative and systematic methods, and lead the research and development of advanced information technologies, so that they can maximize the benefits of mankind, rather than the reverse.

**2.2 What kind of ISD do we need**

Qian Xueshan proposed that the structure of modern science and technology should have three levels: one is the engineering technology which directly transforms the objective world, the other is the technological science which directly provides the theoretical basis for engineering technology, and the third is the basic theory which further generalizes and reveals the objective law of things on the basis of technological science [31]. Today there is a great need for ISDs at the basic science level, specifically meeting the following conditions and content requirements:

First of all, the most concise and universal bridge between information concepts and mathematical theory should be erected. Marx once emphasized that a science is really developed [32] only when it reaches a point where it can use mathematics successfully. The aim of constructing ISD is to support the quantitative, computable and deducible information system analysis and research by using abundant and powerful mathematical tools. Information is the main resource of information system, so the first premise of constructing ISD is to use mathematical method to define the concept and model of information. This concept and model should be very concise, so that people can understand the information at a glance from the perspective of mathematics. Also, it should be universal, including all the content and form of information, so as to meet the needs of analysis and research of information systems in all walks of life.

Secondly, it should be derived from mathematical concepts and models to form a measurement system of information. It can be sure that only information quantity or information entropy is not enough to support the analysis of the mechanism of complex information systems. And any action mechanism must be reflected in the influence and change of the relevant measurement. Therefore, it is necessary to form an information measurement system based on the mathematical concept and model of information, which is closely related to the value of information, generally applicable to information collection, transmission, processing, storage, application and various combinations among them, practical and maneuverable, can guide the analysis and research of information system, then can meet the basic requirements of constructing the theoretical system of ISD.

Third, the components of a general information system should be reasonably deconstructed and summarized. The basic way to construct ISD supporting information system is to analyze and study the movement and function of information flow among the elements of information system. Therefore, only by forming a concept system of information system, elements, structure and function with clear definition, complete elements and progressive decomposition, supporting the relationship between descriptive elements and elements, elements and system, system and environment and other aspects, and fully reflecting the common basic characteristics of information system such as integrity, relevance, timing, hierarchical structure and dynamic balance, can we fully and accurately understand the working mechanism and effect of information system.

Fourthly, we should deconstruct the function of information system according to the measurement system and the elements of information system. The mechanism of information system must be embodied by the measurement of the influence and change of each component on the mutual information. Some aspects of the system may primarily change some metrics that flow through the information, with little or no impact on others. Therefore, it is necessary to pay attention to the impact and degree of change that each system link may have on each measure of the information flowing through it. Because these effects and changes are specific to a particular measure, they can be understood as a measure of efficacy, or simply as measurement efficacy. In fact, the overall function and performance of the information system is the superposition of all these metrics, so the deconstruction analysis of metrics is the basic content of ISD.

Fifth, typical dynamic configurations of the systems of information system should be studied according to common or important application scenarios. If the ISD is compared to a building of great scale, then the measurement efficiency of each system link is the ISD brick by brick, while the dynamic configuration is the whole or part of the ISD structure. The dynamic configuration is composed of various system links and information flows, and any combination may have a completely unique dynamic configuration. Therefore, it is necessary to highlight the key points, choose the typical combination of system links and information flow, use quantitative, computable and deducible methods to study the effectiveness of information measurement from beginning to end, and form a way to analyze and evaluate the effectiveness of the whole system.

Sixth, incorporate a large number of existing classic principles of information technology and commonly used methods. It is not because there is no theoretical or methodological guidance that information technology has been able to bring about such dramatic change. On the contrary, in the course of the rapid development of information technology, new theoretical systems such as information theory, cybernetics and system theory are increasingly enriched, and technical methods such as Internet, big data, cloud computing, supercomputing, Internet of Things, blockchain, artificial intelligence, virtual reality, etc. are emerging one after another, the only lack of which is a unified, rigorous and universal ISD theoretical system. Therefore, based on extensive information measurement efficacy and typical information system dynamic configuration, it can easily absorb the existing information technology theories and methods, enrich the ISD theory system, and provide solid, unified and universal basic science support for engineering technology and technical science.

Finally, explore new theories and methods, with the development of the times to improve the ISD open architecture. The practice of rapid development of information technology has shown that engineering technology, technical science and basic science do not have a strict sequence of development, and practice and theory have always been mutually reinforcing and blending. Therefore, ISD should not be a shackle of closed and rigid theory, but should adopt a completely open system structure, which is based on the theories and methods of mathematics, physics and many other fields, and can absorb the latest achievements of scientific and technological innovation and exploration of all walks of life quickly and fully, so as to make it a mature and complete system of theories and methods in line with the development trend of information technology.

## 2.3 What is the basis to develop the ISD

The great achievements in the development and application of information technology, Shannon's classical research methods, and the mathematical definition, model and measurement of objective information theory have laid a solid foundation for us to construct the theoretical system of ISD.

Thanks to the close combination of computer and digital communication technologies, after the birth of the Internet, information technologies have developed vigorously, and the results of technology application in such aspects as big data, cloud computing, mobile interconnection, supercomputing, block chain, artificial intelligence, Internet of Things, virtual and extended reality emerge in endlessly, and permeate into every corner of human society. Internet technology supports the smooth flow and wide application of objective knowledge and information, big data technology supports the distributed storage, giant capacity management, efficient retrieval, strong association expression and fine image of information, cloud computing technology supports the efficient processing, efficient storage, remote call and remote service of information, mobile interconnection technology supports the extensive collection and casual use of information, supercomputing technology supports the massive, parallel and high-speed processing of information, blockchain technology supports the highly reliable use of information, artificial intelligence technology supports the automated use and high adaptability service of information, Internet of Things technology supports the extensive collection and flow application of objective physical information, and virtual and extended reality technology supports the high fidelity and diversification function of information. The Chinese idiom "true knowledge comes from practice", so abundant information technology application practice and the achievement can certainly produce the broad and profound ISD theory achievement.

Published in 1948, " A Mathematical Theory of Communication" [错误!未定义书签。], Shannon is regarded as the first work of information theory. Because the research object is the communication system, the author first decomposes the communication system into five parts: the source, the transmitter, the channel, the receiver and the signal terminal, and analyzes the message types that flow in the system. Furthermore, it is assumed that the discrete source is a Markov process of selecting successive symbols according to a certain probability value, and three important conditions for information measurement are proposed. Furthermore, the related discussions are extended to the more complicated cases such as the capacity of discrete channel with noise, capacity of continuous channel, generation rate of continuous source, etc. On the basis of Shannon, the results of information entropy research are so magnificent that they fail to break through the basic constraints of Shannon's view of information. There is no doubt that Shannon's information theory itself has epoch-making significance, its theoretical level and practical role are extraordinary. But as a great master of science, he decomposes and induces practical problems, uses mature mathematical definitions and tools to obtain typical mathematical formulas and scientific laws through a series of assumptions, and thus lays the scientific foundation of classical information theory, which should be respected and learned by later generations. Shannon lived in a time when there was no Internet, no large-scale and complex information systems, and no scientific and theoretical tools to solve today's problems could be expected from our predecessors. But as long as we fully understand and inherit the thinking paradigms and scientific methods of our predecessors, we will be able to build an ISD basic science system to solve complex problems in information system engineering.

With regard to the key issues that affect the research paradigm of information science and rationality, such as the lack of universally recognized mathematical basis, the lack of a clear and abundant measurement system for information value, the lack of a scientific and reasonable framework structure for information space, and the lack of a clear and clear function analysis for information, Xu Jianfeng and others, by using the basic mathematical theories such as set theory, mapping theory, measure theory and topology, put forward the definition of information, the six-tuple model and the measurement system [错误!未定义书签。][错误!未定义书签。], among which, the six-tuple model, although simple, deconstructs the information concept in three important ways: firstly, for the binary deconstruction of information subject, based on the characteristic of information reflection, the subject of information is described by binary structure of ontology $o$ and carrier $c$.; secondly, for the time dimension deconstruction of information, the time of occurrence $T_h$ and the time of reflection $T_m$ are introduced to support the analysis and research of information movement from time dimension.; and thirdly, For the deconstruction of information content, two factorization sets, the state set $f(o, T_h)$ and the reflection set $g(c, T_m)$, are introduced to accommodate all information content and form. It can be said that the six-tuple model has built a simple and universal bridge between information concept and mathematical theory. Based on the five basic properties of information, such as objectivity, reducibility, transitivity, combination and relevance, this paper proves the relevant mathematical inference, especially the important inference that the dual state of reducible information can maintain the mathematical isomorphism, which opens the door for extensive information science research with the support of rich mathematical theories. On this basis, according to 11 kinds of information measurement definitions, the information model and measurement system are proposed and proved to be not only consistent with many classical principles of information technology, but also more inclusive and universal, which can provide a solid mathematical theoretical basis for the comprehensive and systematic analysis and study of information science, information technology and information systems. According to the relationship among the real world, information space and information system, this paper also puts forward the integrated framework of the real world and information system, and eleven kinds of information measurement are taken as traction.

# 3 Retrospection of the Development of Information Technology

With the continuous development, information technologies such as networking, Big Data, cloud computing, artificial intelligence and others tend to converge in depth. Today, information technologies and their applications are ubiquitous, and the intelligent stage characterized by deep data mining and fusion applications is opening [33]. In the new world picture of "digitalization of all things", the integration of information space and physical world forms a new human-machine-material integration computing environment, which promotes the application of human-machine-material integration such as intelligent justice, smart cities and intelligent manufacturing.

This section will take the chronological order of the development of information technologies as the main line, and review the development process of major information technologies such as networking and communications, Big Data, blockchains, cloud computing, artificial intelligence, vision computing and extended reality.

### 3.1 Networking and Communications

Networking technology originated from the establishment of Shannon's information theory and the continuous efforts of human beings to connect computers into networks [34] [35] [36]. The theoretical foundation of networking technology was established by Claude Shannon's information theory in the 1940s, which still plays a great role in guiding the basic research and development of network systems. Shannon information theory, developed from the work of Harry Nyquist and Ralph Hartley in the 1920s, describes the trade-off between signal-to-noise ratio, bandwidth, and error-free transmission in the presence of noise in telecommunications technology. In addition, breakthroughs in transistor technology (especially MOS transistors) and laser technology have enabled the rapid growth of bandwidth networking technologies in the past half century [37]. Domain name resolution system and TCP/IP protocol stack are widely used around the world, and computer networks worldwide are constantly integrated and connected, eventually forming the global Internet.

（1）**Early Networking Technologies**

The earliest networking prototypes originated from the mainframe computer architecture in the early 1940s. A mainframe computer has a central mainframe and user terminals that are directly connected by wire for point-to-point communications. During this period, the networking technology was mainly used to solve the time-sharing problem among computer users [38]. However, the capabilities of the point-to-point communication model are limited because it does not allow direct communications between any two arbitrary systems. At the same time, point-to-point communications is strategically and militarily vulnerable, and the interruption of a single link can lead to the overall paralysis of the whole network.

Connecting different physical networks into a logical network, and then improving the coverage of information, is the main problem that network technology has been constantly seeking breakthroughs. Early networks used a message exchange model that required a rigid routing structure and was therefore prone to a single point of failure. With the development of electronic and communication technologies in the 1950s, communications over longer distances (for terminals) or at higher speeds (for local device interconnections), which were necessary for mainframe computer architectures, were gradually realized. In the early 1960s, Paul Baran proposed a distributed network based on message block data [39] when he studied how the US military network could survive in a nuclear war. At the same time, Donald Davies of the National Physical Laboratory in the United Kingdom proposed packet switching, a fast store-and-forward network design that divides messages into arbitrary packets and makes routing decisions for each packet. Compared with the traditional circuit switching technology used for telephony, the packet switching technology can improve the bandwidth utilization on the interconnection links with limited resources, increase the information volume of networks, thereby shortening the response time and improving the delay of network systems [40].

During this period, the United States and the United Kingdom have carried out research and experimental projects on networking technologies. These experimental networks continue to evolve and merge, and the scale of network distribution continues to expand, effectively improving the coverage of information in the information transmission systems. In 1969, the Advanced Research Projects Agency (ARPA) launched the ARPANET project, which used the packet switching technology proposed by Davies and Baran to build the ARPANET network [41]. In 1978, Donald Davies led the research team to adopt the packet switching technology and put forward the concept of "network protocol", and built a practical NPL local area network (Mark I) to realize the interconnection of computers in laboratories in different regions [42], which has developed into a hierarchical network protocol system. The ARPANET network and the NPL network are the earliest operational packet switching networks, which verify the feasibility of this technology under actual working conditions. In addition, Merit network project of Michigan Educational Research Information Research Association in the United States [43], CYCLADES network among French research institutions [44], SERCnet network project among British academic institutions (later JANET network) [45], the extensive research by the UUCPnet project of Duke University based on Usenet [46] gradually promotes the key networking technologies such as Ethernet technology [47], ITU-25 T X. Standard system and virtual telephone line technology [48], international packet switching service, TCP/IP protocol stack [49], etc. These effort enriched the types of information transmission and provides sufficient technical preparation for the final formation of the global Internet.

（2）**Internet**

The term "Internet" originated in the first version of the TCP protocol (RFC 675: Transmission Control Procedures for the Internet) published in 1974 [50]. In general, the Internet is a collection of networks connected by a common protocol. In the late 1980s, the National Supercomputing Center funded by the National Science Foundation (NSF) of the United States carried out network interconnection and built the NSFNET network. In 1981, ARPANET was connected with NSFNET network. From 1984 to 1988, CERN began to build the CERNET network based on the TCP/IP protocols [51], and after 1988, the pan-European CERNET network began to connect to the NSFNET network in the United States [50]. In 1991, JANET in Britain, EARN network, RARE network and EBONE network in Europe successively adopted the TCP/IP protocol architecture and connected with NSFNET network in the United States [49]. After entering the 1990s, Asia, Africa, South America and other countries began to set up TCP/IP networks, and connected with the networks of the United States and Europe, forming a global network, greatly expanding the coverage of information transmission systems of every country. As more and more countries and institutions are connected to the US NSFNET network, the "Internet" is used to refer specifically to this global communication network based on the TCP/IP protocol stack [50].

One of the key technologies of the Internet is optical fiber communication. Network connectivity on a global scale creates challenges in the transmission of very large volumes of data that traditional radio, satellite, and analog copper cables cannot address. In 1995, Bell Labs developed wavelength division multiplexing (WDM) technology based on laser and optical amplifier technology, which effectively solved the problem of insufficient channel bandwidth of traditional information transmission systems, effectively improved the delay performance of information systems, and ensured that the sampling rate of digital signals was not reduced [52]. Since then, optical fiber communication networks have gradually become the main technology for long-distance communication.

The World Wide Web (abbreviated as "www" or "Web") is an Internet-based information space. In which documents and other resources are identified by URIs, linked by hypertext links, and accessible using Web browsers and Web-based applications. Nowadays, the World Wide Web is the main tool for billions of people around the world to interact on the Internet, and it has greatly changed the way people work and live [53]. In the short five years from 2005 to 2010, Web 2.0 has promoted the development of the Internet into a transformative force of social systems. Web 2.0 describes the emphasis on user-generated content (including user-to-user interaction), usability, and interoperability. Typical Web 2.0 applications include social networking services, blogs, wikis, popular categories, video sharing websites, hosting services, Web applications, etc. [54], which greatly enrich the types of information services.

（3）**Mobile Networking Technologies**

With the evolution of Web 2.0, mobile communication technology has undergone a tremendous revolution, which means that smart phones, as an important computing platform, can be carried and used by people, and the concept of computer is not only a fixed environment. The first generation of mobile network technology (1G) is the automatic analog cellular system, which was first used in the NTT system for car phones in Japan in 1979, and later in the NMT system released in the Nordic countries, and the AMPS system widely deployed in North America [55]. The 1G mobile networks use the frequency division multiple access (FDMA) technology scheme, which requires the support of a large amount of wireless spectrum, requires high channel capacity, and has poor security [56].

In the 1990s, mobile communication technology began to realize digitalization and entered the second-generation mobile network era (2G). During this period, the GSM standard developed in Europe and the CDMA standard developed in the United States competed for hegemony in the global market [57]. Unlike the previous generation, 2G uses digital transmission instead of analog transmission, and also provides fast out-of-band networking signals, guaranteeing the sampling rate of digital signals. The 2G era has witnessed the emergence of SMS, MMS, mobile payment and other types of information. In 1999, NTT DoCoMo mobile phone operator in Japan launched the first full mobile Internet service in the world [58].

Driven by more and more mobile applications, people's demand for data is increasing rapidly, which makes 2G technology unable to meet the demand for higher data speed. Mobile networking technology began to use packet switching instead of circuit switching for data transmission, and mobile networks entered the era of 3G technology. In the 3G era, there are many competing technical standards, including EV-DO, CDMA2000, WCDMA, TD-SCDMA, GPRS, EDGE, HSDPA, HSPA, UMTS [59]. Different competitors are pushing their own technologies to improve channel volume, but there has been no unified global standard in the 3G era. The high-speed wireless connectivity of 3G technology has significantly improved the delay of the systems, making it possible to stream radio (and even TV) content by 3G mobile phones, prompting a profound transformation in news, entertainment and other related industries [60].

With the rapid growth of high-bandwidth services such as streaming media in mobile networks, by 2009, the information volume of 3G networks has begun to be unable to meet the needs of intensive applications [61]. Therefore, the industry began to seek the fourth-generation technology (4G) of data optimization, whose main technical goal is to further enhance the frequency of information transmission and increase the information capacity to 10 times that of 3G. The main technologies of 4G mobile networks are WiMAX standard [62] and LTE standard [63]. One of the main technical differences between 4G and 3G is that it no longer uses circuit-switched solutions, but adopts all-IP network solutions. As a result, the 4G network is able to transmit streaming media data via VoIP as it does on the Internet, LAN, or WAN networks.

5G is the next generation version of mobile networking technology, which has entered commercial use. The 5G standard includes millimeter-band radio spectrum, allows bandwidth volume of up to 1 gigabit per second, and reduces the latency (processing time to handle data transmission) between the handsets and the networks to a few milliseconds [64]. The powerful channel volume (up to 10Gb/s) and low delay of 5G open the door for applications that rely on real-time data transmission (AR/VR, cloud games, connected vehicles) [65]. Through the 5G technology, vehicles equipped with powerful computing devices and advanced sensors can achieve high-frequency interaction of information flows, making automatic driving likely to become the most popular scenario in our daily commuting.

（4）**Internet of Things**

With the continuous maturity and integration of the radio frequency identification (RFID) technology, the sensor technology, embedded systems, wireless networks and other technologies, the concept of "Internet of things" (IoT) was proposed by Kevin Ashton in 1999 [66]. Subsequently, this concept has been enriched and improved with the evolution of the information technologies. It is generally acknowledged that the Internet of Things is a new type of communications between people and things and between things by embedding short-range mobile transceivers in various necessities of production and life. Compared with the Internet, IoT further improves the information coverage of networks. At present, the Internet of Things has become a global network infrastructure like the Internet. Its main technical features include standardized interoperability communication protocols, self-configuration capabilities of devices, unique identification mechanisms, integrated interfaces, etc. [67], in order to reduce the networking latency and support different types of IoT information interconnections.

The wide range of applications of IoT technologies typically involves consumer, commercial, industrial, and infrastructure applications [68]. Among them, IoT devices for consumers include smart home, wearable technologies, etc. [69]; the main commercial applications of IoT include Internet of Vehicles (including vehicle-to-vehicle, person-to-vehicle, vehicle-to-road interconnection), smart transportation, smart medical care (health), smart buildings, etc. [70]; The main industrial applications of the IoT include equipment asset registration, smart factories, smart agriculture, and smart marine industries [71] [72]; the main applications of the IoT infrastructure include smart cities, energy management, and environmental monitoring [73] [74] [75]. In addition, the sensor technology will have a great impact on future urban warfare, involving the use of sensors, munitions, vehicles, robots, human wearable biometrics and other intelligent technologies related to the battlefield [76]. The integration of the IoT and augmented reality can combine the real environment with immersive augmented reality content to produce the shocking multi-mode interactive user experiences [77], enhance the ability of information space to reflect the objective world, and present information to the world in an unprecedented scope.

（5）**Internet of Data**

Internet and IoT technologies solve the problem of information transmission between man and machine, and provide basic technical means for human beings to enter the digital age. However, the Internet and the IoT technologies cannot avoid and solve many problems brought about by the continuous development of information technologies and their applications, such as information islands, data out-of-control, data rights and so on. The universality of information's effective reflection of everything in the objective world is still facing bottlenecks. In order to achieve data interconnection and interoperability on the basis of data transmission over Internet, application systems need to coordinate themselves and reach agreement on data syntax, semantics and pragmatics, and address the challenges such as high coordination costs, difficulties of guaranteeing responsibilities and powers, as well as inefficiency, error-prone and difficulty of recovery. Aiming at these problems that the existing platform-based data interconnection technologies are difficult to improve, Chinese researchers learn from the design concept of the Internet and adopt the idea of software definition [78] to connect various heterogeneous data platforms and systems through data-centered open software architecture and standardized interoperability protocols. As a result, a "virtual/data" network is formed on the "physical/machine" Internet, namely the "Internet of Data" (IoD), thus realizing the data interconnection and interoperability of the whole network integration, and greatly improving the coverage of information reflection on the objective world.

The core of the IoD is the Digital Object Architecture (DOA), which was formally proposed by Robert Kahn at the beginning of the 21st century. The conceptual architecture of DOA includes a basic model, two basic protocols, and three core systems. A basic model refers to the digital object, which is used to abstract the data in the Internet, so that heterogeneous data resources can be modeled and described in a unified form, effectively improving the scope, granularity and variance of information. From a technical point of view, a digital object is a bit sequence or a collection of bit sequences, which contains some valuable information for a person or organization. Each digital object must be assigned a globally unique identity. Identity, as a core attribute of a digital object, does not change with changes in the owner, storage location, or access method of the digital object. The Two basic protocols are Digital Object Interface Protocol (DOIP) and Digital Object Identifier Resolution Protocol (DO-IRP), the former is used to interact with digital objects or information systems that manage digital objects, and the latter is used to create, update, delete, and parse globally managed and distributed digital object identities. Three core systems are the repository system, the registry system and the identifier/resolution system. The digital object repository system is mainly responsible for the entity encapsulation, storage, access and destruction of digital objects, the digital object registration system is mainly responsible for the metadata registration, publishing, modification and deletion of digital objects, and the digital object identification system is mainly in charge of the identification creation, parsing, modification and destruction of digital objects. DOA unifies and standardizes the data resources of the Internet through digital objects, adopts two protocols of DOIP and DO-IRP to standardize the data interaction, and realizes the interconnection and intercommunication of heterogeneous, remote and different data based on the open software architecture formed by the three core systems [79]. It greatly improves the ability of information systems to affect the scope, granularity, coverage and variance of information.

DOA has achieved the global large-scale application in the field of digital libraries, namely DOI system. As of May 2021, the DOI system has registered about 257 million digital objects worldwide, covering many internationally renowned academic databases such as IEEE, ACM and Springer. In 2018, China began to build the world's largest DOA application system, the National Industrial Internet Identification Resolution System. As of August 2021, five national top-level nodes and more than 200 secondary nodes have been built in Beijing, Shanghai, Guangzhou, Chongqing and Wuhan, with access to more than 20,000 enterprises in 25 provinces, municipalities and autonomous regions, marking that the information scope of China's digital networking application is at the forefront of the world.

## 3.2 Super-computing

（1）**Supercomputing Technologies**

At the same time, the characteristics of large data volume, heterogeneity and effectiveness also give birth to high-throughput computing technologies. In the face of the rapid and huge amount of data, in order to deal with the insurmountable computing bottleneck faced by the traditional centralized computing architecture, the distributed computing architecture of Massively Parallel Processing (MPP) has emerged. In the 1960s, the concept of "supercomputer" appeared, which achieved lower delay of information flows by highly adjusting the traditional design to obtain faster operation speed than general-purpose computers [80]. In the 1970s, vector processors, the Cray-1, running on large data arrays began to dominate in 1976. Vector-based computers became the mainstream design of supercomputers in the 1990s [81]. Since then, massively parallel supercomputer clusters with tens of thousands of off-the-shelf processors have become the norm. Since 2017, the frequency of information processing has increased rapidly, and supercomputers can perform more than 1017 floating-point operations (100 trillion floating-point operations). In contrast, the performance of desktop computers ranges from hundreds of gigabits to tens of megabits [82].

（2）**High-throughput computing technology**

For massive unstructured text data, Apache Hadoop and Spark ecosystems based on distributed batch computing framework have emerged, whose main goal is to avoid the low performance and complexity problems encountered by traditional technologies in processing and analyzing large-scale data. Hadoop's ability to quickly process large data sets lies in its parallel clusters and distributed file system. Unlike traditional technologies, Hadoop only loads locally stored data in memory to perform computation without loading all remote data sets in memory, effectively reducing the demand for information volume. Therefore, Hadoop effectively reduces the communication loads of the networks and the servers [83]. The power of the Hadoop platform is based on two main components: the distributed file system (HDFS) and the MapReduce framework. In addition, modules such as enhanced capacity, expansibility and security can be added on top of Hadoop to form a richer ecosystem. In view of the limitations of MapReduce cluster computing model, the Apache Foundation developed the Spark system based on elastic distributed data set (RDD) in 2012 [84]. Spark facilitates iterative algorithms (accessing its dataset multiple times in a loop) and interactive/exploratory data analysis (i.e., repeated database-style data queries). Compared to the Apache Hadoop MapReduce implementation, the latency of Spark programs can be reduced by several orders of magnitude [85].

In order to meet the demand of real-time computing for time-sensitive data, Apache Storm, Flink and Spark Streaming, which are based on distributed stream processing computing framework, have emerged. Apache Storm is a distributed real-time data processing framework, which has a topology structure in the form of directed acyclic graph (DAG) through various types of information receiving and processing primitives, so that data can flow from one processing node to another in a pipelined manner. Compared with MapReduce, Apache Storm can process data in real time and at high frequency, rather than in batches [86]. Apache Flink provides a high-volume, low-latency streaming engine [87] to execute arbitrary data streaming in both parallel and pipelined manner, and thus has the ability to support multi-granularity information processing for batch and stream processing. In addition, Flink's runtime supports iterative algorithms such as local execution, as well as event-time processing and state management [88]. Spark Streaming uses Spark Core's fast scheduling capabilities to perform stream analysis, which can receive data in small batches and perform RDD transformations on the batches. This design enables that the application codes written for batch analysis can be used for streaming analysis, simplifying the implementation of the lambda architecture [89]. Storm and Flink process streaming data in the form of events rather than small batches, which makes the granularity of processing information finer. Spark Streaming's method to process streaming data with small batch not only brings convenience, but also address the problem of large delay [90].

## 3.3 Big Data

Big data and the IoT can work together. Data collected from IoT devices providing connections between physical devices can target audiences more accurately, and improve media efficiency. As the IoT technology is increasingly used, the vast amount of sensor data can be used in medical, manufacturing and transportation environments.

（1）**Large-scale Data Storage and Management**

The database technology emerged in the computer field in the mid-1960s with the advent of direct access storage (disks and drums). In contrast to earlier tape-based batch systems, databases allow for the sharing and interactive use of data. As the speed and power of computers increased, general database management systems began to appear around 1966. The early database technology mainly uses two kinds of data models: the hierarchical model and the network model (CODASYL). They are characterized by the use of pointers (usually physical disk addresses) to keep track of relationships from one record to another [91].

In order to solve the problem that the early database model is difficult to support efficient retrieval, Edgar F. Codd proposed the relational model in 1970 [92]. Based on the relational model, databases can be queried with the Structured Query Language (SQL) that has clearer semantics, and significantly improves the integrity and fineness of the data model, refines the granularity of database management information, and realizes efficient data accesses. However, it was not until the mid-1980s that computing hardware have enough storage capacity and computing power to support the wide deployment of relational database systems [93]. Until 2018, relational systems still dominated a large portion of data processing applications. Then, IBM DB2, Oracle, MySQL, and Microsoft SQL Server are the most widely used database management systems in the commercial market.

In the 1990s, with the rise of object-oriented programming, programmers and designers began to view the data in databases as objects. This perspective on data allowed the relationships between data to be relationships of objects and their properties, rather than relationships of individual items. In order to solve the problem of convenient transfer between programming objects and relational database tables, the object-oriented languages (as the extension of SQL) can be used to replace the traditional SQL statements on the database side; on the programming side, this problem can be solved through the object-relational mapping (ORM) call libraries [94].

In the era of big data, the traditional relational database technology is facing the bottleneck of difficult distributed deployment caused by the large data volume, heterogeneous and diverse data sources. After 2000, non-relational NoSQL databases began to appear. Compared with traditional relational database, NoSQL database can be deployed on the basis of distributed file systems (such as HDFS), which improves the coverage of information management systems, and also makes the processing speed of big data faster, ensuring the high sampling rate of digital information [95]. NoSQL databases include several subdivision types such as key-value stores and document-oriented [96]. The types of information that the systems can manage are further improved.

Because the NoSQL databases do not need the fixed table schema, it can realize the horizontal expansion of database design and improve the sope and performance of information processing. Currently, the next-generation database technology that competes with NoSQL databases is the NewSQL database. The NewSQL database is characterized by preserving the relational model (so that SQL technology can be reused), while providing the scalability of NoSQL systems for online transaction processing (OLTP) workloads, achieving high processing performance of NoSQL. Therefore, NewSQL can provides atomicity, consistency, isolation, and durability guarantees as traditional database systems [97] [98].

（2）**Data Security and Privacy Protection**

sharing and circulation are important ways to release the values of big data, whether by directly providing data query services or fusion analysis with external data. In the current situation of increasingly concerning data security, safe and controllable data flows among different organizations are still lack of effective technical supports. At the same time, with the gradual improvement of the rigors of the relevant laws, data circulation is facing more stringent regulatory restrictions. Compliance issues are restricting data circulation among multiple organizations.

Data encryption is a key technology related to data security. Data encryption is the process of encoding information, where the original representation of information (plaintext) is transformed into another form (ciphertext), thus increasing the sampling rate of information to keep the distortion rate of encrypted information unchanged. In 1790, Thomas Jefferson proposed a cryptographic theory called Jefferson Disk to encode and decode information for a more secure military communication [99]. In 1917, U.S. Army Major Joseph Mauborne developed an encryption device similar to Jefferson's Frisbee, the M-94. The equipment was used for US military communications until 1942. In World War II, the Axis powers used an encryption device called the "Enigma Machine," a more complex device that allowed the daily jumble of letters to be converted into a completely new combination. Public key and symmetric key schemes are widely used in modern cryptography [100]. This encryption method can provide higher level of data security because of the high frequency of processing and the lower efficiency required for cracking the information encryption [101].

As a data modeling technique that aims to protect the data itself from external leakage, privacy computing makes it possible to achieve secure and compliant data circulation [102]. At present, under the strong demand for compliant circulation of data, technologies based on privacy computing has become the mainstream for federated data computing. Privacy computing technology can be mainly divided into two schools: multi-part secure computing and trusted hardware. Among them, multi-party secure computing is based on cryptography theory, which can realize multi-part collaborative computing safely without a trusted third party [103]; trusted hardware computing is based on the trust of secure hardware, which constructs a secure area inside hardware, so that data can only be computed in the secure area [104].

By applying the trust mechanism of cryptography or secure hardware, privacy computing technologies can realize the joint modeling of data among multiple organizations on the premise that the data itself is not leaked. In addition, federated learning, shared learning, and other technologies that balance security and computing performance of the privacy protection, are providing new solutions for cross-enterprise machine learning and data mining [105].

（3）**Data Analysis**

Data analysis is a process of examining, cleaning, transforming, and modeling data with the goal of discovering useful information, informing conclusions, and supporting decisions [106]. The goal of data mining is to extract patterns and knowledge from large amounts of data in order to improve the fitness of information. Data analysis involves data management, data preprocessing, modeling and inferencing, measures of interests, complexity, post-processing of discovery structures, visualization, and online updates [107].

Humans have been trying to extract patterns from data for centuries. As early as the 18th century, people began to use Bayes theorem and other methods to identify data patterns, and after the 19th century, regression analysis was used to model data [108]. The proliferation, popularization and increasing power of computer technology have greatly increased the information volume and frequency of operation, enabling semi-automatic or automatic analysis of large amounts of data to extract knowledge of greater scope and finer granularity. With the development of data mining technology, the traditional analysis methods based on independent data sets is becoming more and more mature. The typical data analysis methods include cluster analysis [109], anomaly detection [110], dependencies, etc., which aim to improve the degree of information aggregation. Among them, correlation analysis, including association rule mining [111] and sequential pattern mining [112], is one of the most important applications of current data analysis. Association analysis began in the 1990s, which is derived from the "shopping basket analysis" problem of discovering shopping behavior patterns from customer transaction lists. In classical machine learning, frequent pattern mining algorithms such as Apriori and FP-growth are used to realize the mining and analysis of association rules [113], making information services more suitable for users' needs. Sequential pattern mining is the discovery of sequentially statistical relation patterns between data samples. In general, sequential values are assumed to be discrete, so sequential pattern mining is closely related to time series mining [114].

For the data of social networks, user behaviors, web page links and so on, the association relationship is usually in the form of "graph". The data mining analysis of graph structure is difficult to be realized by traditional data analysis methods such as classification, clustering, regression and frequent pattern mining. Therefore, graph analysis technology for graph structure data has become a new direction of data analysis technology, which can explore and explore the unknown relationship contained in the graph structures and fully obtain the associated relationship in the graph structures [115], which can further improve the degree of information aggregation.

## 3.4 Blockchains

The essence of blockchain is the distributed data storage and management, where data is stored in distributed "blocks" that are linked in chronological order. Data on blockchains is synchronized with other devices on the peer-to-peer network through the consensus mechanism, which makes it open, tamper-resistant, transparent and traceable [116]. In the digital age, the digital twin of the physical world can produce an unimaginable amount of data. Such a huge amount of data is almost impossible to store and manage in a centralized way. Therefore, the distributed networks of trusted collaboration built on blockchains are expected to connect everything in the digital space and become the trust infrastructure of the digital society.

（1）**Distributed Accounting**

The earliest blockchain technology was first proposed by cryptographer David Chaum in 1982 [117]. In 1991, the signed information chain was used as an electronic ledger for digitally signing documents in such a way that all signed documents in the collection could be easily displayed and made unalterable [118]. However, encrypting information in the temporal dimension will increase the frequency of information processing. In 1992, Haber, Stornetta, and Dave Bayer introduced the Merkle Tree into the blockchain design to improve the efficiency of the systems [119]. In 2008, Satoshi Nakamoto (a pseudonym) gave a complete description of "blockchain" in the article "Bitcoin: Peer-to-Peer Electronic Cash System" [120]. In 2009, the modern cryptocurrency scheme described in Satoshi Nakamoto's paper were put into practice, marking the birth of Bitcoin, and later derived many cryptocurrency applications of blockchains.

Unlike traditional relational databases, A blockchain is a distributed digital ledger composed of blocks, which are usually used to record transactions between multiple parties. These transaction data are hashed and encoded into a Merkle Tree. The different blocks are arranged in chronological order, and the cryptographic hashes of the preceding blocks are included in the subsequent blocks, thereby connecting the different blocks to form a chain, viz. a blockchain. In theory, blockchain can grow indefinitely, so the performance of information volume, scope, granularity and duration stored and managed by blockchains can increase indefinitely. In blockchains, the latter block confirms the integrity of the former ones by hashing, and this iterative process goes all the way back to the initial block, thus ensuring that all the information on the blockchain is difficult to change retrospectively. Blockchain is autonomously managed using peer-to-peer networking and distributed timestamp servers, and verified by large-scale consensus. An effective distributed consensus mechanism can eliminate the security uncertainty of the peer-to-peer network users, solve the unavoidable "Byzantine General" problem of distributed databases [121], and ensure the high fidelity of information on the chain.

Currently, there are mainly four types of blockchains: public, private, federated, and hybrid. For the public chains, there are no access restrictions, and consensus verification is provided through proof-of-interest or proof-of-work algorithms. Currently, the most well-known public chains include Bitcoin and Ethereum. The private chains can only be joined by invitation from an administrator, and access to both participants and validators is restricted. The alliance chains are in between the public chains and the private chains, only for the members of a specific group and a limited third party, and a number of pre-selected nodes are designated as bookkeepers. Moreover, the hybrid chains are with the characteristics of centralized management and decentralization at the same time, and have different management and operation modes.

Blockchains provide a secure platform for data interoperability, enabling different trading parties to share data securely. Cryptocurrencies based on interoperable blockchains are also called the fungible tokens (FT). Compared with the homogeneous tokens, NFT without interoperability can also be constructed on the basis of blockchains [122].

（2）**Consensus Technologies**

A fundamental problem in distributed computing and multi-agent systems is how to achieve overall system reliability in the presence of a large number of faulty processes. This usually requires a coordination process to reach consensus, or to agree on certain data values needed in the calculation process, the so-called consensus mechanism. The idea of electronic currency existed before Bitcoin (such as ecash and NetCash), but it could not be widely used because it did not solve the problem of distributed consensus. The core distributed consensus technology concept of blockchain emerged in the late 1980s and early 1990s. In 1989, Leslie Lamport proposed the Paxos protocol and published the consensus model in ACM Transactions on Computer Systems in 1990. Used to agree on results when the computer network or the network itself may be unreliable [123].

Blockchain's consensus mechanism A decentralized self-regulatory mechanism that ensures that only valid transactions and blocks are added to the blockchain. Blockchain mainly uses consensus mechanism based on time stamp, such as proof of workload, proof of rights and interests, to realize the time sequence of information change. However, the main problem of this method is that the computational overhead is too large and the efficiency is low, which leads to the delay of the system is too large, so it is difficult to apply in practical scenarios [124]. In 2021, the Institute of Software of the Chinese Academy of Sciences and the New Jersey Institute of Technology proposed the world's first fully practical "Dumbo" Byzantine fault-tolerant algorithm (DumboBFT), which significantly reduced the latency performance of the blockchain in the actual production environment [125].

In the digital age, around the trusted circulation of data elements, the block chain has entered a new stage of development oriented by "trust chain" and "collaboration chain". At present, the consensus mechanism based on Byzantine Fault Tolerance (BFT) is the mainstream choice for all kinds of blockchain. In order to achieve higher performance, some alliance chains adopt Raft's Crash Fault Tolerance (CFT) consensus to ensure the global consistency of data, and the information processing capacity of the system is greatly improved. At present, the highest transaction performance of this system has exceeded 100,000 [126]. However, the CFT-like consensus further weakens the decentralized degree and consensus fault-tolerant ability of the alliance chain system, which leads to the trend of the chain system towards distributed databases. Most blockchain platforms can support a variety of consensus algorithms and select switching on demand to meet the differentiated needs of capacity, latency, and security.

（3）**Smart Contracts**

A smart contract is a computer program or transaction protocol designed to automate, control, or record law-related events and actions according to the terms of a contract or agreement [122]. The goal of smart contracts is to minimize malicious fraud and enforcement costs of contracts, as well as the need for trusted intermediaries and arbitration. The concept of "smart contract" originated in the early 1990s. In 1998, Szabo proposed that a smart contract infrastructure could be implemented by replicable asset registration and contract execution using cryptographic hashing and Byzantine fault-tolerant replication [127]. Askemos implemented this approach in 2002 using Scheme as the contract scripting language [128]. In 2014, a white paper for the cryptocurrency "Ethereum" described the Bitcoin protocol as a simplified version of the smart contract concept. Since Bitcoin, various cryptocurrencies have supported scripting languages that allow smart contracts between untrusted parties [129].

Similar to value transfer on the blockchain, deploying smart contracts on the blockchain is achieved by sending transactions from the blockchain. The transaction includes the compiled code of the smart contract as well as the recipient address. This transaction must then be included in the block added to the blockchain. At this time, the code of the smart contract will execute to establish the initial state of the smart contract [130]. Distributed Byzantine fault-tolerant algorithms are used to prevent attempts to tamper with smart contracts. Once a smart contract is deployed, it cannot be updated. Smart contracts on blockchain can store arbitrary States and perform arbitrary computations, and terminal clients interact with smart contracts through transactions [131].

Smart contracts enable multi-party transactions and provide provable data and transparency that promote trust, enable better business decisions, reduce reconciliation costs that exist in traditional business-to-business applications, and reduce the time to complete transactions. At present, smart contracts are widely used in bonds, birth certificates, wills, real estate transactions, labor contracts and many other fields.

(4) **Cross-chain Extensions**

With the integration and innovation of block chain technology with 5G, artificial intelligence, big data, cloud computing and other new technologies, the "block chain +" which integrates other new information technologies has increasingly become the consensus of the industry. The large-scale and deep-seated application of block chain technology has strengthened its position as a cross-industry and cross-technology integration hub.

The combination of block chain and the Internet of Things can realize the trusted link between the physical and digital world, and improve the breadth, granularity and universality of the information space reflecting the real world. The secure and trusted execution environment of IOT terminal equipment can solve the problems of IOT terminal identity confirmation and data right confirmation. Meanwhile, the accuracy and non-tampering of blockchain records can promote the standardization of data market transactions and accelerate industry integration and innovation [132]. Based on cloud computing, through integrated development tools, intelligent contract management, automated operation and maintenance, digital identity, cross-chain services, etc., to achieve the bottom of the block chain and application one-stop development and deployment. Reduce blockchain response latency and application development and deployment costs [133]. Block chain and privacy computing integration, block chain can increase credit for multi-party collaboration process, privacy computing can make data available and invisible, the combination of the two can not only ensure that the whole process of data sharing can be verified, traceable and auditable, but also effectively protect data from leakage. It can be applied to data generation and collection validity verification, data processing certification and consensus, data use authorization, data flow, data collaboration, data audit, etc., providing an effective solution for the trusted flow of information flow [134].

With the continuous expansion of the breadth and depth of block chain applications, the problems of cross-chain difficulties between different block chain platforms, the difficulties of switching between the upper application system and the lower chain, and the difficulties of trusted interaction on and off the chain are gradually highlighted. At present, the main challenge of blockchain is the inter-platform interoperability technology that affects the information coverage, including notary mechanism, side chain/relay chain, hash time locking, distributed private key control, etc. [135].

## 3.5 Cloud Computing

Digital transformation has become a "necessary option" for the future of our society. As the service model based on virtualization and networks, cloud computing provides users with distributed resources such as computing, storage, data and applications [136]. Based on cloud computing platforms, the integration of big data, artificial intelligence, blockchains, digital twins and other new generation of digital technologies has become a sufficient and necessary condition for the digital transformation of the economy and society.

**(1) Cloud Native Technologies**

The concept of "cloud computing" first appeared in the literature of Compaq's internal documents in 1996 [137]. In 2002, Amazon began to provide commercial cloud services (Amazon Web Services) to enable developers to build innovative and entrepreneurial applications on their own [138]. In 2008, Google released Google App Engine, a service that allows users to quickly deploy and scale applications on demand. The native technologies of cloud computing include virtualization, multi-tenant supports, automatic deployment of operation and maintenance, microservices, containers, software development-IT operation and maintenance integration (DevOps), etc. Cloud computing can decouple the underlying hardware and operating system through the virtualization technology, which self-maps the physical hardware of the information systems in the information space, and increase the scope and granularity of the system's own information [139]. On this basis, containerization technology uses virtualized hardware resources to achieve elastic expansion of the system's memory, network and other information capacities [140]. Through the introduction and application of the microservice technology, the system downtime caused by a single point of failure is avoided, and the duration capacity of information processing is extended [141]. After containers and microservices respectively realize the fine-grained distinction of resources and applications, the serverless technology refines services to the functional level, and significantly improves the high-speed iteration of application development, rapid product deployment and update capabilities through the encapsulation and orchestration of functions [142].

At the same time, cloud native technologies reconstruct the traditional software development and operation/maintenance mode, enabling continuous delivery that covers the whole process of software development and delivery through the organic combination of the testing infrastructure. Based on the consistent application environment provided by the containers, the loosely coupled application development framework provided by the microservice architecture, the ability of independent iteration and deployment of applications, and the one-stop application development, operation and maintenance platform provided by DevOps, the scope of information processing extends to software coding, hosting, construction, integration, testing, publishing, deployment and operation and maintenance [143]. By integrating testing with development, operation and maintenance, the software development cycle is significantly shortened and the iteration efficiency is improved.

With the rapid development of information technologies, the expansion of information volume, the diversification of information types and the complexity of algorithms put forward higher requirements for computing resources. Cloud computing can quickly meet the requirements of artificial intelligence, big data and other applications, relying on its powerful elastic expansion, high availability and other capabilities. Product systems based on cloud native technologies are effective in simplifying development process and data management, enhancing processing performance, and accelerating the innovation of various applications such as artificial intelligence, big data and blockchains [144].

**(2) Edge Computing**

Mobile Internet has promoted the wide applications of mobile devices. However, the processing power of graphics and chipsets of mobile devices is usually limited, which leads to the increased latency of systems. Load balancing technologies are often used to reduce the computational and memory burden on mobile devices, but at the cost of further network latency [145]. Effective and transparent load balancing is essential for a high-quality user experience. However, due to the highly variable latency and unpredictable nature of cloud computing, the cloud native technologies cannot always achieve the best balance, and even leads to the long tail latency phenomenon [146]. In 2009, it was found that deploying a powerful cloud infrastructure only one wireless hop away from a mobile device can greatly improve the overall latency performance of the system, and many subsequent works have proved that this solution realized by physical deployment is more realistic than the simple arrangement of cloud computing resources [147]. As a result, the concept of "edge computing" has emerged, which aims at computing, storing and transmitting data physically closer to end users and their devices, in order to reduce the user experience latency of traditional cloud computing.

With the rapid popularization of real-time interactive applications that feature large resolution and high sampling rate of visual information, such as ultra-high definition video and virtual reality, system latency is becoming more and more important. At present, the low-latency advantage of edge computing has been able to effectively improve the latency performance of 16K, 24K and even higher resolution streaming media [148] [149] [150]. The MEC standard proposed by the European Telecommunications Standards Institute (ETSI) deploys the edge servers at the same location as the base station or one hop away from the base station, which are operated and maintained by the network operators in the area. The MEC technology can effectively reduce the round-trip time (RTT) of packet delivery, while support the near real-time orchestration of multi-user interaction [151] [152], effectively improve the delay performance of the system, and lay an important foundation for the future development of 6G.

At the same time, because the edge computing architecture increases the scale of information copies, the objects of information increase, thus improving the overall coverage of the information systems.

**(3) Cloud-network Integration Technologies**

The networking virtualization technology integrates cloud computing, edge computing and all kinds of information resources in the wide area networks, and the deployment of computing powers has shifted from the traditional single data center to the cloud-network-edge distributed architecture with three-level integration of center, region and edge. Under this architecture, the computing powers of the supercomputing centers can be allocated to different computing powers according to the volume of information flows and the level of delay through collaborations within the computing infrastructure, so as to realize the optimization of computing power services [153]. For example, compute-intensive applications should perform high-frequency computing tasks through supercomputing center platforms, while data-intensive applications can use edge computing to reduce system latency and efficiently carry out data access and processing.

Driven by the IoT, 5G and other advances of technologies, cloud-edge collaborations have expanded from the initial central cloud and edge cloud collaboration to the cloud-edge architecture that covers the central cloud, edge cloud, edge devices and the IoT devices, where the computing powers sunk to the user side can further provide the overall latency performance of the systems [154]. At the same time, cloud native technologies such as containers and microservices sink to the edge side. By subdividing the types of information and realizing flexible scheduling of resources, this architecture provides a new way for application deployment in the edge environment with limited resources, heterogeneous devices and complex requirements, which enables the cloud and the edge to play their respective advantages, and enhances the collaboration capability of both the cloud and the networks [155].

With the trend of integration of the cloud and the edges, edge computing powers are deeply coupled with the cloud computing powers. The distributed architecture of computing powers across cloud and edges are gradually replacing the centralized computing power configuration, providing more powerful infrastructure for various intelligent applications. The traditional development process for artificial intelligence is complex, involving data processing, model development, training acceleration hardware resources, model deployment service management and other links, which makes the model training and landing face different difficulties. With the expansion of heterogeneous computing powers, such as CPU, DSP, GPU and FPGA from the cloud to the edge, the artificial intelligence technology is experiencing the transformation of collaborative processing mode, in which training is carried out in the clouds and reasoning is performed at edges, thus increasing the coverage of domain information. Such transformation effectively improves the delay of intelligent application systems, and forms a closed loop model of training and deployment [156].

**(4) Zero-trust Security mechanisms**

In the traditional architecture of information systems, which is based on the organization's internal data center, there is a potential default relationship between the network location and trust. In such a situation, the strategy of full trust within the network boundary is adopted. With the continuous deepening of the digitalization process, this traditional architecture is changing to the digital infrastructure, which is based on the cooperation of cloud, network and edge, and integrates the new generation of technologies, such as big data, artificial intelligence and blockchain. Multi-clouds and hybrid clouds have become the main forms of information infrastructures, with frequent cross-cloud interaction of information flows. As a result, the traditional security boundary for information has been broken [157].

In response to this major challenge of information security, cloud computing security based on trust mechanism has begun to rise. The representative technologies include zero-trust network access (ZTNA) [158] and software-defined boundary (SDP) model [159]. These new security concepts break the default relationship between network location and trust in the idea of boundary security. Subsequently, the network boundary is no longer a security boundary, and everything cannot be trusted by default. In order to block the rapid spread of potential risks and make cloud computing adapt to the security needs in the fuzzy state of network boundaries, any object access in the information space must be authenticated and granted minimum permissions.

In addition, the traditional cloud computing platforms often need to collect user data, which are sent to the data center for processing. Therefore, there is a risk of user privacy disclosure [160] [161]. The emerging federated learning technology updates the global model by training and saving user data on local devices without uploading user private data other than local gradient updates. This novel machine learning mode, which is trained at the edge and aggregated by distributed servers in the cloud-network integration environment, can improve the security and privacy of data by limiting the coverage of information while ensuring the fitness of information.

## 3.6 Artificial Intelligence

Artificial intelligence (AI) refers to theories and techniques that enable machines, like intelligent beings, to learn from experience and perform various tasks. AI is a broad concept, including knowledge representation, reasoning, machine learning, swarm intelligence and other sub-areas.

**(1) Symbolic Artificial Intelligence**

From the end of the 19th century, mathematical logic developed rapidly and in 1930s it was used to describe intelligent behavior. In 1950s, Turing, regarded as the "father of computer science", published an article named "Can Machines Think?" In this paper, the concept of machine thinking was put forward, and the famous "Turing Test" was proposed [162]. After the emergence of electronic computers, formal logic deduction can appear in the form of computer programs. In 1952, Arthur Samuel developed a self-learning checkers program that, after training, was able to beat professional human players[163]. In 1956, Herbert Simon and Allen Newell proposed the heuristic artificial intelligence program Logic Theorist, which proved 38 mathematical theorems and showed that computers could be used to simulate human thinking processes [164]. The representatives of artificial intelligence in this period are Newell, Simon and Nilsson. At the Dartmouth Conference in 1956, John McCarthy and Marvin Minsky proposed the concept of "artificial intelligence" to distinguish the concept from connectionism in cybernetics.

The goal of symbolic AI is to put artificially defined knowledge or rules into computer systems, giving machines the abstract and logical capabilities to manipulate systems from a higher level. The main characteristic of symbolism AI is that it endows the computer with an autonomous reasoning space based on the symbol system. The theory of symbolic reasoning space is closely related to cognitive science. Cognitive science holds that the representation of the world within thought and the activity of thought itself can be described and operated by symbols embedded in programs. Therefore, symbolist AI believes that the physical processing of information can be described by symbols, including comparison, inclusion and inference [165]. However, it is difficult to apply in practice because the early symbolist AI greatly simplifies the complex external world into a "toy" form of symbol space [166].

In the early 1980s, there was a second renaissance of symbolic AI: the expert systems. Expert systems, based on more powerful computer processing and storage capabilities, transform the oversimplified "toy" world of earlier times into a knowledge base built by domain experts. When building an artificial intelligence system, domain experts will abstract and decompose the domain knowledge from the highest level to form a conceptual system and refine the granularity of information. At the same time, knowledge is represented as declarative propositions, which can interact with the outside world through natural languages [167] [168]. Take Dendral, the first expert system developed by Edward Feigenbaum to identify the chemical composition of materials, as an example [169], the innovation of the expert systems lies in the knowledge base, the heuristic product system derived from the knowledge base, and the system architecture that separates the knowledge representation from the reasoning engine. Subsequently, knowledge representation has experienced the continuous evolution of semantic network, framework and script, with the continuous optimization of the scope and granularity of information space, and has gradually developed into a complete theory and technology of knowledge engineering. By 1985, the second wave of artificial intelligence represented by expert systems reached its peak. However, in the development and construction of the knowledge bases of the expert systems, it is a very difficult challenge to ensure that the distortion of information is not reduced, that is, to extract, organize and reuse the knowledge of human experts efficiently, which makes it difficult for the expert system to be applied on a large scale [165].

The success of expert systems is of great significance to the application of artificial intelligence from theory to practice. Today, knowledge representation and reasoning are still the important research directions of artificial intelligence. In recent years, deep learning has been widely criticized for its low data efficiency (high sample requirements), poor generalization ability (inadequate adaptation) and lack of interpretability [170], which are the advantages of symbolic AI. Therefore, the combination of symbolism (knowledge base) and connectionism (deep learning) has become one of the current research hotspots in artificial intelligence [171].

**(2) Machine Learning and Deep Learning**

In recent years, thanks to the rapid development of machine learning, especially the deep learning technology, artificial intelligence has achieved excellent performance in many applications, such as natural language processing [172], computer vision [173] and recommendation systems [174]. Machine learning is a widely used AI technique that enables machines to learn and improve performance with knowledge extracted from empirical data. Machine learning falls into three categories: supervised learning, unsupervised learning, and reinforcement learning. Supervised learning requires labeling the training samples to decrease the mismatch of the information space, while unsupervised learning and reinforcement learning are usually applied to unlabeled data, using the inherent aggregation of information to mine the rules and patterns of information movement. Typical supervised learning algorithms include linear regression, random forests, decision trees, etc., which usually require features to be selected by hand; unsupervised learning algorithms include the K-means algorithm, principal component analysis (PCA), singular value decomposition (SVD), etc.; reinforcement learning algorithms include Q-learning, Sarsa and policy gradient [175].

Neural networks, originating from connectionism of early cybernetics, is a mathematical modeling method inspired by human brain nerves. It was proposed by neuroscientist Warren McCulloch and logician Walter Pitts in 1943. In 1949. Donald O. Hebb and other neuroscientists found that the activation of neurons through synaptic information can be regarded as a learning process, thus linking neural networks with machine learning. In the late 20th century, with the development of multi-layer neural network technology, neural networks are also known as "deep learning". Compared with classical machine learning, deep learning can automatically extract features from a large amount of data, so it is also called "representation learning". In deep learning, each layer of neural networks receives the input from the previous layer and outputs the processed data to the subsequent layer. Compared with traditional machine learning algorithms, deep learning can fully discover the relevance of the latent layer of large data, provide satisfactory prediction accuracy, and effectively improve the scope, granularity, delay and other capabilities of information, but the disadvantage of deep learning is that it requires massive training data and requires too much information volume. At present, Convolutional Neural Network (CNN) and Recurrent Neural Network (RNN) are two typical and widely used deep learning algorithms [176], which can process text, image, video, audio and other kinds of information, and can be used in face recognition, image classification, target tracking, image semantic segmentation and other computer vision fields, as well as in semantic analysis, information extraction, text mining, information retrieval, machine translation, question-answering system and dialogue system and other natural language processing fields.

**(3) Intelligent Agents and Robotics**

In AI, an intelligent agent is anything that can sense its environment, act autonomously to achieve a goal, and improve its performance by learning or using knowledge. In the intelligent agent, there is a pre-designed "objective function", which makes the intelligent agent create or execute related tasks to maximize the expected value of the objective function, so as to enhance the adaptability of information. For example, reinforcement learning shapes the expected behavior of intelligent agents through the reward mechanism, while evolutionary computing adjusts the behavior of intelligent agents through the fitness function.

Intelligent agents are closely related to the concept of robots in artificial intelligence. The ideal of creating robots capable of autonomous operation can date back to classical times, but the study of robotics and the potential uses did not advance substantially until the 20th century. In 1948, the cybernetics principle proposed by Norbert Wiener laid a theoretical foundation for practical robotics. Fully autonomous robots did not appear until the second half of the 20th century. In 1961, the first digitally operated and programmable robot, Unimate, lifted hot metal sheets from a die casting machine and stacked them [177]. In recent years, robots have been widely used in manufacturing, assembly, packaging, mining, transportation, earth and space exploration, surgery, weapons, laboratory research, security, and mass production of consumer and industrial products [178] [179].

In digital games or social scenes, intelligent agents are also called "non-player characters" (NPCs), which refer to characters that are not controlled by human players. NPCs are used to remotely and online reflect features such as the presence, face, and movement of a person's body [180]. Finite state machine (FSM), a NPC technology easy to implement, is widely used in the early days. However, as the scalability performance of FSM is very poor [181], it is difficult for this technique to support the expansion of information scope and granularity. At the beginning of the 21st century, as the classifier with maximum margin between different classes, support vector machines (SVMs) were applied to control NPCs in games. The advantage of SVMs is that it effectively improves the fitness of information, but the main disadvantage is that the technique lacks the flexibility to simulate human behavior and the process of decision-making [182]. In recent years, reinforcement learning has been widely used because it enables agents to learn automatically from the interactive experience within the surrounding environment, thus increasing the scope, granularity and coverage of information, which makes NPC have better adaptability than ever before. Among the NPCs, the most famous one is the AlphaGo based on deep reinforcement learning developed by DeepMind in 2015. AlphaGo can make decisions that are most likely to win through the neural network process [183].

**(4) Digital Twins**

The conceptual model of "Digital Twins" first appeared in 2003, proposed by Professor Grieves of the University of Michigan in the United States. It was initially called "Mirror Space Model" and later evolved into "Information Mirror Model" and "Digital Twins" [184]. In 2010, the National Aeronautics and Space Administration (NASA) introduced the concept of digital twins for the first time in the space technology roadmap, using digital twins to achieve comprehensive diagnosis and prediction of flight systems. After that, NASA and the US Air Force jointly proposed a digital twin paradigm for future aircraft, which defined digital twin as the simulation process that integrates multi-physical fields, multi-scales and probabilities. The essence of digital twin is to enhance the scope, granularity and coverage of information through information systems. In industry, GE uses digital twins for total life-cycle management of assets [13]. With the intensification of dynamic changes in the operating environment of complex equipment in the industrial field, the amount of equipment monitoring data has doubled, showing typical industrial data characteristics such as high-speed, multi-source, heterogeneous and changeable. The traditional PHM technology is limited in information volume, scope, granularity, delay and acquisition rate, which is difficult to meet the accuracy and adaptability requirements of real-time assessment and prediction of the states of complex equipment in the dynamic and changeable operating environment. The digital twin technology provides a new way to solve the above problems. Through digital twins, Siemens helps manufacturing enterprises to build and integrate manufacturing processes in the information space to realize the digitalization of the whole process from product design to manufacturing execution in the physical space [185], which comprehensively improves the scope, granularity, delay, duration and variance of the physical environment of industrial production in the information space.

The sensor technologies bring the fusion of models with multi-domains and multi-scales; big data and artificial intelligence bring the data-driven models, the fusion of physical models, and life-cycle data management; Internet and IoT bring data acquisition and transmission; and the cloud computing technologies bring the high-performance computing. As these various information technologies continue to develop and integrate, the concept of digital twin further refers to the simulation of physical entities, processes or systems of the real-world in the information space through information technologies. With the help of digital twins, human beings can understand the states of physical entities inside the information space, and control the elements of the physical entities through the predefined interfaces, under the circumstance of the improved mapping of the overall states of the physical world to the information space. Based on the feedback data of physical integration, digital twins can automatically adjust to the changes of physical entities through the artificial intelligence technology, and substantially improve the delay efficiency of the systems. Ideally, digital twins can learn from the feedback data of multiple sources, present the actual situation of physical entities in the digital world almost in real time, and improve the fidelity of information. The self-learning of digital twins (that is, the extension of machine learning in information space) can carry out rapid deep mining and accurate simulation according to the feedback of massive information, thus improving the fitness of information. Data twins use the existing cognitive and knowledge from human beings to deduce and understand the problems in the real-world through high-performance computing. In the future, with the continuous enrichment and improvement of information space, the volume, delay, scope, granularity, duration, variance, distortion and fitness of information space will be improved in an all-round way. The cognitive activities of human exploration and discovery of new knowledge may migrate from "physical world as the center" to "digital information space as the center", entering the so-called digital native generation.

## 3.7 Visual Computing and Extended Reality

Augmented reality originated from Milgram and Kishino's concept of "reality-virtual continuum" [186], and the latest augmented reality technology has been developing in the direction of deep integration with physical reality [187], viz. mixed reality and future holograms [188]. In this section, we'll start with visual computing and work up to emerging areas such as virtual reality (VR), augmented reality (AR), and the relevant advanced variant, mixed reality. Also, we will discuss how the augmented reality technology can connect virtual entities to the physical environments.

**(1) Visual Computing**

Computer vision is an interdisciplinary field of computer science that studies how computers gain a high level of understanding from digital images or videos. Computer vision tasks include the acquisition, processing, analysis, and understanding of digital images, as well as methods for extracting high-dimensional data from the real-world to produce numerical or symbolic information [189]. The computer vision techniques include scene reconstruction, object detection, event detection, video tracking, object recognition, 3D pose estimation, 3D scene modeling, and image restoration [190].

In the early 1960s, the purpose of image processing was the improvement of image quality so as to improve the visual effect of photographs [191]. In 1972, Nasir Ahmed first proposed the lossy compression technology for images based on discrete cosine transform (DCT) [192]. On this basis, the International Image Experts Group proposed the JPEG digital image standard in 1992 [193]. The efficient compression capability of DCT promotes the wide spread of digital images and digital photographs, and improves the information volume of images. As of 2015, billions of JPEG images are produced daily, making JPEG the most widely used format for image files today. In 1977, researchers developed the motion compensated DCT (MC-DCT) coding technology by combining the discrete cosine transform (DCT) coding in the spatial dimension with the predictive motion compensation algorithm in the temporal dimension, which enables the compression of videos to be used, thus improving the sampling rate of visual information and the volume of image information in the time domain. Nowadays, MPEG, a video compression technology based on MC-DCT, has become the most widely used information technology standard in the world [194]. Research in the 1970s has laid an important foundation for many computer vision algorithms today, including edge extraction from images, line labeling techniques, non-polyhedral and polyhedral modeling techniques, optical flows and motion estimation, etc. [195]. In the 1980s, visual computing was developed towards the direction with more rigorous mathematical analysis and quantitative techniques, including the concept of scale space, shape inference based on shadows, textures, focus and so on. At this stage, many mathematical optimization problems of visual computing are addressed within the framework of regularization or Markov random fields [196]. In the 1990s, with the emergence of the camera calibration optimization methods, important progress has been made in the reconstruction of three-dimensional scenes from multiple sparse images with the multi-view stereo technology, which significantly enhances the scope and granularity of visual information. In the late 1990s, the deep integration of computer graphics and computer vision has led to the emergence of complex visual computing technologies such as image rendering, image deformation, view interpolation, panoramic image mosaic and early light field rendering [195], which can support the processing of different kinds of visual information.

In the 21st century, the application of machine learning and complex optimization techniques, especially the deep learning technology, has brought great leaps to visual computing. In order to realize the interoperability between physical environment and digital space in digital twins, the visual computing technology needs to deeply understand human activities and behaviors. In recent years, the simultaneous localization and mapping (SLAM) technology reconstructs the three-dimensional structure of the unknown environments through motion estimation of mobile devices, which lays an important foundation for mapping the three-dimensional structure of the physical world to the digital space [197], and enhances the granularity of information space that reflects the physical space. At the same time, the interaction between objects in the physical and virtual worlds requires the support of the overall scene understanding technology [198], including semantic segmentation technology that classifies images into different categories according to pixel information, and the target detection technology that aims to locate objects in images or scenes and identify the category information of each object, which improves the granularity and fitness of information. In the immersive environments, the position of a virtual object determined by the stereo depth estimation technology is the key to the interchange of physical objects and virtual objects. In recent years, by combining deep learning with the stereo camera technology, high-accurate depth estimation performance has been achieved [199], which means distortion is improved by increasing information volume. In many augmented reality applications, it is often necessary to generate action-specific feedback in the 3D immersive environment by observing and recognizing the user's actions. In machine vision, understanding the person's actions is called motion recognition, which involves locating and predicting the human behaviors. In recent years, deep learning of pure RGB image data or multi-modal data of the sensor fusion can process different kinds of information, which has been applied to action recognition in augmented reality, and such technology also has the potential of emotion recognition in virtual reality [200].

In computer vision, problems are mainly studied from two aspects: image restoration and image enhancement. The purpose of image restoration is to reconstruct a clean image from a degraded image (e.g., noise, blurred image) to improve the fidelity of the image information. In contrast, image enhancement focuses on improving image quality, that is, improving the volume and coverage of visual information. Among them, image restoration techniques have been used to restore texture details of virtual images in virtual reality and remove artifacts [201]. In a fully immersive environment, super-resolution display can affect the perception of the 3D virtual world, which requires not only image super-resolution in optical imaging, but also in the imaging process. At present, thanks to the development of the optical and display technologies with high sampling rate, the image super-resolution technology has been directly applied to the ultra-high definition display screen.

**(2) Virtual Reality**

The outstanding technical feature of virtual reality is the full synthetic view. Commercial virtual reality headsets provide common ways of the user-interaction techniques, including head tracking or tangible controllers [202]. Thus, the user is located in a fully virtual environment and interacts with the virtual objects through the user-interaction techniques. In addition, the virtual reality technology is at the farthest end of the virtual continuum from reality, which means that users using VR headsets must be completely focused on the virtual environment and separate from the physical real-world [203]. Currently, the commercial virtual reality technologies enable users to create contents (such as VR paintings) in a virtual environment. At the same time, the heuristic exploration can be achieved by users interacting with virtual entities in the virtual environments. For example, modifying the shape of virtual objects, and creating new art objects. In this virtual environment, multiple users can collaborate in real time. The main characteristics of this virtualized collaborative environment include shared sense of space, shared presence, shared sense of time (real-time interaction), communication methods (through gestures, text, voice, etc.), and ways of sharing information and manipulating objects [204]. Therefore, the scope, granularity, delay, duration and variance of information are highly required. The key technical challenges in this highly virtualized space are how users control virtual objects and how multiple users collaborate in a virtual shared space.

**(3) Augmented Reality**

Augmented reality, on the basis of the VR technologies, further provides users with alternate experiences in the physical environments. Its purpose is to enhance our physical world through the information space, thus expanding the scope and granularity of the information space. In theory, any computer-generated virtual content can be presented through various perceptual information channels, such as audio, visual, olfactory and tactile [205]. The types of information and the related coverage is the key metrics of augmented reality. The first generation of the augmented reality technologies only considered visual enhancement, and its purpose was to organize and display digital overlays superimposed on top of the physical environments. For example, in the work of the early 1990s, large see-through displays did not allow for user mobility, requiring users to interact with text and 2D images in a static posture through a tangible controller [206], with high latency of information.

Therefore, how to ensure seamless and real-time interaction between users and digital objects in augmented reality is the key challenge for the augmented reality technologies [207]. Since the advent of the augmented reality technologies, there have been a lot of work and research devoted to improving the user interactive experiences with digital objects in augmented reality. For example, the hand-drawn interaction technology provides an intuitive and easy-to-use interaction interface for the augmented reality users. Through this technology, users can select and process objects in virtual space with the gesture of pinch [208], which improves the scope, granularity and coverage of information. In other words, the users using augmented reality technology can simultaneously interact with virtual objects in the digital space of a physical working environment to improve the delay performance of information.

In order to realize the seamless connection between the objects in physical world and virtual digital objects, it is necessary to break through the machine vision detection and tracking technology, so as to map the virtual contents displayed visually with the corresponding position in the real environment, so as to enhance the scope, granularity and fitness of information. Augmented reality head display technology has been significantly improved in recent years. By embedding head display technology into glasses, the mobility of lightweight augmented reality is effectively improved, and users can recognize different kinds of objects in augmented reality by receiving visual and audio feedback from the head display [209]. Although augmented reality can also be realized by handheld touch screen, ceiling projector, Pico (wearable) projector and other types of devices, head display technology has the advantages of freely switching the user's attention and releasing the user's hands, increasing the delay performance of information. Thus, it is generally considered to be the main means for future universe users to interact with the virtual world [210].

**(4) Mixed Reality**

Augmented reality and virtual reality describe two ends of the reality-virtual continuum, while mixed reality is an alternate reality between the two [211]. Augmented reality usually simply displays information superimposed on the physical environments, and does not consider the interoperability between the physical world and the virtual world. In contrast, mixed reality focuses more on how the physical environments interacts with virtual entities. Therefore, many researchers believe that mixed reality is an enhanced scene of augmented reality. Therefore, there is a closer connection and collaboration between physical space and virtual objects for mix reality [212], which comprehensively enhances the scope, granularity, coverage, fitness and variance of the physical world reflected in the information space.

Many researchers believe that digital twin connecting the physical world to the virtual world is the starting point of the future metaverse, while mixed reality provides users with a window to achieve seamless interaction between the physical world and the virtual world in the metaverse [213]. In the metaverse, physical objects, digital avatars of physical objects, and the interaction between objects are integrated, thus forming a huge virtual shared space. Activities in all virtual environments should be synchronized and reflect changes in the state of motion in the virtual space [214]. Through mixed reality technology, human users can make creations with digital twins [215] [216]. The content created in the digital space can be presented in the physical environment at the same time, and can be integrated with the physical environment across time and space [217]. Even if we can't predict exactly how the metaverse will eventually affect our physical world in the future, we can get a glimpse of it through the current prototype of mixed reality technology. For example, highly realistic scenes in mixed reality, realistic sense of existence, empathic physical space, etc. [218] [219] are all the basic characteristics of the metaverse that people imagine, where physical space and multiple virtual worlds will complement each other in the future [213].

In the metaverse, augmented reality-based technologies provide solutions for open communication between robots and virtual environments due to their significant visual content characteristics [220]. In addition, the integration of virtual environments in tasks, such as scenario analysis and safety analysis, enables human users to understand robot operations, thus building trust and confidence in robots, which will lead to a paradigm shift in human-robot collaboration [221]. At the same time, robots will serve as physical containers for digital substitutes of people in the real-world, and virtual environments in the metaverse can also change human perception of cooperative robots. In the future, digital twin technologies and the metaverses will be the virtual testing grounds for new robot designs.

In the future metaverse, the virtual world is seamlessly connected with the physical environment in real time. In this scenario, 3D vision technologies with less noise, blur, and high resolution become very important. Whether it is the features based on manual composition, or the semantic segmentation technology or target detection technology based on deep learning, there is a problem of excessive computational overhead, and it is still difficult to support the real-time overall scene understanding required in the metaverse. Therefore, vision research that merges image restoration and image enhancement methods [222] can realize the metaverse vision of seamless connection between virtual and real. At the same time, the metaverse will enable real-time multimedia applications to grow exponentially, requiring a large amount of bandwidth to transmit very high-resolution content in real time, while requiring very high network latency. However, many interactive applications regard motion as photonic latency, viz. the delay between user actions and on-screen responses [223]. Therefore, the current 5G can hardly meet the latency requirements of multimedia applications such as perspective augmented reality or virtual reality in the metaverse. In order to make the users experience a truly ubiquitous metaverse, outdoor seamless network access based on wireless mobile technology becomes essential. At present, last mile access is still the bottleneck of LTE5G network [224]. To overcome this technical challenge, multi-access edge computing (MEC) is expected to reduce 5G latency to 1 millisecond by providing standard, universal edge load balancing services one hop away from wirelessly connected user equipment. However, given the infinite number of concurrent users working together on virtual objects and interacting with each other in the metaverse, especially the possible delays that can negatively affect the user experience, managing and synchronizing the state of information movement on such a very large scale is a huge challenge.

# 4 The Framework of Information Space

The wide application of a series of information technologies and systems, such as Internet, mobile Internet, cloud computing, big data, supercomputing, artificial intelligence, block chain, Internet of Things and virtual reality, has profoundly changed the way of human production and life. In recent years, the term metaverse has attracted the attention of the field of information technology, including all sectors of society [225]. As far as the meaning of the word is concerned, the metaverse has a broader meaning than the previous achievements of information technology. Therefore, according to the important achievements of information technology, combined with the objective law of information movement and utilization, from the overall perspective of the real-world, human society and information system, it is urgent to comprehensively examine the framework of information space and strive to achieve the unity of cognitive basis.

## 4.1 The Concept and Nature of Metaverse

The concept of metaverse appears in the critical period when the achievements of information technology are constantly emerging and a new round of scientific and technological revolution is in the ascendant. Some people think deeply about the great changes it may bring, some people use it to package their business and products, and some people regard it as a gimmick or joke in the history of technological development [226]. However, in any case, after decades of vigorous development of information technology, we really need to use broader concepts to fully accommodate, integrate and integrate all kinds of rich and colorful achievements in order to promote faster, more orderly and more predictable development of information technology.

The metaverse can be regarded as a phased integration of various information technologies in the present era, which can absorb the achievements of digital technology so far, and may also significantly change the paradigm of scientific research, and promote a more comprehensive interaction between information science and life science, social science, quantum science and other fields. In fact, so far, all kinds of information technology achievements have promoted the movement and utilization of various forms of information with different effects, and the information flow which is closely combined and interacted with material flow and energy flow runs through it. Therefore, based on the existing representative views [225] [226] [227] [228], it can be considered that the metaverse is the sum of the ubiquitous, diverse and endless information flows in the real-world and information space.

**(1) Basic Elements of Metaverse**

From its concept, we can see that the metaverse includes the basic elements of information, information movement, information systems and information applications.

The metaverse is the sum of all information flows in the world, and the main body of information flows is information, so the most fundamental element of the metaverse is information. If the origin of the universe is matter, then the origin of the metaverse is information.

On the other hand, information that does not move does not become information flows, and therefore is not part of the metaverse. Therefore, another basic element of the metaverse is the movement of information. The information flows formed by the information starting from the source and reaching the destination through various links is the constituent element of the metaverse. Information movement can be either a photoelectric conversion within a single device, or a global journey of thousands of turns and mountains, which are important factors that the metaverse must pay attention to.

In today's information age, almost all information flows cannot be formed only through natural objects in nature, but can only be realized with the support of information infrastructure, human-computer interaction, media centers, space-time computing, creative economy, innovative discovery, in-depth experience and other information systems throughout the world. Therefore, information systems are the main carrier of carrying information and driving information flows, and it is also an important factor in the development of information technologies and the birth of the concept of metaverse. The study of the metaverse cannot be separated from the full understanding of the existing information systems, but also requires insight and foresight into the future development of information systems.

Similarly, in today's information age, the fundamental significance of people's attention to information flows is to make the best use of information and serve mankind. So, the metaverse must focus on a higher level of information utilization. An example of this is that deep experience technologies [228] will naturally and accurately simulate human behavior in the real-world through the interconnection, integration, sharing and learning of identity, friends, immersion, pluralism, anytime, anywhere, economy and civilization, so that people can switch their identities and shuttle between the real-world at anytime and anywhere. Enter the metaverse composed of one of the space and time nodes at will, and study, work, make friends, shop and travel in it, so as to achieve "immersion". In this scenario, the representational metaverse is based on the abstract information space, which supports people to use information more freely. There are countless similar scenarios, which provide infinite possibilities for the technological development and system development of the metaverse, and then promote a higher level of information utilization.

**(2) Technical Characteristics of Metaverse**

Information is the blood of the metaverse, and the flows of information gives vitality to the metaverse. The metaverse takes the information systems as the carriers to realize the integration and organic combination of the real-world and the information space. Therefore, the metaverse is not static, but dynamic; not local, but global; not a single form, but diverse; not only belongs to the real world or information system, but also a close bridge between the two.

Movement is the main technical feature of the metaverse. The continuous flows of information drive the metaverse to serve human beings day and night. The source and destination of information movement, the way, speed and acceleration of movement, and the links and processes of movement are all related to the concrete realization of the metaverse. It is necessary to study and establish the corresponding dynamic mechanism in order to truly understand and master the basic laws of the metaverse.

Globality is an important technical feature of the metaverse. Since we hope to accommodate all kinds of achievements of information technologies through the concept of metaverse, we should not only consider the development of metaverse in a certain link or scene, but also from the various links of information perception, transmission, knowledge and use, all kinds of users all over the world, science, economy, society, humanities, nature and so on. In view of the metrics of information utilization, such as broadness, universality and authenticity, the comprehensive development of metaverse is planned.

The shape of information flows will also be an important technical feature of the metaverse. After decades of rapid development, information technologies have been able to provide people with digital signals, data, text, audio, video, multimedia and other forms of information flows to meet the diversified needs of people's production and life. With the mature application of 5G mobile communication and high-performance artificial intelligence, the use of virtual reality, augmented reality, hybrid display and other technologies to enhance the user's sense of presence, immersion and substitution, greatly enhance the user's experience of information flows and information space, is the proper meaning of the development of the metaverse. Therefore, the metrics of information flows, such as meticulousness, timeliness, richness and adaptability, will become important features of evaluating metaverse technologies.

Another important technical feature of the metaverse is to realize the seamless connection between the real-world and the information system more closely. Human beings have achieved great success in improving work efficiency and optimizing life style by using information technologies and information resources. However, just as we have not yet opened a reasonable paradigm for information science research, the full utilization of information resources in the real-world is still at a very low level. Many fields are not without more profound practical application needs, but there are still technical bottlenecks in the collection, transmission, processing, convergence and action of information, and even a gap in understanding. In a word, there is still a serious disconnection and division between the information systems and the real-world. Therefore, we should drive the full integration of the real-world and information systems through the information flows full of the metaverse, so as to achieve a higher level of information utilization.

**(3) Social Form of Metaverse.**

The metaverse will use a variety of new technologies to integrate the social form of virtual and real integration, enrich the digital economic model, and promote the breakthrough of traditional philosophy, sociology, and even the humanities system. In the process of human attention and participation in the formation and development of the metaverse, the traditional concepts of life, time and space, energy, ethnic groups, economy and values may be changed and subverted, forcing people to rethink the basic philosophical concepts: transcendental knowledge, existence and existentialism, empiricism, dualism, the nature of language, surreal society and so on. From the perspective of information flows in the metaverse [228], real human beings and the virtual humans they create, including biological humans, electronic humans, digital humans, virtual humans and information humans, as well as their offspring with different personalities, skills, knowledge and experience, will form new social relations and emotional connections, and eventually evolve into organisms. It has become a pioneer in opening up the boundaries of the metaverse and building a "post-human society" on the new digital continent. If the formation process of "post-human society" is regarded as the transition process of life form from "carbon-based life" to "silicon-based life", then there will be the evolution of biology, information theory, technology, as well as the evolution of ethics, culture and society, full of expectations and risks. The new form of human beings between reality and virtual depends on the information flows, and will walk in organisms and machines in the future.

## 4.2 Real-World and Information Space

Reviewing the development of information technology, studying information flow, and then exploring information system dynamics, we have to consider the real-world, information space and the relationship between them. It can be considered that the real-world is the root and the information space is the form.

**(1) Information Space in the Real-world**

Wiener, the founder of cybernetics, pointed out that "information is information, not matter or energy" [229]. On this basis, Steucke, a German philosopher, proposed that "information is the third thing juxtaposed with matter and energy" [230]. The most famous information theory is Shannon's information theory, followed by Zhong Yixin's full information theory, Burgin's general information theory, Vigo's representation information theory and Fleissner and Hofkirchner's unified information theory. Almost all the representative achievements of the theoretical system of information geometry follow the basic law of information entropy [231]. According to Shannon's information theory, information can only be meaningful when it is received by the sink. This implies that the information must be objective and real. In fact, people can recognize information, describe information and use information, but they cannot change information with their subjective will. Even though there are different interpretations of information based on their respective disciplines and knowledge backgrounds, the objective existence of information cannot be denied. Just as matter and energy are not transferred by human will, the objectivity of information should not be transferred by human subjective consciousness, especially for information systems. Therefore, we define information in the objective category, and regard material, energy and information as the three major elements of the objective world. Among them, information objectively reflects things and their state of motion in the real-world through the medium of matter and energy in the real-world.

Popper, a British scientist, put forward the division of the real-world [232]: the first is the world of physical objects or physical States; the second is the world of conscious States or mental States, or the world of behavioral intentions about activities; The third is the world of the objective content of thought, especially the world of scientific thought, poetic thought and artistic works, with special emphasis on the objectivity of the third world. Among them, the first world and the third world are actually a further division of the objective world, while the second world is the subjective world in the usual sense. According to this view, the real-world includes the subjective world and the objective world, and the objective world is divided into the objective knowledge world and the objective physical world. Among them, the subjective world refers to the world of consciousness and ideas, which is the sum of spiritual and psychological activities to understand and grasp the whole world, not information, but knowledge and perception. The objective world refers to the material and perceptible world, which is the sum of all substances and their movements other than conscious activities, relying on physical carriers and existing on physical media such as books, tapes and CD-ROMs.

It can be seen that there has been information space in the real-world since ancient times. Sound, light and electricity in nature provide information to observers, and language, words and images also express information to people. These are the contents of information space in the real-world, which belong to the first world and the third world in Popper's "three worlds". To some extent, the two worlds themselves are the carriers of information. It can be seen as a collection of information space in the real-world. However, until the emergence of large-scale information systems, the scope and effectiveness of these information are very limited, so that people in today's information age tend to ignore their existence.

**(2) Information Space in the Information Age**

In the information age, various information systems emerge as the times require. The information in the real-world enters the information system through digitization, which objectively reflects the real-world. With the increasing accumulation of information in the information system, the role of information flow has become increasingly prominent, and it has increasingly become the main carrier and focus of attention in the information space. For example, the information system uses digitalization to configure the physiological existence, cultural existence, psychological and spiritual existence of real human beings to form a virtual human in the information space. For another example, the digital twin in the information system is to construct an object entity completely expressed by data in the information space, and use the core of the information space to reflect the shell of the real-world. At this stage, people do not pay attention to how to process and store, but only to the accuracy of its expression of the real-world. It can be seen that the real-world contains information space, and information space is more based on information systems. The real-world is the essence and connotation of information space, which provides the source of information. Information space reflects the real-world, but also deduces and simulates things that are difficult to achieve in the real-world, and feeds back the real-world.

Furthermore, the interaction between the information space containing information systems and the real-world is conducive to the realization of complementarity and balance from the conceptual, technological and cultural levels. It can be imagined that in the real-world and information space, the real and virtual human individuals who inhabit at the same time are not single identities but multiple identities, at this time, human beings and their virtual lives in the social activities and lifestyles of information space, self-learning, self-adapting, self-interacting, self-evolving, get more happiness, and bring such feelings and experiences back to the real-world. It is conducive to the change of the real-world to the good, the formation of a new "human community" civilization ecology, and life may also expand from physiological limitation to digital infinity in the information space.

（3）**Elements of Information Space**

It can be seen that information space spans the real-world and information systems, including the first world in the real-world, that is, the world of objective physics, and the third world, that is, the world of objective knowledge. At the same time, it also includes all the information in the information system, including all kinds of digital information in the process of collection, transmission, processing and action, as well as all kinds of information gathered and precipitated and expressed in the form of data. The metaverse is the complete set of all kinds of information flows in the information space.

Further analysis shows that information space includes three basic elements: natural information, behavioral information and media information.

1）Natural information

Natural information is the direct manifestation of the state of motion of matter and energy in the objective world, and it is the first basic form of information. The sun, moon and stars, mountains and rivers, city streets and rural fields show natural information all the time. The buds on the branches of trees in spring, the red leaves on the hills in autumn, the grey hair on the heads of the old people coming towards them, and the strong and handsome figure of the young people reflected in the mirror also express natural information. It can be seen that natural information makes us feel the weather of nature most directly. The main characteristics are as follows:

First, the noumenon reflected by natural information is the things in the objective world, such as geographical and astronomical phenomena, architectural objects, animals and plants, etc. Even if human beings still have a complex subjective world, natural information reflects only the external image of people; Secondly, the performance of natural information is time-varying, many landscapes may remain unchanged for a long time on the macro level, while on the micro level, they are likely to change rapidly; Thirdly, natural information can be displayed not only through the noumenon itself, but also through other carriers. The towering mountain can not only make the climber feel its existence, but also show its grandeur through other carriers such as water, air and so on in the way of reflection in the water or even mirage.

2）Behavioral information

Behavior information is the indirect reflection of consciousness and state of mind in the subjective world acting on material and energy in the objective world. Human laughter and scolding reflect the mood of the subjective world through expression or language, birds flutter their wings and sing through posture or voice to express their conscious desire to escape or courtship, and elephants will stay around the body of the deceased for a long time to express their inner sadness when their companions pass away. It can be seen that behavioral information indirectly reflects the consciousness and state of mind of the subjective world of human beings and other creatures through actions, expressions, languages and voices. The main characteristics are as follows: First, the noumenon reflected by behavioral information is the consciousness and state of mind in the subjective world, such as the instinct, desire, emotion, judgment and decision-making of human beings, animals and even some plants and microorganisms; Secondly, the performance of behavioral information is also time-varying, because subjective consciousness is complex and changeable, so the behavior dominated by it will show a variety of forms due to the change of time; thirdly, behavioral information can only be indirectly reflected through other carriers such as body, voice or tools, but it is difficult to be directly presented in the form of noumenon, that is, subjective state itself. The author believes that although lie detectors and other instruments that try to peep at the subjective world of human beings have been born for more than a hundred years, no matter how advanced science and technology are, the subjective world of human beings will never be directly presented in broad daylight, because the subjective world and the objective world are two distinct categories after all. Only in this way can the existence and development of the world be more exquisite.

3）Media information

Media information is the image of natural information and behavioral information stored in the form of material and energy after collection, transmission or processing. A large number of social news, people's comments and encyclopedic knowledge recorded in newspapers, periodicals and books are media information that can be read repeatedly, while audio and video broadcast in radio, movies and television are media information that can be listened to or watched repeatedly; Internet servers all over the world store data codes with huge capacity and various forms, which are media information that can display or support the operation of various information systems to users. It can be seen that media information records or replicates all kinds of natural information or behavioral information, which indirectly makes the state of motion of things in the objective and subjective world last or last for a long time. The main characteristics are as follows: first, the noumenon of media information can be things in the objective world as well as the consciousness and state of mind in the subjective world; second, media information is stable in time, which is convenient for users to feel or process repeatedly; third, media information only reflects the movement and changing state of noumenon through other carriers such as paper, bamboo slips, stone, disk, circuit, screen and so on.

Table 1: Feature Characteristics of Information Space

| Classification | Connotation | Instance | Subject | Time-varying characteristics | Carrier |
|---|---|---|---|---|---|
| Natural information | The direct manifestation of the state of motion of things in the objective world. | Scenery of mountains and rivers, city style, figure, etc. | Things in the objective world. | Change with time. | The body itself or another carrier. |
| Behavioral information | The indirect reflection of consciousness and state of mind in the subjective world acting on the objective world. | The expression, language and singing of the characters, as well as the body movements and singing sounds of the animals. | Consciousness and state of mind in the subjective world. | Change with time. | Other carriers than the body. |
| Media information | Storage mapping of natural and behavioral information. | Books and materials, calligraphy and painting products, audio-visual media, databases, etc. | Things in the objective world or States of consciousness and thought in the subjective world. | It doesn't change with time. | Other carriers than the body. |

## 4.3 Information Space Framework Based on Information Flows

According to the relationship between the real-world and information space, it is an important prerequisite to classify the basic components of information systems reasonably and describe the framework of the whole information space based on the driving force of information flows to support the formation of the theoretical system of information systems dynamics.

（1）**Construction Principle**

In information space, information provides resources, information flows give vitality, and information systems achieve efficacy value. The framework of information space should fully integrate the real-world and information systems, absorb the achievements of information science and technology, cover all information processes, present the role of information movement, and support the research, analysis and evaluation of information systems. Therefore, the following four principles should be followed to construct the framework of information space:

1) Fusion of virtuality and reality: the framework of information space should be the comprehensive fusion of the real-world and information system;

2) Flows of information: the framework of information space depends on the driving role of information flows for vigorous vitality;

3) Coverage of processes: the framework of information space must include all the important processes of information movement;

4) Inclusion of achievements: The framework of information space must contain as many important results of information technology as possible.

（2）**Structural Framework**

Based on the above principles, we proposed the information space framework as shown in Figure 1.

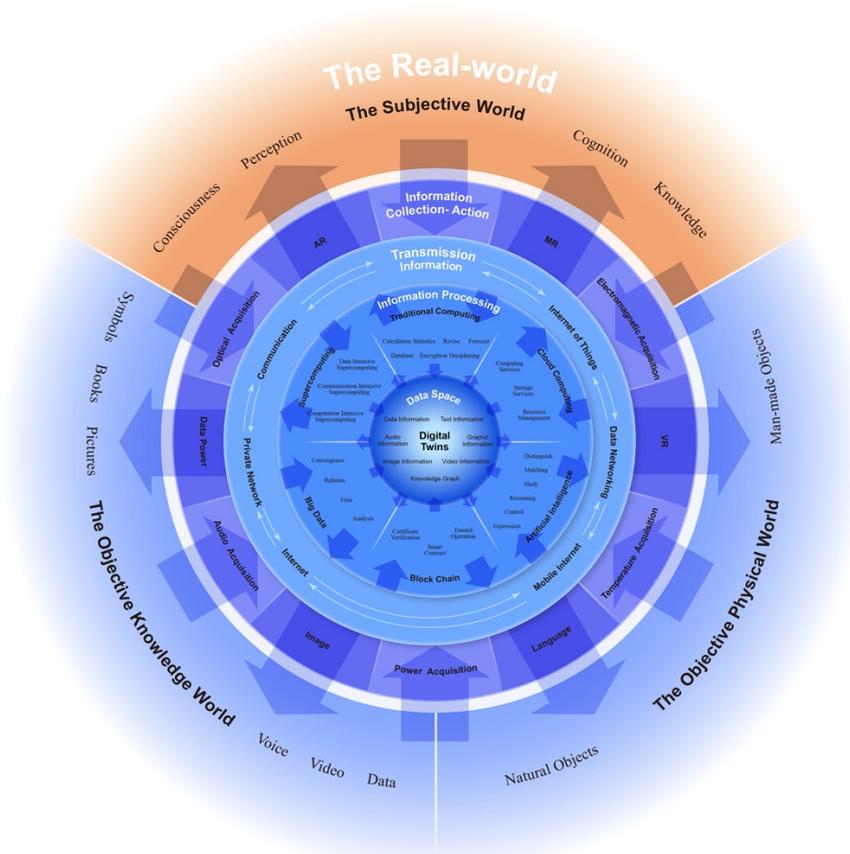

**Figure 1: Framework of the information space**

Wherein, the orange area is the subjective world in the real-world; All the blue areas are carriers of information space, covering not only the objective physical world and the objective knowledge world in the real-world, but also all information systems such as information collection/exertion, transmission, processing and data space from the second ring to the outside. The arrows shuttling through various components/parts represent various information flows moving in the real-world and the information systems. In the real-world, they go through the three rings of information collection, information transmission and information processing from outside to inside, and finally converge and deposit to the core area-data space. Subsequently, from inside to outside, they function on the real-world through the three rings of information processing, transmission and exertion. As information collection/exertion are directly faced to the real-world, we present information collection/exertion in the same loop and distinguish them only by color shades and different directions of information flows. In theory, information systems are parts of the real-world, but because of their special roles in information, information flow and information space, and also the main research subject of information system dynamics, we separate them from the real-world in the framework of information space, and place them in the center as the main body of our attention and research. We consider that the totality of all information flows in Figure 1 constitutes the metaverse, which drives the realization of various interactions between the real-world and the information systems.

**(3) The Real-world**

Following Popper's theory, the real-world consists of three parts: the subjective world, the objective physical world and the objective knowledge world. Amongst them, the subjective world points to the information system to send and receive information from it, and does not store information itself, but only contains subjective contents such as consciousness, cognition, perception and knowledge, which conforms to the basic orientation that information only belongs to the objective category. The objective physical world consists of natural objects and man-made objects. The objective knowledge world includes knowledge products such as symbols, data, pictures, books, voice and video. Both the objective physical world and the objective knowledge world can not only produce or receive information, but also store information, so they are also important carriers of information space.

**(4) Information Collection/Action Ring**

The information collection/exertion ring is directly oriented to the real-world. Information acquisition systems collect all kinds of information from the real-world through optical acquisition, audio acquisition, power acquisition, temperature acquisition, electromagnetic acquisition and other ways and means, and then sends it to other components of the information systems. Information actions in the opposite direction, receiving information from other components of the information system and exerting on the real-world through language, image, data, power, temperature, virtual reality, augmented reality, mixed reality and other ways.

**(5) Information Transmission Ring**

The information transmission ring mainly uses communication, private network, Internet, mobile Internet, data networking, Internet of Things and other ways to realize end-to-end transmission of various types of information within the information systems or information exchanges between them.

**(6) Information Processing Ring**

The information processing ring mainly uses traditional computing, supercomputing, cloud computing, big data, artificial intelligence, blockchains and other methods, and can be further subdivided into several sub-methods and algorithms with specific functions to process all kinds of information in the information systems according to business needs, so as to meet the information needs of all kinds of users.

**(7) Data space**

Data space is the core of information systems. After processes of collection, transmission and processing, all kinds of information in the real-world are converted into huge amounts of data with rich types, large scale and close correlation, which are fed into the data space in the form of data information, text information, audio information, graphic information, image information, video information, knowledge graph and digital twins, so as to form a holographic mirror representing the real-world. Compared with the real-world itself, such holographic images obviously have many incomparable conditions and conveniences in transmission and processing, which can make people use them almost at will, repay the real-world and promote the progress of human civilization. Therefore, to some extent, the central task of information systems is to construct a data space that can represent the holographic image of the real-world.

# 5 Rethinking of the Model, Properties and Measures of Information

In order to study the dynamic mechanism of information system, Xu Jianfeng et al. Proposed the mathematical definition of objective information theory and information space for the ubiquitous information in the objective world[25], and then made a slight improvement[26], and also made an example analysis in combination with the air traffic control system[错误!未定义书签。]. The definition, mathematical expression, basic properties and measurement system of information play an important role in the design, analysis and evaluation of information systems. The information space discussed in this paper mainly involves the collection, transmission, processing, aggregation and function of information, which naturally requires the use of the definition, model, nature and measurement of information to support the research on the functional performance and mechanism of the whole system and its components. According to the author's recent research results and system engineering practice, a more thorough understanding of the relevant content has been formed. In order to facilitate readers' thorough understanding, the basic theory system of objective information theory, which has been revised and supplemented, is comprehensively introduced here.

## 5.1 Model of Information

**Definition 1** Definition $O$ represents the objective world set, $S$ represents the subjective world set, $T$ is the time set, and the elements in $O$、$S$ and $T$ can be divided appropriately according to the specific requirements of the domain of discourse, the subject $o \in 2^{O \cup S}$、the state occurrence time $T_h \in 2^T$、the state set $f(o, T_h)$ of $o$ on $T_h$, the carrier $c \in 2^O$, the reflection time $T_m \in 2^T$ and the reflection set $g(c, T_m)$ of $c$ on $T_m$ are all non-empty sets, and the information $I$ is the full mapping from $f(o, T_h)$ to $g(c, T_m)$, that is:

$$I: f(o, T_h) \to g(c, T_m), \tag{5.1.1}$$

or

$$I(f(o, T_h)) = g(c, T_m). \tag{5.1.2}$$

The set of all information $I$ is called information space, denoted as $\mathfrak{I}$, which is one of the three elements of the objective world.

It should be emphasized that in order to accommodate the widest range of possible situations, [3]states that the mapping in the definition of information "is not limited to single-valued mapping, but can also be multi-valued mapping". After many years of research and analysis, the author has not found any examples that must be explained by multi-valued mapping. Because multi-valued mappings are difficult to understand and easy to lead to confusion in practical applications, the revised definition limits information to single-valued mappings, which will bring obvious convenience to many subsequent studies, while not affecting most or even all applications.

**Definition 2** Call $\langle o, T_h, f, c, T_m, g \rangle$ the six-tuple model of information $I$, also denoted by $I = \langle o, T_h, f, c, T_m, g \rangle$.

Although the six-tuple model of information is simple, it deconstructs the concept of information in three important ways. One is the dualistic deconstruction of information subject: according to the reflection characteristics of information, the dualistic structure of subject $o$ and carrier $c$ is used to describe the information subject; the other is the time dimension deconstruction of information: according to the time-varying characteristics of information, two parameters, state occurrence time $T_h$ and reflection time $T_m$, are introduced to support the analysis and research of information movement from the time dimension; The third is the deconstruction of the information content: the state set $f(o, T_h)$ and the reflection set $g(c, T_m)$ are introduced to accommodate all the information content and form. Through these three important deconstructions, we can make a more profound and comprehensive analysis of information in addition to the amount of information, and provide a sufficient mathematical basis for the establishment of information system dynamics.

Figure 2 shows the whole picture of information flows during the news interviewing and releasing process that people are familiar with in every-day life. This intuitive and vivid scenario can help understanding the sextuple model of information. In Figure 2, the information collection link mainly collects the state information of the interviewees through video, audio, text and other collection means; the information transmission link transmits the collected information to the corresponding processing system through the Internet and other wide area networks; the information processing link carries out video, audio, text and mutual fusion processing to form various news materials; These news materials are gathered into a more comprehensive news database to support more extensive access and application; then in the information processing link, the news information with richer contents and forms is distributed and arranged to meet the publishing conditions; then through the information transmission link, all types of media news information are transmitted to various information terminals through the Internet; Finally, in the information action link, all kinds of terminal devices around the world directly display the corresponding news information to various kinds of audiences or readers in a variety of forms.

According to the analysis of information space framework, the whole process of news gathering and releasing mainly includes seven important links, in which the information in each link has six elements, such as noumenon, occurrence time, state set, carrier, reflection time and reflection set. Table 2 shows the specific contents.

The information noumenon and carrier at each link in Table 2 are different. Especially, the noumenon, occurrence time and state set of the next link are the carrier, reflection time and reflection set of the previous link, respectively, reflecting the concept of *information flow*, an important characteristic of information transmission. On the other hand, because the news information itself reflects the subjective and objective state of the interviewee, the noumenon and state of all links can be understood as the subjective and objective content of the interviewee's image, voice and text during the interview period, which is also the basic characteristic of information transmission.

**Table 2: Information elements in main links of the news gathering and releasing process**

| No. | Links in Information Systems | Noumenon ($o$) | Occurrence Time ($T_h$) | State Set ($f$) | Carrier ($c$) | Reflection Time ($T_m$) | Reflection Set ($g$) |
|---|---|---|---|---|---|---|---|
| 1 | Information collection ($I_1$) | Interviewees ($o_1$) | From the beginning to the end of the news interview ($T_{h1}$) | The images, voice and text of the interviewees and the interview scene, as well as the subjective consciousness of the interviewees ($f_1$) | Video camera, camera, voice recorder, notebook, etc. ($c_1$) | From the beginning to the end of the news interview ($T_{m1}$) | Image, voice, text and other data and text collection of the interviewee and the interview site ($g_1$) |
| 2 | Information transmission ($I_2$) | Video camera, camera, | From the beginning to the end of the | Image, voice, text and other data and text | Transmission links such as the Internet | From the beginning of news data | Digital coding of images, voice |

| # | | | | | | |
|---|---|---|---|---|---|---|
| | | voice recorder, notebook, etc. ($o_2$) | news interview ($T_{h2}$) | collection of the interviewee and the interview site ($f_2$) | ($c_2$) | transmission to the end of transmission ($T_{m2}$) | and text of the interviewees and the interview site ($g_2$) |
| 3 | Information processing ($I_3$) | Transmission links such as the Internet ($o_3$) | From the beginning of news data transmission to the end of transmission ($T_{h3}$) | Images, voice, text and other digital contents of the interviewees and the interview site ($f_3$) | Video processor, audio processor, codec, news editing and clipping subsystem, etc. ($c_3$) | From the beginning of digital decoding to the completion of audio, video, text and fusion processing ($T_{m3}$) | News video, audio, text and fusion material information ($g_3$) |
| 4 | Data space ($I_4$) | video processor, audio processor, codec, news editing and clipping subsystem, etc. ($o_4$) | From the beginning of digital decoding to the completion of audio, video, text and fusion processing ($T_{h4}$) | News video, audio, text and fusion material information ($f_4$) | News database ($c_4$) | From entering news video, audio, text and fusion information into the information database to deleting the corresponding content or disabling the database ($T_{m4}$) | News video, audio, text and fusion information ($g_4$) |
| 5 | Information processing ($I_5$) | News database ($o_5$) | From entering news video, audio, text and fusion information into the information database to deleting the corresponding content or disabling the database ($T_{h5}$) | News video, audio, text and fusion information ($f_5$) | News Media Production Subsystem ($c_5$) | From the news media began to accept video, audio, text and fusion information to the completion of the production of publishing content editing. ($T_{m5}$) | News video, audio and text content with broadcasting and publishing conditions ($g_5$) |
| 6 | Information transmission ($I_6$) | News media production subsystem ($o_6$) | From the news media began to accept video, audio, text and fusion information to the completion of the production of publishing | News video, audio and text content with broadcasting and publishing conditions ($f_6$) | Internet transmission link ($c_6$) | From the beginning to the end of the transmission of news articles ($T_{m6}$) | Digital coding of news manuscript ($g_6$) |
| | | | content editing. ($T_{h6}$) | | | | |
| 7 | Information action ($I_7$) | Internet transmission link ($o_7$) | From the beginning to the end of the transmission of news articles ($T_{h7}$) | Digital coding of news manuscript ($f_7$) | Television, radio, newspapers, mobile phones, etc. ($c_7$) | From the beginning to the end of the news broadcast and reading ($T_{m7}$) | News video, audio, web pages, text, etc. ($g_7$) |

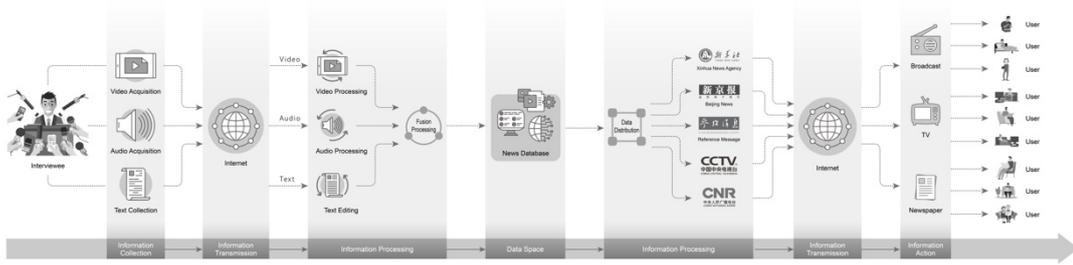

**Figure. 2: Information flows during the news gathering and releasing process**

## 5.2 Primary Properties of Information

Several particularly important fundamental properties of the information can be further discussed according to the sextuple model.

**（1）Objectivity**

By definition of 1, The information $I = \langle o, T_h, f, c, T_m, g \rangle$ is a full mapping of $f(o, T_h)$ to $g(c, T_m)$, Thus the information $I$ is only passed through $g(c, T_m)$. Instead, $g(c, T_m)$ Is the set of States of the carrier $c$ in the objective world at time $T_m$, So information can only be reflected through the objective world. Therefore, we say that information must belongs to the objective world, which is the objectivity of information. Due to the objectivity of information, people can collect, transmit, process, gather, and apply information through a large number of determined methods.

**Corollary 1** Information can be perceived and measured through things in the objective world.

**Proof**: For information $I = \langle o, T_h, f, c, T_m, g \rangle$, the set of reflections $g(c, T_m)$ is the set of the states of carrier $c$ in the objective world at time $T_m$. According to the division of the real-world by British philosopher Popper [233], the objective world consists of two parts, one is the world of physical objects or physical states, and the other is the world of the objective content of thought, especially the world of scientific thought, poetic thought and artistic works. The former exists in the form of nature or behavior, while the latter exists in the form of books, paintings, audio and video, data, etc. Therefore, whatever form the carrier $c$ belongs to, it is obvious that its state set $g(c, T_m)$ can be sensed and measured.

At the same time, because $g(c, T_m)$ is the mapping $I(f(o, T_h))$ of the state set $f(o, T_h)$ of the noumenon $o$ at the occurrence time $T_h$, as long as certain conditions are met, the content of $f(o, T_h)$ can always be perceived or measured from $g(c, T_m)$, indicating that the main content of information $I$ can be perceived and measured through things in the objective world.

The corollary is proved.

It is because of the objectivity of information that we call the theoretical systems derived from definition 1 and definition 2 *Objective Information Theory* (OIT). Also, because of the objectivity of information, people can collect, transmit, process, aggregate and apply information through a large number of definite methods. In fact, a large number of information systems that promote human society to enter the information age are objective systems. Despite the rapid development of new technologies such as artificial intelligence, brain-like systems and brain-computer interfaces, their main role is to use advanced equipment to fully simulate human's way of thinking or accept human's subjective consciousness, and transform them into and objectively existing information that can be processed by information systems. Therefore, OIT plays a fundamental supporting role in the analysis and research of information systems and technologies.

（2）**Reducibility**

Information $I = \langle o, T_h, f, c, T_m, g \rangle$ i a surjective mapping of $f(o, T_h)$ onto $g(c, T_m)$, If this mapping is still injective that is for any $o_\lambda \in o, T_{h\lambda} \in T_h, f_\lambda \in f, o_\mu \in o, T_{h\mu} \in T_h, f_\mu \in f$, if
$$f_\lambda(o_\lambda, T_{h\lambda}) \neq f_\mu(o_\mu, T_{h\mu})$$

There must be
$$I(f_\lambda(o_\lambda, T_{h\lambda})) \neq I(f_\mu(o_\mu, T_{h\mu})),$$

In this case, $I$ is an invertible mapping. That is, there is an inverse map $I$ of $I^{-1}$, To make any group $c_\lambda \in c, T_{m\lambda} \in T_m, g_\lambda \in g$, There is a unique set of $o_\lambda \in o, T_{h\lambda} \in T_h, f_\lambda \in f$, such that
$$I^{-1}(g_\lambda(c_\lambda, T_{m\lambda})) = f_\lambda(o_\lambda, T_{h\lambda}),$$
thus
$$I^{-1}(g(c, T_m)) = f(o, T_h). \tag{5.2.1}$$

It can be seen that according to $g(c, T_m)$ and $I^{-1}$ can be reduced to the state $f(o, T_h)$ of $o$ on $T_h$, In this case, we call the information $I$ reducible. $f(o, T_h)$ is also called the reduction state of $I$ information.

**Corollary 2** For reducible information $I = \langle o, T_h, f, c, T_m, g \rangle$, if the set of the states $f(o, T_h)$ is a mathematical object, then the mathematical structure on the set of reflections $g(c, T_m)$ can be defined such that $g(c, T_m)$ is isomorphic to $f(o, T_h)$.

**Proof**: Since $I = \langle o, T_h, f, c, T_m, g \rangle$ is reducible information, $I$ is a one-to-one surjective mapping from $f(o, T_h)$ to $g(c, T_m)$. Therefore, for any set $o_\lambda \in o, T_{h\lambda} \in T_h, f_\lambda \in f$, there is a unique set $c_\lambda \in c, T_{m\lambda} \in T_m, g_\lambda \in g$, such that $I(f_\lambda(o_\lambda, T_{h\lambda})) = g_\lambda(c_\lambda, T_{m\lambda})$. Thus, for any subset $\{I_\lambda = \langle o_\lambda, T_{h\lambda}, f_\lambda, c_\lambda, T_{m\lambda}, g_\lambda \rangle | \lambda \in \Lambda$ is the index set$\}$ in $f(o, T_h)$, if and only if all elements of $\{f_\lambda(o_\lambda, T_{h\lambda}) | \lambda \in \Lambda$ is the index set$\}$ in $f(o, T_h)$ form a mathematical structure, $\{g_\lambda(c_\lambda, T_{m\lambda}) | \lambda \in \Lambda$ is the index set$\}$ in $g(c, T_m)$ is defined to form the same mathematical structure. Obviously, the isomorphism of $g(c, T_m)$ and $f(o, T_h)$ can be realized at this time.

The corollary is proved.

The isomorphism between the set of reducible information noumenon states and the set of carrier reflection states is of great significance. Therefore, the same mathematical method can be applied to two different sets of information noumenon states and carrier states, the objects in the two sets have the same attributes and operations, and the proposition established on one set can also be established on the other set, which opens a convenient door for using abundant mathematical theories to support extensive information science research. The detailed proof of the internal relations of sets can show the special role in this respect later in this paper.

（3）**Transitivity**

Information $I = \langle o, T_h, f, c, T_m, g \rangle$ is a surjective mapping of $f(o, T_h)$ to $g(c, T_m)$, Of course, it does not exclude the existence of the set $c'$ in the objective world, the time set $T'_m$ and the set $g'(c', T'_m)$ of all the States of $c'$ on $T'_m$. And form a surjective map of $g(c, T_m)$ onto $g'(c', T'_m)$. By definition, this map

$$I': g(c, T_m) \to g'(c', T'_m) \qquad (5.2.2)$$

It's also information, and

$$I'(g(c, T_m)) = I'(I(f(o, T_h))), \qquad (5.2.3)$$

In fact, $f(o, T_h)$ is the composite map from $g(c, T_m)$ to $g'(c', T'_m)$.

It can be seen that the composite mapping $I'(I(f(o, T_h)))$ completes the transmission from $o$ to $c$ and then to $c'$. $T_h$ to $T_m$ then to $T'_m$, $f(o, T_h)$ to $g(c, T_m)$ then to $g'(c', T'_m)$. This is the transmission of information. It is precisely because of the transmission of information that people can realize the effective transmission of information through various media.

**Corollary 3 (Serial information transmission chain)** Let a set $\{I_i = \langle o_i, T_{hi}, f_i, c_i, T_{mi}, g_i \rangle | i = 1,,, n\}$ be a set of reducible information, if for any $i < n$, there is:

$$c_i = o_{i+1}, T_{mi} = T_{h(i+1)}, g_i = f_{i+1}.$$

Then $\{I_i = \langle o_i, T_{hi}, f_i, c_i, T_{mi}, g_i \rangle | i = 1,,, n\}$ is the serial information transmission chain between information $I_1$ and $I_n$, and there exists $\{I'_i = \langle o_1, T_{h1}, f_1, c_i, T_{mi}, g_i \rangle | i = 1,,, n\}$, which are both reducible information, and they have the same reduction state $\{I'_i = \langle o_1, T_{h1}, f_1, c_i, T_{mi}, g_i \rangle | i = 1,,, n\}$.

**Proof**: For $i = 1$, it is obvious that $I'_1 = \langle o_1, T_{h1}, f_1, c_1, T_{m1}, g_1 \rangle = I_1$ is reducible information, and its reductive state is $f_1(o_1, T_{h1})$;

Let $i < n$, $I'_i = \langle o_1, T_{h1}, f_1, c_i, T_{mi}, g_i \rangle$ is reducible information, then since
$$c_i = o_{i+1}, T_{mi} = T_{h(i+1)}, g_i = f_{i+1}$$
and $I_{i+1} = \langle o_{i+1}, T_{h(i+1)}, f_{i+1}, c_{i+1}, T_{m(i+1)}, g_{i+1} \rangle$ is also reducible information, so there exist inverse maps $I'^{-1}_i$ and $I^{-1}_{i+1}$, respectively, such that
$$I'^{-1}_i\big(g_i(c_i, T_{mi})\big) = f_1(o_1, T_{h1})$$
$$I^{-1}_{i+1}\big(g_{i+1}(c_{i+1}, T_{m(i+1)})\big) = f_{i+1}(o_{i+1}, T_{h(i+1)}) = g_i(c_i, T_{mi})$$

It can be seen that as long as $I'_{i+1} = I_{i+1}I'_i$ is set, then $I'_{i+1} = \langle o_1, T_{h1}, f_1, c_{i+1}, T_{m(i+1)}, g_{i+1} \rangle$ is the reducible information whose reduction state is $f_1(o_1, T_{h1})$.

The corollary is proved.

Serial information transmission is a very common form of information movement in information systems, and it is of great significance to analyze many mechanisms of serial information transmission chain for the construction of theoretical system of information systems dynamics.

（4）**Compositionality**

$o, T_h, f, c, T_m, g$ in information $I = \langle o, T_h, f, c, T_m, g \rangle$ is a variety of sets with different roles. Naturally, it can be decomposed or combined into several new sets, so the information is combinatorial. approach:

**Definition 3** (Sub-information) For information $I' = \langle o', T'_h, f', c', T'_m, g' \rangle$ and $I = \langle o, T_h, f, c, T_m, g \rangle$, if
$$o' \subseteq o, T'_h \subseteq T_h, f' \subseteq f, c' \subseteq c, T'_m \subseteq T_m, g' \subseteq g, \tag{5.2.4}$$
And
$$I'\big(f'(o', T'_h)\big) = I\big(f'(o', T'_h)\big), \tag{5.2.5}$$
Then $I'$ is called a subinformation of $I$, Record as:
$$I' \subseteq I. \tag{5.2.6}$$
It is called $I'$ contained in $I$. At the same time
$$o' \subset o, T'_h \subset T_h, f' \subset f, c' \subset c, T'_m \subset T_m, g' \subset g \tag{5.2.7}$$
When at least one of the formulas is valid. Call $I'$ a proper subinformation of $I$. Record as $I' \subset I$, $I'$ is said to be properly contained in $I$.

**Definition 4 (Combined information)** For message $I = \langle o, T_h, f, c, T_m, g \rangle$ and its two proper sub-messages $I' = \langle o', T'_h, f', c', T'_m, g' \rangle$ and $I'' = \langle o'', T''_h, f'', c'', T''_m, g'' \rangle$, if
$$o = o' \cup o'', T_h = T'_h \cup T''_h, f = f' \cup f'', c = c' \cup c'', T_m = T'_m \cup T''_m, g = g' \cup g'', \tag{5.2.8}$$
And for any $o_\lambda \in o, T_{h\lambda} \in T_h, f_\lambda \in f, c_\lambda \in c, T_{m\lambda} \in T_m$,
$$I\big(f(o_\lambda, T_{h\lambda})\big) = I'\big(f'(o_\lambda, T_{h\lambda})\big) 或 I''\big(f''(o_\lambda, T_{h\lambda})\big), \tag{5.2.9}$$
Then $I$ is said to be a combination of $I'$ and $I''$, record as
$$I = I' \cup I''. \tag{5.2.10}$$

The combinatorial of information determines that information can be flexibly split and can be arbitrarily combined, which creates sufficient conditions for people to determine the objects of information processing according to actual needs.

**Corollary 4** Let information $I' = \langle o', T_h', f', c', T_m', g' \rangle$ be the sub-information of information $I = \langle o, T_h, f, c, T_m, g \rangle$. If $I$ is reducible information, then $I'$ is also reducible information.

**Proof**: To prove that information $I' = \langle o', T_h', f', c', T_m', g' \rangle$ is reducible, it is enough to prove that $I'$ is a one-to-one mapping from $f'(o', T_h')$ to $g'(c', T_m')$. In fact, if $I'$ is not a one-to-one mapping, then there are two distinct sets of $o_1', o_2' \in o', T_{h1}', T_{h2}' \in T_h', f_1', f_2' \in f'$ and a set of $c_1' \in c', T_{m1}' \in T_m', g_1' \in g'$ such that
$$I'(f_1'(o_1', T_{h1}')) = I'(f_2'(o_2', T_{h2}')) = g_1'(c_1', T_{m1}')$$
However, according to the definition of sub-information, there is
$$I(f_1'(o_1', T_{h1}')) = I'(f_1'(o_1', T_{h1}')) = g_1'(c_1', T_{m1}') = I'(f_2'(o_2', T_{h2}')) = I(f_2'(o_2', T_{h2}'))$$
This is a contradiction to $I$ reducible. So $I'$ is also reducible information.

The corollary is proved.

（5）**Relevance**

The association of the information is shown in at least three ways.

First, for information A $I = \langle o, T_h, f, c, T_m, g \rangle$, $o$ and $c$, $T_h$ and $T_m$, $f(o, T_h)$ and $g(c, T_m)$ both come in pairs. In particular $f(o, T_h)$ and $g(c, T_m)$, $I$ as a surjective map of $f(o, T_h)$ onto $g(c, T_m)$, A particular connection is established between the States in which a pair of things, $o$ and $c$, are situated. And because of the transmission, information can also connect more things together, which is an important embodiment of information relevance, so people often say that information is the bridge of all things.

Second, because information can be decomposed into several sub-information, there may be a relationship between different information, or jointly contained in another information. It can be seen that various mutual relationships can be established between information and each other, which will be another form of information correlation, by which people can analyze and utilize the various relationships between information.

Then, by analyzing the internal structure of the information, the most important embodiment of the information correlation is that it can reflect the various relationships within its reduced state. It can be proved that the restored information can completely retain the internal correlation structure of its reduced state, which provides an important prerequisite for people to process, analyze and utilize the internal structure of the information.

**Corollary 5** For reducible information $I = \langle o, T_h, f, c, T_m, g \rangle$, if $R$ is an equivalence relation on the set of states $f(o, T_h)$, then there must be an equivalence relation $Q$ on the set of states $g(c, T_m)$, such that for any two sub-information $I_\lambda = \langle o_\lambda, T_{h\lambda}, f_\lambda, c_\lambda, T_{m\lambda}, g_\lambda \rangle$ and $I_\mu = \langle o_\mu, T_{h\mu}, f_\mu, c_\mu, T_{m\mu}, g_\mu \rangle$ of $I$, if $f_\lambda(o_\lambda, T_{h\lambda}) R f_\mu(o_\mu, T_{h\mu})$, there must be $g_\lambda(c_\lambda, T_{m\lambda}) Q g_\mu(c_\mu, T_{m\mu})$.

**Proof**: For any two groups of $c_\lambda \in c, T_{m\lambda} \in T_m, g_\lambda \in g$ and $c_\mu \in c, T_{m\mu} \in T_m, g_\mu \in g$, since $I = \langle o, T_h, f, c, T_m, g \rangle$ is reducible information, when $g_\lambda(c_\lambda, T_{m\lambda}) \neq g_\mu(c_\mu, T_{m\mu})$, there must be $o_\lambda \in o, T_{h\lambda} \in T_h, f_\lambda \in f, o_\mu \in o, T_{h\mu} \in T_h, f_\mu \in f$ such that $I(f_\lambda(o_\lambda, T_{h\lambda})) = g_\lambda(c_\lambda, T_{m\lambda})$, $I(f_\mu(o_\mu, T_{h\mu})) = g_\mu(c_\mu, T_{m\mu})$, and $f_\lambda(o_\lambda, T_{h\lambda}) \neq f_\mu(o_\mu, T_{h\mu})$. From this, the relation $R$ can be defined in terms of the relation $Q$, that is, if and only if $f_\lambda(o_\lambda, T_{h\lambda}) R f_\mu(o_\mu, T_{h\mu})$, $g_\lambda(c_\lambda, T_{m\lambda}) Q g_\mu(c_\mu, T_{m\mu})$.

Because $R$ is an equivalence relation, it has reflexivity, that is, $f_\lambda(o_\lambda, T_{h\lambda}) R f_\lambda(o_\lambda, T_{h\lambda})$. According to the definition of $Q$, when $I(f_\lambda(o_\lambda, T_{h\lambda})) = g_\lambda(c_\lambda, T_{m\lambda})$, $g_\lambda(c_\lambda, T_{m\lambda}) Q g_\lambda(c_\lambda, T_{m\lambda})$ holds. This shows that $Q$ is also reflexive.

Similarly, $R$ is of symmetry, that is, $f_\mu(o_\mu, T_{h\mu}) R f_\lambda(o_\lambda, T_{h\lambda})$ can be derived from $f_\lambda(o_\lambda, T_{h\lambda}) R f_\mu(o_\mu, T_{h\mu})$. As $I(f_\lambda(o_\lambda, T_{h\lambda})) = g_\lambda(c_\lambda, T_{m\lambda})$, $I(f_\mu(o_\mu, T_{h\mu})) = g_\mu(c_\mu, T_{m\mu})$, according to the definition of $Q$, when $g_\lambda(c_\lambda, T_{m\lambda}) Q g_\mu(c_\mu, T_{m\mu})$, there must be $f_\lambda(o_\lambda, T_{h\lambda}) R f_\mu(o_\mu, T_{h\mu})$, so $f_\mu(o_\mu, T_{h\mu}) R f_\lambda(o_\lambda, T_{h\lambda})$, and then $g_\mu(c_\mu, T_{m\mu}) Q g_\lambda(c_\lambda, T_{m\lambda})$. This shows that $Q$ is also of symmetry.

In addition, $R$ is transitive, that is, there is another group $o_\delta \in o, T_{h\delta} \in T_h, f_\delta \in f$ such that $f_\lambda(o_\lambda, T_{h\lambda}) R f_\mu(o_\mu, T_{h\mu})$ and $f_\mu(o_\mu, T_{h\mu}) R f_\delta(o_\delta, T_{h\delta})$ entail $f_\lambda(o_\lambda, T_{h\lambda}) R f_\delta(o_\delta, T_{h\delta})$. In this case, let $c_\delta \in c, T_{m\delta} \in T_m, g_\delta \in g$, and $I(f_\delta(o_\delta, T_{h\delta})) = g_\delta(c_\delta, T_{m\delta})$, when $g_\lambda(c_\lambda, T_{m\lambda}) Q g_\mu(c_\mu, T_{m\mu})$ and $g_\mu(c_\mu, T_{m\mu}) Q g_\delta(c_\delta, T_{m\delta})$, according to the definition of $Q$, there must be $f_\lambda(o_\lambda, T_{h\lambda}) R f_\mu(o_\mu, T_{h\mu})$ and $f_\mu(o_\mu, T_{h\mu}) R f_\delta(o_\delta, T_{h\delta})$, and thus $f_\lambda(o_\lambda, T_{h\lambda}) R f_\delta(o_\delta, T_{h\delta})$, from which $g_\lambda(c_\lambda, T_{m\lambda}) Q g_\delta(c_\delta, T_{m\delta})$ can still be obtained according to the definition of $Q$. This shows that $Q$ is also transitive.

Because $Q$ has the properties of reflexivity, symmetry and transitivity, $Q$ is also an equivalence relation on the reflection set $g(c, T_m)$.

The corollary is proved.

## 5.3 The Metrics System of Information

In our previous work, we proposed the information measure metrics with the following principles [25][26]:

1) Traceability: the specific definitions and mathematical expressions of various metrics should be derived from the mathematical definition of information;

2) Integrity: the complete measure metrics system should be formed closely related to its value based on the actual connotation of information;

3) Generality: the information measure metrics should be applicable to all kinds of information systems for information acquisition, transmission, processing, application and their combinations, not limited to a specific field;

4) Practicality: the information measure metrics should be able to guide the design, implementation and analysis of practical information systems.

In this paper, according to theoretical research and practical experience, a fifth principle are added, namely,

5) Openness: due to the complex characteristics of information, it is difficult to come up with a complete set of measure metrics of information, and it is necessary to reasonably supplement, revise and improve the measure metrics of information according to the needs of theoretical research and engineering application.

Based on the six-tuple model and primary properties of information, [25] [26] proposed the specific definitions of nine kinds of information measures, and obtained the related basic propositions according to the properties of measure, potential and distance of sets. On this basis, this paper emphasizes that all metrics are based on reducible information, and thus revises the names and definitions of some metrics according to the latest research results. Moreover, the sampling rate and aggregation degree of information are added in this paper, the former is used to characterize the density of information state in the time domain, and the latter is used to measure the closeness of the relationship between the various components within the information, both of which are of great guiding significance in the design and implementation of information systems. On the whole, the order of the metrics is adjusted to put the most understandable information metrics such as volume and delay in the first place, so that readers can understand the connotation and significance of the whole information metrics system from shallow to deep.

**Definition 5 (Volume of information)** Let $O$ denote the set of objective world, $(O, 2^O, \sigma)$ be a measure space, and $\sigma$ be some measure on the set $O$. The capacity $\text{volume}_\sigma(I)$ of reducible information $I = \langle o, T_h, f, c, T_m, g \rangle$ with respect to the measure $\sigma$ is the measure $\sigma(c)$ of $c$, that is

$$\text{volume}_\sigma(I) = \sigma(c). \tag{5.3.1}$$

For the same set of objects, different measures can be defined mathematically according to different concerns. Therefore, the capacity of information defined here is not unique, but can be defined differently according to different needs. For the same reason, the definitions of each of the following metrics can change as concerns change.

In information systems, the capacity of information is usually measured in bits, which is the most understandable measure of information.

**Corollary 6 (Minimum reducible capacity of random event information)** Suppose that the probability of a random event $x$ taking the value $X_i$ is $p_i$ （$i = 1, \ldots, n$）, $n \geq 2$, and $\sum_{i=1}^{n} p_i = 1$，and the information $I = \langle o, T_h, f, c, T_m, g \rangle$ represents the value of $x$. In this case, $I$ is called a random event information，where the noumenon $o$ itself is the random event $x$, the state occurrence time $T_h$ is the time when the random event $x$ takes the value $X_i$ （$i = 1, \ldots, n$, the state set $f(o, T_h)$ is the value $X_i$ （$i = 1, \ldots, n$）, the carrier $c$ is the object or medium that records the value of $x$, and the state reaction time $T_m$ is the time when the carrier $c$ records the value of $x$, the reflection set $g(c, T_m)$ is the specific form of the value of $x$ recorded by the carrier $c$. If $\sigma$ represents the number of bits of the carrier $c$, according to Shannon's information entropy theory, $I$ is the minimum capacity of reducible information

$$\text{volume}_\sigma(I) = -\sum_{i=1}^{n} p_i \log_2 p_i$$

and

$\text{volume}_\sigma(I) \leq \log_2 n \leq n\text{-}1$

**Proof:** For the random event information $I = \langle o, T_h, f, c, T_m, g \rangle$, the carrier $c$ can be regarded as a binary code because $f(o, T_h) = x$ can take the value $X_i$ ($i = 1, \ldots, n$).

For example, the number of bits of $c$ may be $n$, and it is defined that when $x$ takes the value $X_i$, the bit $i$ of $c$ is 1, and the other bits are 0 ($1 \leq i \leq n$), at this time, the value of $x$ can be derived according to the value of $c$, so $I$ is reducible information, and $\text{volume}_\sigma(I) = n$.

In addition, it can also be defined that when $x$ takes the value of $X_i$, $c$ takes the value of $i$ ($1 \leq i \leq n$), at this time, the value of $x$ can also be derived according to the values of $c$, and $I$ is also reducible information, and

$$\text{volume}_\sigma(I) = \begin{cases} \log_2 n, & \text{(when } \log_2 n \text{ is an integer)} \\ [\log_2 n] + 1, & \text{(when } \log_2 n \text{ is not an integer)} \end{cases}$$

In order to simplify the formula without loss of generality, it is assumed in the following discussion that $\log_2 n$ and $-\sum_{i=1}^{n} p_i \log_2 p_i$ are integers.

It can be seen that as long as $\text{volume}_\sigma(I)$ is greater than a certain value, the reducibility of $I$ can always be maintained. However, if $\text{volume}_\sigma(I)$ is less than the required value, the reducibility of $I$ cannot be maintained. According to Shannon's information entropy theory, when the probability of $x$ taking the value of $X_i$ is $p_i$ ($i = 1, \ldots, n$), $n \geq 2$, and $\sum_{i=1}^{n} p_i = 1$, the information content of $x$ is $-\sum_{i=1}^{n} p_i \log_2 p_i$, that is, the minimum reducible capacity of the information $I$ is:

$$\text{volume}_\sigma(I) = -\sum_{i=1}^{n} p_i \log_2 p_i$$

Therefore, if $\text{volume}_\sigma(I) < -\sum_{i=1}^{n} p_i \log_2 p_i$, the information $I$ is irreducible. The following proof reflects the volume-size ordering of the above three cases, respectively.

First, solving the maximum value of the strictly concave function

$$h_n: \Omega_n \to \mathbb{R}, \quad (p_1, \ldots, p_n) \mapsto -\sum_{i=1}^{n} p_i \log_2 p_i$$

where

$$\Omega_n := \{(p_1, \ldots, p_n) \mid \sum_{i=1}^{n} p_i = 1, p_j > 0, j = 1, \ldots, n, n \geq 2\}$$

Since the function $h_n(p_1,\ldots,p_n)$ is strictly concave, if the function $h_n(p_1,\ldots,p_n)$ can reach its maximum in $\Omega_n$, then the point at which it reaches its maximum is unique.

The extreme point of the function $h_n(p_1,\ldots,p_n)$ in $\Omega_n$ can be obtained by using the Lagrange multiplier method. Assume

$$F(p_1,\ldots,p_n,\lambda) = -\sum_{i=1}^{n} p_i \log_2 p_i - \lambda\left(\sum_{i=1}^{n} p_i - 1\right)$$

and $\tilde{p} = (\tilde{p}_1,\ldots,\tilde{p}_n,\lambda_0)$ is a extreme point of $F(p_1,\ldots,p_n,\lambda)$, then we have

$$\frac{\partial F}{\partial \lambda}(\tilde{p}) = 0, \qquad \frac{\partial F}{\partial p_i}(\tilde{p}) = 0, i = 1,\ldots,n.$$

It can be obtained that

$$\tilde{p}_i = \frac{1}{n}, i = 1,\ldots,n$$

that is, $\left(\frac{1}{n},\ldots,\frac{1}{n}\right)$ is the only extreme point of the function $h_n(p_1,\ldots,p_n)$, where the extreme value is $\log_2 n$.

The maximum of $h_n(p_1,\ldots,p_n)$ over $\Omega_n$ can be obtained by mathematical induction. Observing that the set $\Omega_n$ is bounded by

$$\partial\Omega_n := \{(p_1,\ldots,p_n) | \sum_{i=1}^{n} p_i = 1, p_j \geq 0, j = 1,\ldots,n, \text{there exists at least one } p_i = 0 \text{ or } 1\}$$

as

$$\lim_{x \to 0+0} x\log_2 x = 0, \qquad \lim_{x \to 1-0} x\log_2 x = 0 \tag{5.3.2}$$

Therefore, the function $h_n(p_1,\ldots,p_n)$ can be extended to $\partial\Omega_n$.

When $n = 2$, $\partial\Omega_2 = \{(1,0),(0,1)\}$ is homeomorphic to the disjoint union $\Omega_1 \sqcup \Omega_1$ of two $\Omega_1$. By (5.3.2), the function $h_2(p_1,p_2)$ attains its maximum $\log_2 2 = 1$ at the extreme point $\left(\frac{1}{2},\frac{1}{2}\right)$.

Suppose that the function $h_n(p_1,\ldots,p_n)$ attains its maximum value $\log_2 n$ at the extreme point $\left(\frac{1}{n},\ldots,\frac{1}{n}\right)$ when $n \leq k(k \geq 2)$.

When $n = k+1$, the extreme value of the function $h_{k+1}(p_1,\ldots,p_{k+1})$ at the extreme point $\left(\frac{1}{k+1},\ldots,\frac{1}{k+1}\right)$ in $\Omega_{k+1}$ is $\log(k+1)$. Since $\partial\Omega_{k+1} = \bigcup_{i=1}^{k} K_i$, where $K_i$ is homeomorphic to the disjoint union of $C_{k+1}^i$ $\Omega_i$, it is known from the induction hypothesis and (5.3.2) that the maximum value of the function $h_{k+1}(p_1,\ldots,p_{k+1})$ on $\partial\Omega_{k+1}$ is $\log k$. Therefore, the function $h_{k+1}(p_1,\ldots,p_{k+1})$ attains its maximum $\log_2(k+1)$ at the extreme point $\left(\frac{1}{k+1},\ldots,\frac{1}{k+1}\right) \in \Omega_{k+1}$.

It follows from the inductive principle that for any given natural number $n \geq 2$, the function $h_n(p_1, \ldots, p_n)$ attains its maximum $\log_2 n$ at the extreme point $\left(\frac{1}{n}, \ldots, \frac{1}{n}\right) \in \Omega_n$.

Then we prove $\log_2 n \leq n-1$. Assume that
$$f(x) = \log_2 x - x + 1, \quad x > 0$$

Since $f'(x) = \frac{1}{x \ln 2} - 1$, combined with the monotonicity of the function, one infers that $f(x)$ attains its maximum at $x = \frac{1}{\ln 2}$. Since the variable $x$ in this corollary can only take positive integers and $\frac{1}{2} < \ln 2 < 1$, we have
$$\log_2 n - n + 1 \leq \max\{f(1), f(2)\} = 0$$

The corollary is proved.

**Definition 6 (Delay of information)** The set of all atomic information in the reducible information $I = \langle o, T_h, f, c, T_m, g \rangle$ is denoted by $A = \{I_\lambda = \langle o_\lambda, T_{h\lambda}, f_\lambda, c_\lambda, T_{m\lambda}, g_\lambda \rangle\}_{\lambda \in \Lambda}$, and $\Lambda$ is the index set. Let $\mu$ be the measure of the index set $\Lambda$, and let $\mu(\Lambda) \neq 0$, without loss of generality, then the delay($I$) of information $I$ is the ratio of the integral of the difference between the supremum of the reflection time and the supremum of the occurrence time of all atomic information in $A$ to the measure of the index set $\Lambda$, that is
$$\text{delay}(I) = \frac{\int_\Lambda (\sup T_{m\lambda} - \sup T_{h\lambda}) \, d\mu}{\mu(\Lambda)},$$

Where $\sup T_{m\lambda}$ and $\sup T_{h\lambda}$ denote the supremum of $T_{m\lambda}$ and $T_{h\lambda}$ in the set $A$, respectively. Since the information processed in an information system is generally a finite set, it is usually most appropriate to take the counting measure for $\mu$. If there exists some $T_{h\lambda}$ whose supremum is $+\infty$, then define
$$\sup T_{m\lambda} - \sup T_{h\lambda} = 0. \tag{5.3.3}$$

Delay is also an easily understood information measure, because it intuitively depicts the speed of the information carrier's response to the ontology state. It should be emphasized that the definition of delay allows it to have positive and negative properties. In particular, when
$$\sup T_{m\lambda} < \sup T_{h\lambda}, \quad \text{delay}(I) < 0, \tag{5.3.4}$$

This represents the prediction of the future relevant state by the carrier prior to the ontic state occurrence time $T_h$. For example, the motion of related objects and the occurrence of related events are predicted in the information system.

The definition of delay proposed by [26] is corrected here. Compared with the definition of delay($I$) = $\max_{\lambda \in \Lambda}\{\inf T_{m\lambda} - \sup T_{h\lambda}\}$ in [26], the current $(\sup T_{m\lambda} - \sup T_{h\lambda})$ represents the delay from the end of state occurrence time to the end of reaction time in atomic information. Compared with the original $(\inf T_{m\lambda} - \sup T_{h\lambda})$, which represents the delay from the start of state occurrence time to the end of reflection time in atomic information, the former can better reflect the objective characteristics of information without considering the state occurrence time itself. In addition, the delay is defined by the ratio of the integral of the delay of all atomic information in $A$ to the measure of the index set $\Lambda$, which has more average and universal significance.

**Corollary 7 (Serial information transfer delay)** Let $\{I_i = \langle o_i, T_{hi}, f_i, c_i, T_{mi}, g_i \rangle | i = 1,,,n\}$ be the serial information transmission chain between information $I_1$ and $I_n$. According to Corollary 3, $I = \langle o_1, T_{h1}, f_1, o_n, T_{mn}, g_n \rangle$ are also reducible information. Its delay $\text{delay}(I)$ is the sum of the delays of all information $I_i$.

**Proof:** For the serial information transmission chain $\{I_i = \langle o_i, T_{hi}, f_i, c_i, T_{mi}, g_i \rangle | i = 1,,,n\}$, by its definition, for any $i < n$, we have $T_{mi} = T_{h(i+1)}$,

Therefore,

$$\sum_{i=1}^{n} \text{delay}(I_i) = \sum_{i=1}^{n} (\sup T_{mi} - \sup T_{hi})$$

$$= \sup T_{mn} - \sup T_{hn} + \sum_{i=1}^{n-1} (\sup T_{h(i+1)} - \sup T_{hi}) = \sup T_{mn} - \sup T_{h1} = \text{delay}(I)$$

the corollary is proved.

**Definition 7 (Scope of information)** Let $O$ and $S$ be the sets of the objective world and the subjective world respectively, $(O \cup S, 2^{O \cup S}, \sigma)$ be a measure space, and $\sigma$ be some measure on the set $(O \cup S)$. The breadth $\text{scope}_\mu(I)$ of reducible information $I = \langle o, T_h, f, c, T_m, g \rangle$ with respect to the measure $\sigma$ is the measure $\sigma(o)$ of $o$, that is,

$$\text{scope}_\mu(I) = \sigma(o). \tag{5.3.5}$$

**Corollary 8 (The extent of radar detection information)** Let the reducible information $I = \langle o, T_h, f, c, T_m, g \rangle$ be the radar detection information, the noumenon $o$ be the object detected by the radar, the state occurrence time $T_h$ be the time when the radar beam shines on the detected object, and the state set $f(o, T_h)$ be the state of the detected object itself and its motion. The carrier $c$ is the radar, the reflection time $T_m$ is the time for the radar to receive, process, store and display the echo signal and/or data of the detected object, and the reflection set $g(c, T_m)$ is the echo signal and/or data of the detected object for the radar to receive, process, store and display. In this case, we define the measure $\sigma$ of the noumenon $o$ as its reflection area. According to the radar equation [234], when the radar transmitting power, antenna gain, antenna effective aperture and minimum detectable signal are determined, the maximum detection range of the radar depends on the $\text{scope}_\sigma(I)$ of the information $I$ and is proportional to its quartic square root.

Proof: According to the definition of radar detection information $I = \langle o, T_h, f, c, T_m, g \rangle$, the object $o$ is the object detected by radar, and the $\text{scope}_\sigma(I)$ of $I$ is the reflection area $\sigma$ of $o$.

In the radar equation

$$R_{max}^4 = \frac{P_t G_t A_e \sigma}{(4\pi)^2 S_{min}}$$

$R_{max}$ is that maximum detection range of the radar, $P_t$ is the radar transmitting power, $G_t$ is the radar antenna gain, $A_e$ is the effective aperture of the radar antenna, $S_{min}$ is the minimum detectable signal of the radar, and $\sigma$ is the reflection area of the detected object. It can be seen that when the important parameters $P_t$、$G_t$、$A_e$ and $S_{min}$ of the radar itself are determined, the maximum detection range of the radar is completely determined by σ and is proportional to the fourth root of the scope of the information $I$.

The corollary is proved.

**Definition 8 (Atomic information)** For reducible information $I = \langle o, T_h, f, c, T_m, g \rangle$ and $I' = \langle o', T_h', f', c', T_m', g' \rangle$, if $I'$ is a proper subinformation of $I$, and there is no other proper subinformation $I'' = \langle o'', T_h'', f'', c'', T_m'', g'' \rangle$ of $I$, such that

$$I'' \subset I', \tag{5.3.6}$$

Then $I'$ is called the atomic information of $I$.

**Definition 9 (Granularity of information)** Let $O$ and $S$ be the objective world and subjective world sets respectively, $(O \cup S, 2^{O \cup S}, \sigma)$ be a measure space, $\sigma$ be some measure on the set $(O \cup S)$, the set of all atomic information in the information $I = \langle o, T_h, f, c, T_m, g \rangle$ be denoted as $A = \{I_\lambda = \langle o_\lambda, T_{h\lambda}, f_\lambda, c_\lambda, T_{m\lambda}, g_\lambda \rangle\}_{\lambda \in \Lambda}$, and $\Lambda$ be the index set. Let $\mu$ be the measure of index set $\Lambda$ and $\mu(\Lambda) \neq 0$, then the granularity granularity$_\sigma(I)$ of reducible information I with respect to measure $\sigma$ is the ratio of the integral of all atomic information ontology measures in $A$ to the measure of index set $\Lambda$, that is,

$$\text{granularity}_\sigma(I) = \frac{\int_\Lambda \sigma(o_\lambda) d\mu}{\mu(\Lambda)}, \tag{5.3.7}$$

Among them, $\mu$ is usually the most appropriate counting measure. The definition of fineness proposed in [3] is modified here, because the current criterion of using the average value to define the granularity is more average and universal than the way of using the minimum value to define the fineness in [26].

**Corollary 9 (Resolution of optical imaging information)** Let the reducible information $I = \langle o, T_h, f, c, T_m, g \rangle$ be the optical imaging information, the noumenon $o$ be the object to be photographed or photographed, the state occurrence time $T_h$ be the time when the shutter is opened or the camera is recorded, and the state set $f(o, T_h)$ be the state of the photographed object itself and its motion. The carrier $c$ is a camera or a video camera, the reflection time $T_m$ is the time for the camera or the video camera to shoot, process, store and display the photo or image of the shot object, and the reflection set $g(c, T_m)$ is the photo or image of the shot object for shooting, processing, storing and displaying. Here, we define the measure $\sigma$ of the noumenon $o$ to be the minimum distinguishable angle at the time of being photographed. According to the Rayleigh criterion [235], the resolution of the optical imaging information $I$, i.e. the granularity$_\sigma(I)$, is proportional to the light wavelength and inversely proportional to the width of the photosensitive unit.

**Proof**: For an object $o$ to be shot, each frame in the optical imaging information $I = \langle o, T_h, f, c, T_m, g \rangle$ contains a large number of pixel points, and each pixel point is a local image of $o$ that cannot be subdivided, so the pixel is the atomic information $I_\lambda = \langle o_\lambda, T_{h\lambda}, f_\lambda, c_\lambda, T_{m\lambda}, g_\lambda \rangle$, where $\lambda \in \Lambda$, and $\Lambda$ is the index set. By the definition, the granularity$_\sigma(I)$ of information $I$ is the average of the measure $\sigma(o_\lambda)$ of the noumenon $o_\lambda$ of all atomic information $I_\lambda$. From the principle of optical imaging, it can be derived that $\sigma(o_\lambda)$ is the same for all $\lambda \in \Lambda$, and according to the Rayleigh criterion,

$$\sigma(o_\lambda) = \frac{l}{a}$$

where $l$ is the wavelength of light and $a$ is the width of the photosensitive unit. It is thus clear that the granularity$_\sigma(I) = \sigma(o_\lambda)$ (for any $\lambda \in \Lambda$) of the information $I$ is proportional to the wavelength of the light and inversely proportional to the width of the photosensitive unit.

The corollary is proved.

**Definition 10 (Variety of information)** For reducible information $I = \langle o, T_h, f, c, T_m, g \rangle$, let $\Re$ be an equivalence relation on the state set $f(o, T_h)$, and the set of equivalence classes of elements in $f(o, T_h)$ with respect to $R$ is $[f(o, T_h)]_R$, then the variety variety$_R(I)$ of information $I$ with respect to $R$ is the cardinality of the set $[f(o, T_h)]_R$, that is,

$$\text{variety}_R(I) = \overline{[f(o, T_h)]_R}.$$

In this paper, we always use $\bar{X}$ to denote the cardinality of a set $X$. Reducible information can transfer the equivalence relation within the state set to the reflection set, so its information carrier can fully reflect the class of information measurement.

**Corollary 10** For reducible information $I = \langle o, T_h, f, c, T_m, g \rangle$, let $R$ be the equivalence relation on the state set $f(o, T_h)$, and the set of equivalence classes of the elements in $f(o, T_h)$ with respect to $\Re$ is $[f(o, T_h)]_R$, then there must exist an equivalence relation $Q$ on the reflection set $g(c, T_m)$ such that the set $[g(c, T_m)]_Q$ and $[f(o, T_h)]_R$ of the equivalence classes of the elements of $g(c, T_m)$ with respect to $Q$ form a one-to-one surjective relation by the information $I$, so that the cardinalities of the two equivalence classes are equal, that is

$$\text{variety}_R(I) = \overline{[f(o, T_h)]_R} = \overline{[g(c, T_m)]_Q} \qquad (5.3.8)$$

**Proof**: Corollary 5 has proved that the equivalence relation $Q$ on the reflection set $g(c, T_m)$ can be established according to the equivalence relation $R$ on the state set $f(o, T_h)$. For any two pieces of sub-information $I_\lambda = \langle o_\lambda, T_{h\lambda}, f_\lambda, c_\lambda, T_{m\lambda}, g_\lambda \rangle$ and $I_\mu = \langle o_\mu, T_{h\mu}, f_\mu, c_\mu, T_{m\mu}, g_\mu \rangle$ of $I$, if $f_\lambda(o_\lambda, T_{h\lambda}) R f_\mu(o_\mu, T_{h\mu})$, there must be $g_\lambda(c_\lambda, T_{m\lambda}) Q g_\mu(c_\mu, T_{m\mu})$. Therefore, there is also an equivalence class $[g(c, T_m)]_Q$ on $g(c, T_m)$. Because the equivalence relation $Q$ is established completely based on the mapping relation of the information $I$, the one-to-one surjective relation between the two equivalence classes $[f(o, T_h)]_R$ and $[g(c, T_m)]_Q$ can also be established completely based on the information $I$,

$$\overline{[f(o,T_h)]_R} = \overline{[g(c,T_m)]_Q}$$

The corollary is proved.

**Definition 11 (Duration of information)** The duration $\text{duration}(I)$ of reducible information $I = \langle o, T_h, f, c, T_m, g \rangle$ is the difference between the supremum and the infimum of $T_h$, that is,

$$\text{duration}(I) = \sup T_h - \inf T_h. \qquad (5.3.9)$$

**Corollary 11 (Average duration of continuous monitoring information)** The average duration of continuous monitoring information is equal to the mean time between failures (MTBF) of the information collection device.

**Proof**: Let the reducible information $I = \langle o, T_h, f, c, T_m, g \rangle$ be the continuous monitoring information, where the noumenon $o$ can be regarded as the monitored object, the state occurrence time $T_h$ can be regarded as the time period when $o$ is in the monitored state, the state set $f(o, T_h)$ can be considered as the state when $o$ is in the monitored time period, and the carrier $c$ can be considered as the information collection device; The reflection time $T_m$ can be regarded as the working period of $c$, and the reflection set $g(c, T_m)$ can be regarded as the information collected and presented by $c$. The continuous monitoring information systems is generally required to maintain continuous and uninterrupted monitoring of the monitored object, so its duration is often equal to the working period of the information collection equipment, that is,

$$\text{duration}(I) = \sup T_h - \inf T_h = \sup T_m - \inf T_m$$

However, any equipment has the possibility of failure, so it is necessary to specify the mean fault-free MTBF [236] of systems in engineering practice, indicating the duration of normal operation of the system without failure in the whole life-cycle. In continuous monitoring systems, if the MTBF of information acquisition equipment $c$ is the average of $\sup T_m - \inf T_m$ in the full life-cycle, then it can be seen that the average of continuous monitoring information $\text{duration}(I)$ is also the same.

The corollary is proved.

**Definition 12 (Sampling rate of information)** For reducible information $I = \langle o, T_h, f, c, T_m, g \rangle$, without loss of generality, let $\inf T_h < \sup T_h$, let $\{U_\lambda\}_{\lambda \in \Lambda}$ be a family of pairwise disjoint connected sets, and satisfy: for any $\lambda \in \Lambda$, there are $U_\lambda \subseteq [\inf T_h, \sup T_h]$, $T_h \cap U_\lambda = \emptyset$, where $\Lambda$ is the index set. Then the sampling rate $\text{sampling\_rate}(I)$ of information $I$ is just the ratio of the cardinality of $\Lambda$ to the Lebesgue measure $|U|$ of $U = \bigcup_{\lambda \in \Lambda} U_\lambda$, that is

$$\text{sampling\_rate}(I) = \frac{\overline{\Lambda}}{|U|}.$$

In particular, if the Lebesgue measure of $U$ is $|U| = 0$, then $\text{sanmple rate}(I) = \infty$ is defined. This indicates that the set of States of information $I$ is completely continuous in time.

**Corollary 12 (Minimum reducible sampling rate of periodic information)** For reducible information $I = \langle o, T_h, f, c, T_m, g \rangle$, if $f(o, T_h)$ is a numeric set, and there is a minimum of $T$ such that for $\forall x \epsilon o, t, t + T \epsilon T_h$, there are

$$f(x,t) = f(x, t+T)$$

In the meantime, let $\{U_\lambda\}_{\lambda \in \Lambda}$ be a family of pairwise disjoint connected sets such that for $\forall \lambda \in \Lambda$, $U_\lambda \subseteq [\inf T_h, \sup T_h]$, $T_h \cap U_\lambda = \emptyset$ and the Lebesgue measure $|U_\lambda|$ of $U_\lambda$ is the same, where $\Lambda$ is an index set. Then $I$ is named as the periodic information, and the lowest reducible sampling rate of information $I$ is equal to $1/(2T)$.

**Proof**: For periodic information $I = \langle o, T_h, f, c, T_m, g \rangle$, it is clear that $\inf T_h \neq \sup T_h$. Since $f(x, t) = f(x, t+T)$ for $\forall x \epsilon o, t, t+T \epsilon T_h$, and $T$ is the smallest value satisfying this condition. Then, for $\forall x \in o$, $f(x, t)$ contains no frequency higher than $1/T$ with respect to time $t$. According to the Nyquist sampling theorem [7], $f(x, t)$ will be completely determined with respect to time $t$ by a series of values that are no more than $T/2$ apart. In the definition of information $I$, $\{U_\lambda\}_{\lambda \in \Lambda}$ is a series of sampling intervals with equal measure, and the cardinal number $\bar{\Lambda}$ is the number of sampling intervals. Therefore, the Lebesgue measure $|U| = |U_\lambda| \bar{\Lambda}$ of $U = \bigcup_{\lambda \in \Lambda} U_\lambda$ holds for $\forall \lambda \in \Lambda$, and then the sampling rate of information $I$

$$\text{sampling\_rate}(I) = \frac{\bar{\Lambda}}{|U|} = \frac{\bar{\Lambda}}{|U_\lambda|\bar{\Lambda}} = \frac{1}{|U_\lambda|} \quad (5.3.10)$$

Holds for $\forall \lambda \in \Lambda$.

$|U_\lambda| \leq T/2$ holds for $\forall \lambda \in \Lambda$ if and only if $\text{sampling\_rate}(I) \geq 1/(2T)$, Hence, for $\forall x \in o$, the values of $f(x, t)$ are completely determined and $I$ has a definite reduction state.

The corollary is proved.

**Definition 13 (Aggregation of information)** For reducible information $I = \langle o, T_h, f, c, T_m, g \rangle$, let $\mathfrak{R}$ be the set of relations between all elements on the state set $f(o, T_h)$. If the cardinality of the set $f(o, T_h)$ is $\overline{f(o, T_h)} \neq 0$, then the aggregation degree $\text{clustering}(I)$ of $I$ is the ratio of the cardinality of the set $\mathfrak{R}$ to that of the set $f(o, T_h)$. That is

$$\text{aggregation}(I) = \frac{\overline{\mathfrak{R}}}{\overline{f(o, T_h)}} \quad (5.3.11)$$

The degree of aggregation characterizes the closeness of the relationship between the elements of the state set $f(o, T_h)$. In general, the closer the relationship between the internal elements of the state set $f(o, T_h)$, that is, the higher the degree of aggregation, the higher the value of information.

**Proof**: Because $I = \langle o, T_h, f, c, T_m, g \rangle$ is reducible information, there is a surjective map from the state set $f(o, T_h)$ to the reflection set $g(c, T_m)$, and therefore the cardinalities of the two sets are necessarily equal, that is, $\overline{g(c, T_m)} = \overline{f(o, T_h)}$. In the meantime, we can define the set of relations $\mathfrak{Q} = \{Q_R | R \in \mathfrak{R}\}$ on $g(c, T_m)$, such that for any $R \in \mathfrak{R}$, if $o_\lambda, o_\mu \in o$, $T_{h\lambda}, T_{h\mu} \in T_h$, $f_\lambda, f_\mu \in f$, and $f_\lambda(o_\lambda, T_{h\lambda}) R f_\mu(o_\mu, T_{h\mu})$, then $I(f_\lambda(o_\lambda, T_{h\lambda})), I(f_\mu(o_\mu, T_{h\mu})) \in g(c, T_m)$, and define $f_\lambda(o_\lambda, T_{h\lambda})Q_R f_\mu(o_\mu, T_{h\mu})$ such that $Q_R$ is the relation on $g(c, T_m)$. The set $\mathfrak{Q} = \{Q_R | R \in \mathfrak{R}\}$ obviously has a one-to-one surjective relationship with the set $\mathfrak{R}$, so the cardinalities are exactly equal, that is, $\overline{\mathfrak{Q}} = \overline{\mathfrak{R}}$. From this we can obtain

$$\text{aggregation}(I) = \frac{\overline{\mathfrak{R}}}{f(o, T_h)} = \frac{\overline{\mathfrak{Q}}}{g(c, T_m)}$$

The corollary is proved.

**Definition 14 (Copy of information)** For reducible information $I = \langle o, T_h, f, c, T_m, g \rangle$, if $I' = \langle o, T_h, f, c', T'_m, g' \rangle$ is also reducible information, then there are inverse mappings $I^{-1}$ and $I'^{-1}$, so that

$$I^{-1}(g(c, T_m)) = I'^{-1}(g'(c', T'_m)) = f(o, T_h) \quad (5.3.12)$$

Then the two pieces of information $I$ and $I'$ are said to be copies of each other.

**Definition 15 (Coverage of information)** Let $\{I_\lambda = \langle o_\lambda, T_{h\lambda}, f_\lambda, c_\lambda, T_{m\lambda}, g_\lambda \rangle\}_{\lambda \in \Lambda}$ be a set containing reducible information $I = \langle o, T_h, f, c, T_m, g \rangle$ and all its copies, $\Lambda$ be an index set, and $\mu$ be a measure on the index set $\Lambda$. Suppose that the target set $C$ is a subset of the objective world, $(C, 2^C, \sigma)$ is a measure space, $\sigma$ is a measure on the set $C$, $\sigma(C) \neq 0$, and $c_\lambda \subset C$ for any given $\lambda \in \Lambda$, then the pervasiveness $I$ of information coverage$_\sigma(I)$ on $C$ with respect to the measure $\sigma$ is the ratio of the integral of the measures of all $c_\lambda$ to the measure of $C$, that is

$$\text{coverage}_\sigma(I) = \frac{\int_\Lambda \sigma(c_\lambda) d\mu}{\sigma(C)}. \quad (5.3.13)$$

**Corollary 14 (The value of a network system is equal to the product of the maximum scope and the maximum coverage of the information it carries)** For reducible information $I = \langle o, T_h, f, c, T_m, g \rangle$, if its carrier $c$ is a network system composed of finite nodes, its value is equal to the product of the maximum possible values of the scope and coverage of $I$.

**Proof**: Given that carry $c$ in the reducible information $I = \langle o, T_h, f, c, T_m, g \rangle$ is a network system composed of finite node, if the number of nodes in $c$ is $n$, according to Metcalfe's law [8], the value of the network system $c$ is equal to the square of the number of nodes $n^2$. On the other hand, we can regard $I$ as the information from all the nodes in a network, so the noumenon $o$ is the network system $c$, and its measure $\sigma$ is the number of nodes. In this case, the maximum scope of information $I$ is $\text{scope}_\sigma(I) = n$. The measure of carrier $c$ is also the number of nodes, and the maximum value of the coverage of information $I$ is also $\text{coverage}_\sigma(I) = n$. It can be seen that the value of this network system is equal to the product of the maximum possible values of the scope and the coverage of information $I$.

The corollary is proved.

Also, we can conclude that the value of a network system is equal to the square of the maximum possible scope of information $I$, or the square of the maximum possible coverage of information $I$. However, from the perspective of information movement, the value of a network system is best expressed by the product of the scope of the information it carries and the maximum possible value of coverage.

**Definition 16 (Reflection and reflection state of information)** For reducible information $I = \langle o, T_h, f, c, T_m, g \rangle$, if there is a mapping $J$ such that $J(g(c, T_m)) = \tilde{f}(\tilde{o}, \widetilde{T_h})$, where $\tilde{o} \in 2^{O \cup S}$, $\widetilde{T_h} \in 2^T$, and $\tilde{f}(\tilde{o}, \widetilde{T_h})$ are a certain set of States of $\tilde{o}$ on $\widetilde{T_h}$, then $J$ is called a reflection of $I$, and $\tilde{f}(\tilde{o}, \widetilde{T_h})$ is the reflection state of $I$ based on $J$.

It can be seen that when $J = I^{-1}$ is mapped, $\tilde{f}(\tilde{o}, \widetilde{T_h})$ is the reduced state of $I$.

**Definition 17 (Distortion of information)** For reducible information $I = \langle o, T_h, f, c, T_m, g \rangle$, let the set of States $f(o, T_h)$ and the reflected state $\tilde{f}(\tilde{o}, \widetilde{T_h})$ based on $J$ be elements in the distance space $\langle \mathcal{F}, d \rangle$, where $d$ is the distance on $\mathcal{F}$. Then the distortion $\text{distortion}_J(I)$ of the reflection $J$ of information $I$ in the distance space $\langle \mathcal{F}, d \rangle$ is the distance between $\tilde{f}(\tilde{o}, \widetilde{T_h})$ and $f(o, T_h)$, that is,

$$\text{distortion}_J(I) = d(f, \tilde{f}). \tag{5.3.14}$$

It can be seen that the degree of distortion is the degree of deviation between the reflected state and the reduced state. The reflected state of the information $I$ is its reduced state if and only if the degree of distortion $\text{distortion}_J(I) = 0$.

**Corollary 15 (Minimum distortion estimation method for discrete linear stochastic systems)** Let $I = \langle o, T_h, f, c, T_m, g \rangle$ be the state information of a discrete linear stochastic system, of which the motion and measurement are both affected by Gaussian white noise, then the minimum distortion estimation of $I$ can be obtained based on the reflection $J$ of the Kalman filter.

**Proof**: For the state information $I = \langle o, T_h, f, c, T_m, g \rangle$ of a discrete linear stochastic system, in which the motion and measurement are both affected by Gaussian white noise, the noumenon $o$ is the system itself, and the set of occurrence times $T_h$ is a series of time sequences with equal intervals, which can be written as $1, 2, \ldots, k$, the state set $f(o, T_h)$ can be written as $x(k), k = 1, 2, \cdots$, and

$$x(k) = Ax(k-1) + BU(k) + W(k)$$

holds, where $x(k)$ is the system state at time $k$ and $U(k)$ is the system input at time $k$. $A$ and $B$ are system parameters, which are matrices for a multi-model system, $W(k)$ is the motion noise of the system, and $Q$ is its covariance.

The carrier $c$ is a measuring system, the reflection time set $T_m$ is the same as the occurrence time set $T_h$, which is also denoted by $1, 2, \cdots k, \cdots$, and the reflection set $g(c, T_m)$ is a series of measured values on $T_m$, which is denoted by $z(k), k = 1, 2, \ldots$, and

$$z(k) = Hx(k) + V(k)$$

holds, where $z(k)$ is the measured value at time $k$, $H$ is the measuring system parameter, and for a multi-measurement system, $H$ is a matrix, $V(k)$ is the measuring noise at time $k$, and $R$ is its covariance.

If $J$ consists of the following five formulas:

$$x(k|k-1) = Ax(k-1|k-1) + BU(k)$$

where $x(k|k-1)$ is the result predicted by the previous state, $x(k-1|k-1)$ is the optimal result of the previous state, and $U(k)$ is the current system input state.

$$P(k|k-1) = AP(k|k-1)A^T + Q$$

where $P(k|k-1)$ is the covariance corresponding to $x(k|k-1)$, $P(k-1|k-1)$ is the covariance corresponding to $x(k-1|k-1)$, $A^T$ is the transpose matrix of $A$, and $Q$ is the covariance of the motion of the system.

$$x(k|k) = x(k|k-1) + G(k)(z(k) - Hx(k|k-1))$$

where $G(k)$ is the Kalman gain.

$$G(k) = P(k|k-1)H^T / (HP(k|k-1)H^T + R)$$
$$P(k|k) = (I - G(k)H)P(k|k-1)$$

where $I$ is the identity matrix.

It is obvious that $J$ can be recursively inferred to be the surjective map from $z(k)$ to $x(k)$, that is, from $g(c, T_m)$ to $f(o, T_h)$, so $J$ is a reflection of $I$. Consequently, according to the Kalman filtering principle [9], $x(k|k)$ is the optimal estimation of $x(k)$, that is, the minimum distortion estimation of $I$ can be obtained based on the reflection $J$ of the Kalman filter.

The corollary is proved.

**Definition 18 (Mismatch of information)** Let the target information $I_0 = \langle o_0, T_{h0}, f_0, c_0, T_{m0}, g_0 \rangle$ be reducible information, and for reducible information $I = \langle o, T_h, f, c, T_m, g \rangle$, let $o_0$ and $o$, $T_{h0}$ and $T_h$, $f_0$ and $f$, $c_0$ and $c$, $T_{m0}$ and $T_m$, $g_0$ and $g$ be elements in the set $\mathcal{P}_o, \mathcal{P}_{T_h}, \mathcal{P}_f, \mathcal{P}_c, \mathcal{P}_{T_m}, \mathcal{P}_g$ and $I_0$ and $I$ be elements in the distance space $\langle (\mathcal{P}_o, \mathcal{P}_{T_h}, \mathcal{P}_f, \mathcal{P}_c, \mathcal{P}_{T_m}, \mathcal{P}_g), d \rangle$, respectively. Then the mismatch degree $\text{mismatch}_{I_0}(I)$ of the information $I$ to the target information $I_0$ is the distance between them in the distance space $\langle (\mathcal{P}_o, \mathcal{P}_{T_h}, \mathcal{P}_f, \mathcal{P}_c, \mathcal{P}_{T_m}, \mathcal{P}_g), d \rangle$, that is,

$$\text{mismatch}_{I_0}(I) = d(I, I_0). \tag{5.3.15}$$

**Corollary 16 (Average search length of minimum mismatch information for search algorithms)** Let the target information $I_0 = \langle o_0, T_{h0}, f_0, c_0, T_{m0}, g_0 \rangle$ and the set $\{I_i = \langle o_i, T_{hi}, f_i, c_i, T_{mi}, g_i \rangle | i = 1,,,n\}$ be reducible information, $o_0$ and $o_i$, $T_{h0}$ and $T_{hi}$, $f_0$ and $f_i$, $c_0$ and $c_i$, $T_{m0}$ and $T_{mi}$, $g_0$ and $g_i$ are elements of the sets $\mathcal{P}_o, \mathcal{P}_{T_h}, \mathcal{P}_f, \mathcal{P}_c, \mathcal{P}_{T_m}, \mathcal{P}_g$, respectively, and $I_0$ and $I_i(i = 1,,,n)$ are elements of the distance space $\langle (\mathcal{P}_o, \mathcal{P}_{T_h}, \mathcal{P}_f, \mathcal{P}_c, \mathcal{P}_{T_m}, \mathcal{P}_g), d \rangle$. Let $1 \leq m \leq n$ such that

$$mismatch_{I_0}(I_m) < mismatch_{I_0}(I_i), \quad 1 \leq i \leq n \text{ and } i \neq m$$

Then the average search length (ASL) from $\{I_i | i = 1,,,n\}$ to $I_m$ is not only related to the mismatch $mismatch_{I_0}(I_m)$, but also to the different search algorithms.

**Proof**: With the increasingly abundant information content of the Internet and the increasing number of information query tools used by people, there have been a large number of complex information queries. Not only do users have requirements for the noumenon, occurrence time and state of information, but also the requirements for the carrier, reflection time and mode of information, while it is difficult to clearly describe all these requirements. Therefore, it is difficult to find the target information that completely matches the needs of users. Advanced retrieval or intelligent recommendation systems often analyze and estimate the target information $I_0$ that meets the user's needs for a specific user's application scenario, and then search and calculate the information $I_m$ with the minimum mismatch degree $mismatch_{I_0}(I_i)(i = 1,,,n)$ from the limited information set $\{I_i | i = 1,,,n\}$, which is finally pushed to the end users.

According to the ASL principle [10], the definition of ASL is:

$$ASL = \sum_{i=1}^{n} p_i c_i$$

where $p_i$ is the probability of finding the information $I_i$. In general, we assume that the probability of finding each information is the same, that is, $p_i = 1/n$; $c_i$ is the number of comparisons to find the information $I_i$. Then there are two situations to be considered respectively:

The first one is $mismatch_{I_0}(I_m) = 0$. If the sequential search method is used, the mismatch degree $mismatch_{I_0}(I_i)$ is calculated incrementally from the information $I_1$ until $I_m$ is found. So

$$ASL = \sum_{i=1}^{n} p_i c_i = \frac{1}{n} \sum_{i=1}^{n} i = \frac{1}{n} \cdot \frac{n(n+1)}{2} = \frac{n+1}{2}$$

If the bisection search method is adopted, the middle serial number information is always used as the root to divide the left and right subtrees, and each subtree is still used as the root of the middle serial number information to continuously divide in a progressive manner until the subtrees cannot be further divided. For each subtree, information $I_i$ is searched from the root, and the mismatch $mismatch_{I_0}(I_i)$ is calculated until $I_m$ is found. so

$$ASL = \sum_{i=1}^{n} p_i c_i = \frac{1}{n} \sum_{i=1}^{h} 2^{i-1} i = \frac{n+1}{n} log_2(n+1) - 1$$

where $h = log_2(n+1)$ is the height of the n information discrimination trees.

The second one is $mismatch_{I_0}(I_m) \neq 0$. In this case, the ASL is always n because we need to compare the mismatch degree, $mismatch_{I_0}(I_i)(i = 1,,,n)$, of all information and select the minimum one to obtain $I_m$. When $n$ is very large, the amount of searching computation is also very large owing to ASL. For this reason, an appropriate threshold value can be set, and the search is completed when the mismatch degree $mismatch_{I_0}(I_i)$ is less than or equal to the threshold value, so as to sufficiently reduce the ASL.

The corollary is proved.

[25][26] have pointed out that Shannon information entropy is actually the information volume required by a communication system to transmit discrete messages. In fact, it is not difficult to provide that the eleven types of information measures defined in this paper can find corresponding instances from classical or common information science theories or principles (Table 3).

**Table 3: Corresponding common instances of information measure metrics**

| Metrics | Classical/common theories | Basic inference |
|---|---|---|
| Volume | Shannon information entropy | The minimum reducible volume of random event information is its information entropy. |
| Delay | Whole and partial delay principle | The overall delay of serial information transmission is equal to the sum of the delays of each link. |
| Scope | Radar equation [237] | The extent of radar detection information is directly proportional to the square root of transmitting power, antenna aperture and antenna gain, and inversely proportional to the square root of detection sensitivity. |
| Granularity | Rayleigh criterion for optical imaging [238] | The granularity of the optical imaging information is proportional to the wavelength of the light and inversely proportional to the width of the sampling pore. |
| Variety | Invariance principle of reducible information type | Reducible information can keep the type of information unchanged. |
| Duration | Continuous monitoring information average duration criteria [239] | The average time of information collection of the continuous monitoring system is equal to the mean time between failures of the system. |
| Sampling-rate | Nyquist sampling theorem [240] | The lowest reducible sampling rate of the periodic function information is equal to half of its frequency. |
| Aggregation | Invariance principle of reducible information aggregation degree | The reducible information can keep the aggregation degree of the information unchanged. |
| Coverage | Metcalfe's law [241] | The value of a network system is equal to the product of the maximum scope and the maximum coverage of all the information inside. |
| Distortion | Kalman filtering principle [242] | A minimum distortion estimation method for linear systems with known measurement variances. |
| Mismatch | Average search length principle [243] | The shortest search path for minimum mismatch information in a finite set of information. |

# 6 Measured Efficacies and Dynamics Configurations of Information Systems

Information, amongst matter and energy, is one of the three elements that constitute the objective world. The dynamics theory of matter and energy has existed for a long period of time and has been deeply rooted and flourishing, which has strongly promoted the development and progress of industrial civilization. Many theoretical achievements of information science and technology, such as Nyquist sampling theorem [244], Shannon information entropy [3], Kalman filtering method [245]and etc., have revealed the profound mathematical rules in the process of information collection, transmission and processing, which can be regarded as the information dynamics principles of a certain process, and have played an extremely significant role in the development and application of information technology. However, the scope of the previous theories is limited to individual processes or tasks, from which it is difficult to fully grasp the overview of information dynamics. Since the information dynamics was put forward as a research field, there have been many efforts in this filed. However, as stated in [20], the studies in the previous work are mainly qualitative and lack of quantitative investigations. At the same time, as the dynamics system is a mathematical expression of fixed rules and mechanics, demonstrated by the experimental science that studies the motion and change of objects in space and time, is inseparable from measure and its units. Therefore, "mathematical expression" and "measure" are the imperative conditions for the study of the dynamics mechanism in a specific filed. The investigation on the formal model, primary properties and measure metrics of information in this paper lays a solid mathematical foundation to carry out the investigations on dynamics of information systems profoundly and quantitatively.

## 6.1 Measured Efficacies of Information Systems

Any information system can be simplified as a basic process of receiving input information, executing some operations, and finally producing output information. Therefore, a major significance of information systems lies in the various efficacies on the input information, which can be expressed through the output information. Without comprehensive analysis, reasonable deconstruction and quantitative expression of these functions, it is difficult to deeply understand the operation mechanism and inherent laws of information systems, and to build a theoretical system of information systems dynamics that can lead the construction and development of information systems. Therefore, an accurate understanding of the various efficacies om information systems is of decisive significance for the in-depth study of information system dynamics. In Section 3, a mathematical six-tuple model of information and an eleven-class measure metrics system are proposed, which provides the key to the theory of information systems dynamics. As any type of efficacy cannot be expressed quantitatively without measurement indicators, and there must be actual efficacy behind any kind of the information measures, it is natural to apply the information measure metrics to describe and analyze the efficacies of information systems comprehensively and quantitatively. Therefore, there are eleven types of measure efficacies on information systems, including volume-efficacy, delay-efficacy, scope-efficacy, granularity-efficacy, variety-efficacy, duration-efficacy, sampling-rate-efficacy, aggregation-efficacy, coverage-efficacy, distortion-efficacy and mismatch-efficacy.

Figure 2 reflects the information measure efficiency distribution on each part in information systems. The position of ★ indicates that the ring representing some part in information systems has the information measure efficiency of the corresponding sector. The information collection and information exertion are in the same ring at the periphery, which are distinguished by two sectors in two colors: dark color represents the information collection part, and light color represents the information exertion part. Therefore, the function and performance indicators of the whole information system can be deconstructed by the information measure efficacy distribution, which provides sufficient and quantitative basis for system design, analysis, testing and integration.

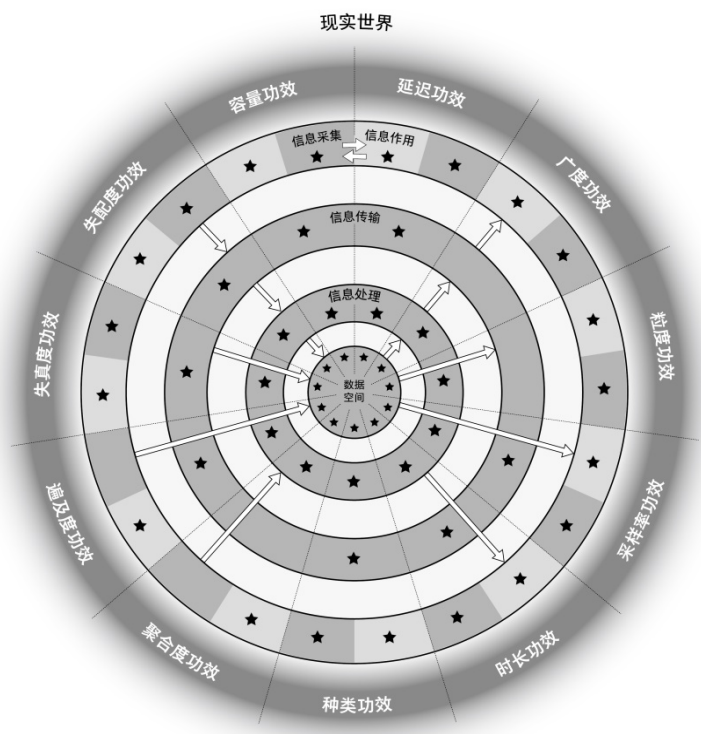

**Figure 2: Distribution of measure efficacies on each part of information systems**

Amongst the measured efficacies, volume-efficacy is the functional effect of information systems exerting on information, which leads to the change of the information volume measure. Because the volume measure depends on the carrying capacity of the information carrier, in actual information systems, each step in the process of information collection, transmission, processing, data space and exertion can affect the volume measure caused by the changes of the carrying capacity of the system (Figure 2). For example, information collection, data space and information exertion will discard part of the information and reduce the volume of information because of the insufficient storage capacity of the system. The information transmission system will discard part of the information due to the lack of channel bandwidth, which will also reduce the volume of the information. Shannon information entropy is actually the minimum information volume required by a communication system to transmit discrete messages on the premise that the information can be restored [3]. Information processing also needs enough storage space to support, so it will also affect the volume of information. In particular, information processing systems can also reduce the information volume requirement through the data compression processing, thereby actually increasing the information volume as a whole whereas the data decompression processing will reduce the information volume of the whole system. Consequently, different data compression and decompression processing have different volume-efficiencies.

Delay-efficacy is the functional effect of information system exerting on information, which leads to the change of information delay measure. In fact, all information flows and processing require a period of time, so each step of information collection, transmission, processing, data space and information exertion will certainly affect the delay measure information (Figure 2). However, through the improvement of equipment or algorithms, each step can also achieve the lowest possible delay, so as to optimize the delay efficiency of the whole system. In particular, in the information processing step, the state set of the subjects in the future can be predicted by the extrapolation algorithm in time dimension, which can actually reduce the delay measure of information and thus improve the delay-efficiency of the information system.

Scope-efficacy is the functional effect of an information system exerting on information, resulting in changes in information scope measure. Scope measure characterizes the scope of an information subject. Therefore, the information collection step will affect the scope measure of information acquisition due to the energy, distribution and other physical attributes of the collection device. For example, the most basic radar equation in the field of radio detection expresses the decisive role of physical parameters such as antenna aperture, transmitter power, receiver sensitivity on radar detection range, that is, the scope of information acquisition. The information processing step will also affect the scope-efficacy of the output information due to the differences in the means, methods and interfaces of the equipment. It is noted that the typical information processing step does not directly involve the subject of information, and consequently it could not affect the scope-efficacy of information. However, through the extrapolation algorithm in space dimension, it is possible to extend the scope of information subjects, thereby improving the scope-efficacy of information. As data space is the concrete reflection of the real-world in information systems, it will also affect the scope-efficacy of information by the integrity of the data model, database capacity and other factors. Remarkably, the volume measure of information transmission step can certainly affect the scope-efficacy of information, but in an indirect manner. In addition, there is hardly any other direct relationship between information transmission and the scope of information. In order to focus on the essential issues, we can assume that the information transmission does not directly have the scope-efficacy in the study of information systems (Figure 2).

Granularity-efficacy is the functional effect of information system acting on information, which leads to the change of information granularity measure. Granularity characterizes the meticulousness of information subjects. Therefore, the information acquisition will also affect the granularity measure of information by the aperture area of the acquisition device, the number of sensors and other physical attributes. For example, the number of photoelectric sensors integrated in the video information acquisition device determines the resolution or pixels of the video picture, which is the granularity effect of information acquisition. Information processing will also affect the granularity-efficiency of information due to the differences in the means, methods and interfaces of output devices. Similarly, through interpolation algorithms in space dimension, it is possible to increase the scope of information subjects, thereby optimizing the granularity of information. Data space can also affect the granularity of information by factors such as the model integrity, granularity, and database capacity. Similar to the analysis of scope-efficacy, the information transmission does not directly have the granularity-efficacy on information in the study of information systems (Figure 2).

Variety-efficacy is the functional effect of an information system exerting on information, resulting in changes in information variety measure. Variety measure characterizes the richness of the state set types of information subjects, and each major part of information systems can affect the variety measure of information, resulting in the variety-efficacy of the system (Figure 2). Specifically, information acquisition and processing can obtain and output different types of information due to the differences in input and output means and methods. For example, microwave acquisition and audio acquisition devices can obtain different input information, and optical output and audio output devices can also produce different output information. It can be seen that information acquisition and exertion can affect the variety-efficacy of information. With the development of network technology, communication transmission protocols for various information types have been designed and developed in the information transmission systems, which can standardize and simplify the system interface while ensuring the transmission efficiency, and has become a common way to implement the Internet and the Internet of Things. Therefore, the information transmission obviously has the variety-efficiency. Different information types naturally require different information processing methods, so the information processing obviously affects the variety-efficacy of information. The internal structure, model design and storage capacity of data space can directly affect the richness of information types, and thus affect the variety-efficacy.

Duration-efficacy is the functional effect of information system exerting on information, which leads to the change of information duration measure. Duration represents the time span of information. Therefore, the duration of information collection directly determines the duration measure of information, and the duration of information processing also affects the duration measure of output information. Although in many cases, the duration of information transmission does not necessarily affect the duration measure of output information, the duration of information transmission will directly affect the duration measure of broadcast information in the case of live broadcasting, which is often used in radio and television. In general, information processing does not directly affect the duration measure of information, but by extrapolating information processing, it can also expand the state set of information in time dimension, which consequently affects the duration measure of information. Obviously, the storage capacity and structural design of the data space will directly affect the duration measure of information, so it can be considered that all parts of information systems have duration-efficacy (Figure 2).

Sampling-rate-efficacy is the functional effect of an information system exerting on information, resulting in a change in information sampling-rate measure. The sampling-rate characterizes the occurrence density of the set of information states per unit time. Therefore, the density of information collection directly determines the sampling-rate measure of information. Nyquist sampling theorem shows that for periodic sine function curve, as long as the sampling rate is higher than half of its frequency, the original function curve can be restored by sampling information. Similarly, the frequency with which the information is processed obviously affects the intensity of the output information, which is the sampling-rate measure. In information transmission, if the bandwidth of the communication system is higher than the sampling rate of the input information, it will not affect the sampling rate of the output information, otherwise it will inevitably reduce the sampling rate. In general, information processing does not directly affect the sampling-rate measure of information, but through interpolation information processing, it can also increase the state set of information in time dimension, thus affecting the sampling-rate measure of information. Similarly, the storage capacity and structure design of data space will directly affect the sampling-rate measure of information, therefore each part in information systems has sampling-rate-efficiency (Figure 2).

Aggregation-efficacy is the functional effect of information system exerting on information, which leads to the change of information aggregation measure. Aggregation characterizes the closeness of the relationship between the elements within the information state set, as many features cannot be obtained from the fragment or local information dominance. Therefore, information collection and transmission does not directly affect the information aggregation. Through the analysis, association and fusion of information processing, the internal relationship of information state set can be found, established and expanded, which improves the information aggregation measure. The internal structure and model design of data space will directly determine the aggregation measure of information. The information exertion based on information processing and data space certainly has the aggregation measure of information to achieve a more comprehensive effect. Therefore, these three parts of information processing, data space and information exertion in information systems have the aggregation-efficacy (Figure 2).

Coverage-efficacy is the functional effect of information system exerting on information, which leads to the change of information coverage measure. Coverage reflects the pervasiveness of the carrier of the information and its copies in the target set. In general, information collection does not involve the formation of copies, so it has nothing to do with the coverage measure. Information exertion ultimately produces output information, and the scope of exertion directly reflects the coverage measure of information. In information transmission, the distribution of communication networks is the prerequisite for determining the coverage of information, so it will directly affect the coverage metrics of information. Although information processing does not directly connect or exert on end users, its gating or distribution processing can determine and control the objects of information, so it also affects the coverage metrics of information. The distributed structure design and replica distribution range in the data space are directly related to the coverage metric of information, therefore the four parts of information transmission, processing, data space and information function in information systems have the coverage-efficacy (Figure 2).

Distortion-efficacy is the functional effect of the information system acting on the information, resulting in changes in information distortion measure. It is obvious that most of information collection are physical or human-in-loop processes, which often produce errors due to various reasons, thus increases the distortion of information. Similarly, most of information exertion are also physical or human-in-loop processes in the loop, which will also affect the distortion of information. Information transmission increases the distortion of information due to the limitation of communication bandwidth, error code, packet loss and other reasons. On the one hand, information processing can increase the distortion of information due to computation errors, while it can also improve the processing accuracy through filtering, smoothing and other algorithms, thereby reducing the distortion of information. Information representation and storage in the data space affect the degree of distortion of information as well, therefore all parts in information systems have the distortion-efficacy (Figure 2).

Mismatch-efficacy is the functional effect of information system exerting on information, which leads to the change of information mismatch measure. The degree of mismatch reflects the degree to which the information deviates from the needs of a particular user. It is clear that the distortion measure should be the one that all users will pay attention to. As each part in information systems has distortion-efficiency, it can be simply inferred that each part also has mismatch-efficiency (Figure 2), and the same conclusion can be more clearly obtained by detailed analysis of each part as well.

## 6.2 Single-ring Dynamic Configuration of Information Systems

Figure 2 presents the complete framework structure of information systems and distribution of measured efficacies across it. However, in actual system construction, not all situations need to include all links, especially in order to focus on key issues, system designers often adopt simplified solutions for many mature or unnecessary technologies and products, so we can analyze several typical dynamic configurations of information systems according to Figure 2. It can be briefly divided into single-ring, double-ring, triple-ring, and triple-ring & one-core dynamic configurations.

Figure 3 shows the dynamic configuration of a single-ring of information systems, which is the simplest and probably the earliest information system configuration.

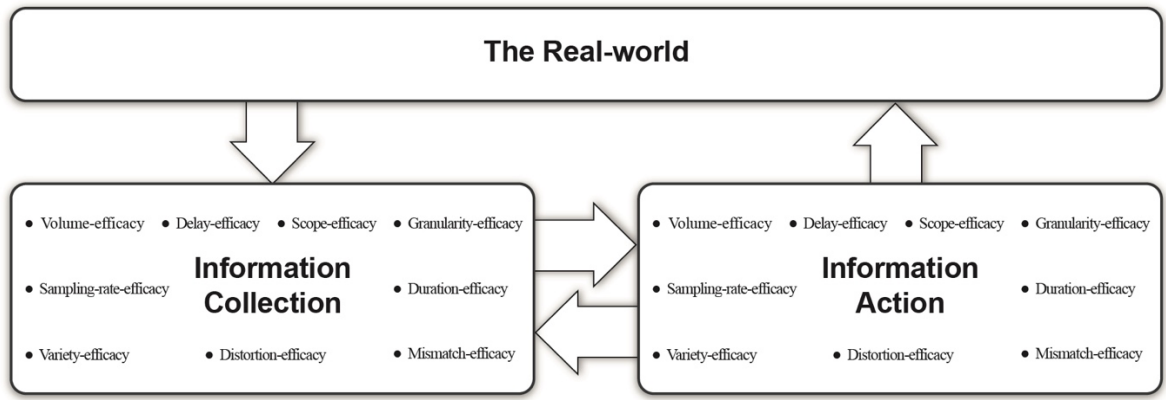

**Figure 3: Dynamical configuration of a single-ring of information systems.**

The single ring dynamic configuration only includes two links of information collection and information action in the same ring. Because of its simple structure, it may often be overlooked, but in fact it is the most classical and most common way of information system application. Typical scenarios include holding a telescope to observe the distant scenery or sitting in front of a microscope to observe the structure of cells, etc. At this time, the telescope or microscope is the simplest information system to collect the information of the observed objects in the real world and act on the observed information to the observers in the real world in time. If there is a certain difference between a simple optical telescope or microscope and the concept of modern information system, then the digital camera or video camera can better reflect the single-ring configuration in which information acquisition directly acts on the real world. Travelers hold digital cameras, digital video cameras or mobile phones owned by almost everyone to take pictures of the surrounding landscape. The lens is a part of the real world. The photos or images taken are the information collected. The photographer himself operates the equipment and enjoys the photos or images according to the lens picture. In this process, the physical parameters of the shooting equipment and the various operations of the photographer determine or adjust the capacity, delay, breadth, granularity, type, duration, sampling rate, distortion and mismatch indicators of information collection, and these corresponding indicators also directly determine the measurement efficacy of information. It should be noted that advanced shooting equipment may already have strong information processing, storage and even transmission functions, but there is no doubt that these functions are in a subordinate position in the shooting process, which is significantly different from the concepts of information transmission, information processing and data space of general information systems. Therefore, we still incorporate the shooting process of advanced digital equipment into the single-ring information system configuration. Because there is no information processing or data space link to provide the relationship structure within the information in the single-ring configuration, the information action in this configuration will not produce the effect of aggregation, and because there is no information transmission link, the information action will not have a significant impact on the effect of coverage. A very important example in the history of scientific development is that Nobel laureates Penzias and Wilson observed the microwave background radiation with a radio detection device in 1964, which is a typical application of single-ring information system configuration. The scope of the information obtained covers the entire universe, and the delay is almost as long as the life of the Universe。

## 6.3 Double-rings Dynamic Configuration of Information Systems

Because the double-ring configuration involves three different information system links, there are many different information movement modes. Fig. 4 presents the collection-transmission-action double-ring dynamic configuration of the information systems.

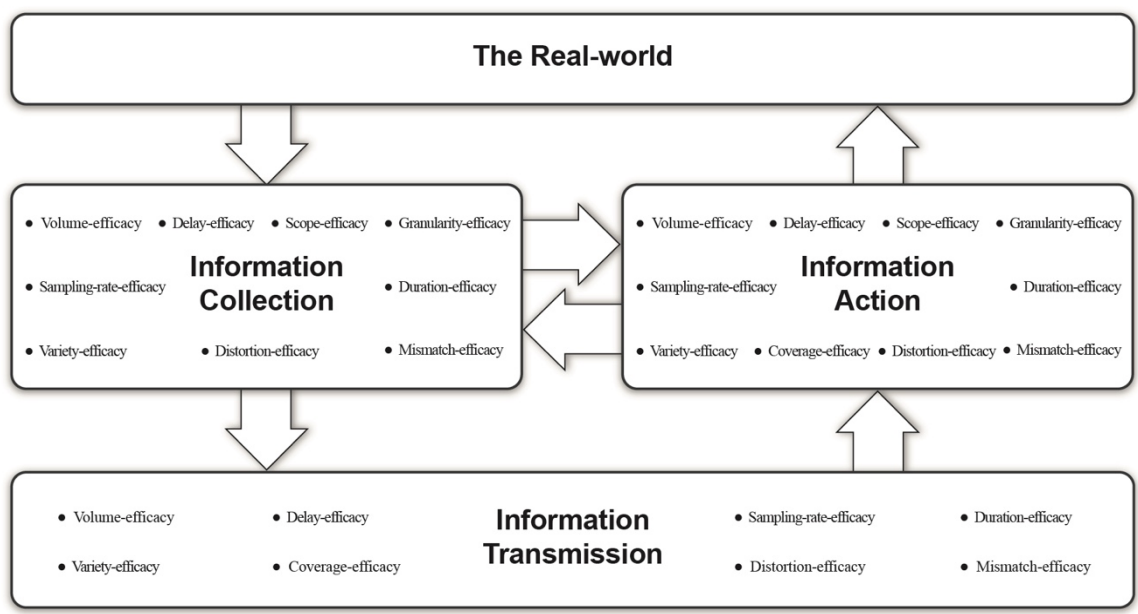

**Figure 4: Two-ring dynamic configuration of information acquisition-transmission-action systems**

A typical scenario for double-ring configuration of acquisition-transmission-action is a live broadcast of radio or television. In this case, audio or video information is collected by recording or camera equipment, transmitted to millions of households through a wide area communication network, and acted on a large number of listeners or viewers through radio or television equipment. As with the single-ring configuration, the physical characteristics of the acquisition equipment, as well as the various operations of field personnel, determine or adjust the volume, delay, scope, granularity, type, duration, sampling rate, distortion, and mismatch metrics of information acquisition. The information transmission link will affect the volume, delay, type, duration, sampling rate, coverage, distortion and mismatch of the transmitted information due to the physical characteristics of the channel, communication protocol and network distribution. In view of the fact that the transmission link is not directly related to the specific content of information, it can be considered that it does not affect the scope and granularity of live broadcasting. The final information action link, facing tens of millions of listeners or viewers, will obviously produce different volume, delay, scope, granularity, type, duration, sampling rate, coverage, distortion and mismatch efficiency for all kinds of listeners and viewers based on the received information metrics, combined with the physical characteristics of audio or video output devices.

The information system can also fully process the collected information to obtain the required results and directly act on the real-world, thus forming a double-ring dynamic configuration of collection-processing-action (Figure 5).

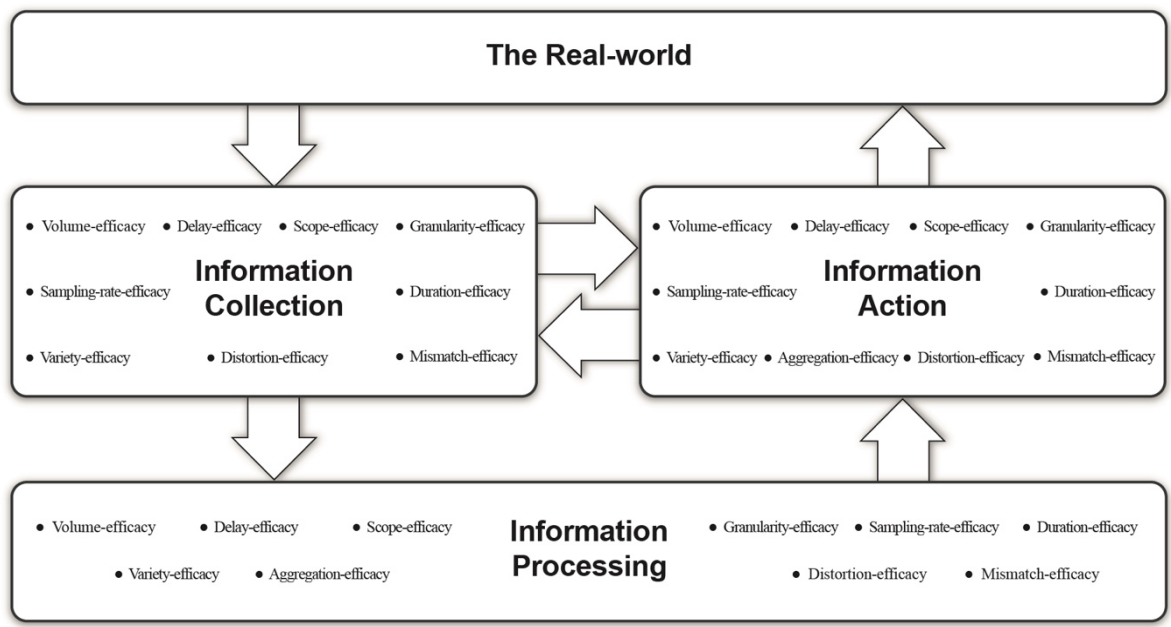

**Figure 5: Two-ring dynamic configuration of information acquisition-processing-action systems**

A typical scenario of the acquisition-processing-action double-loop configuration is a supercomputer, which collects input information in a specific field through a variety of peripherals, performs high-speed, large-capacity intensive computing or massive data processing, and generates the required result information for specific users. Among them, the physical characteristics of peripheral acquisition equipment and various operations of staff obviously affect the volume, delay, scope, granularity, variance, duration, sampling rate, distortion and mismatch of information collected by supercomputers. In the process of supercomputer processing, huge CPU processing units, high-speed computing algorithms, parallel processing software and massive data processing methods will also have a comprehensive impact on these metrics. In addition, in the process of processing, the internal relationship of information can be found and associated through complex computation process, and the aggregation metric of information can be changed. Finally, the result data of supercomputing naturally depends on the various capabilities of peripheral output devices, which affect the information volume, delay, scope, granularity, variance, duration, sampling rate, aggregation, distortion and mismatch efficiency of the user.

Although supercomputers themselves may have powerful information transmission and storage capabilities, because their core mission is supercomputing, other capabilities are subordinate to it, so we can still incorporate them into the double-ring configuration of collection-processing-action.

The information system can also directly input the collected information into the data space, and use the powerful information resources of the data space to directly act on the real-world, thus forming a double-ring dynamic configuration of collection-data space-action (Figure 6).

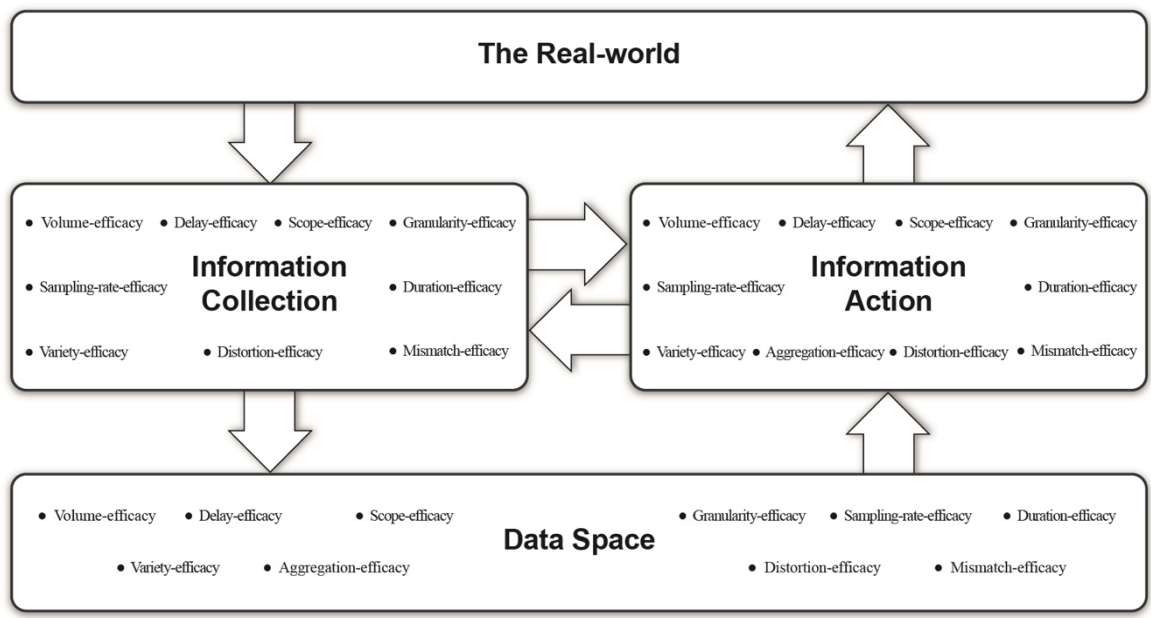

**Figure 6: Double ring dynamic configuration of information acquisition-data space-action systems**

A typical scenario for double-ring configuration of acquisition-data space-action is the construction of a data center. With the wide application of big data technologies, all kinds of data centers play an increasingly important role in industry and regional informatization. Its construction process can be simplified as that all kinds of information are gathered into the data space through peripheral devices and interfaces, and the functional performance of these peripheral devices and interfaces will obviously affect the volume, delay, scope, granularity, variance, duration, sampling rate, distortion and mismatch of information collection. The structure design, model type, resource accumulation and storage capacity of the data space itself will also affect these corresponding metrics. At the same time, because the structural model of data space itself often implies the internal relationship of information aggregation, data space will also affect the aggregation metric of all information. Therefore, the final output information of the acquisition-data space-action double-ring configuration has the functions of volume, delay, scope, granularity, type, duration, sampling rate, aggregation degree, distortion degree and mismatch degree, and naturally these metrics will also be affected by various capabilities of the peripheral output equipment and interfaces of the data center. Similarly, any data center itself has the necessary information transmission and processing capabilities, but its core mission is to build a data space that echoes the real-world. In order to simplify the problem, other capabilities can be ignored and incorporated into the double-ring configuration of acquisition-data space-action.

## 6.4 Triple-rings Dynamic Configuration of Information Systems

The triple-ring configuration often involves four different information system links, and there are also three typical information movement modes. Fig. 7 presents the acquisition-transmission-processing-transmission-action triple-ring dynamic configuration for information systems.

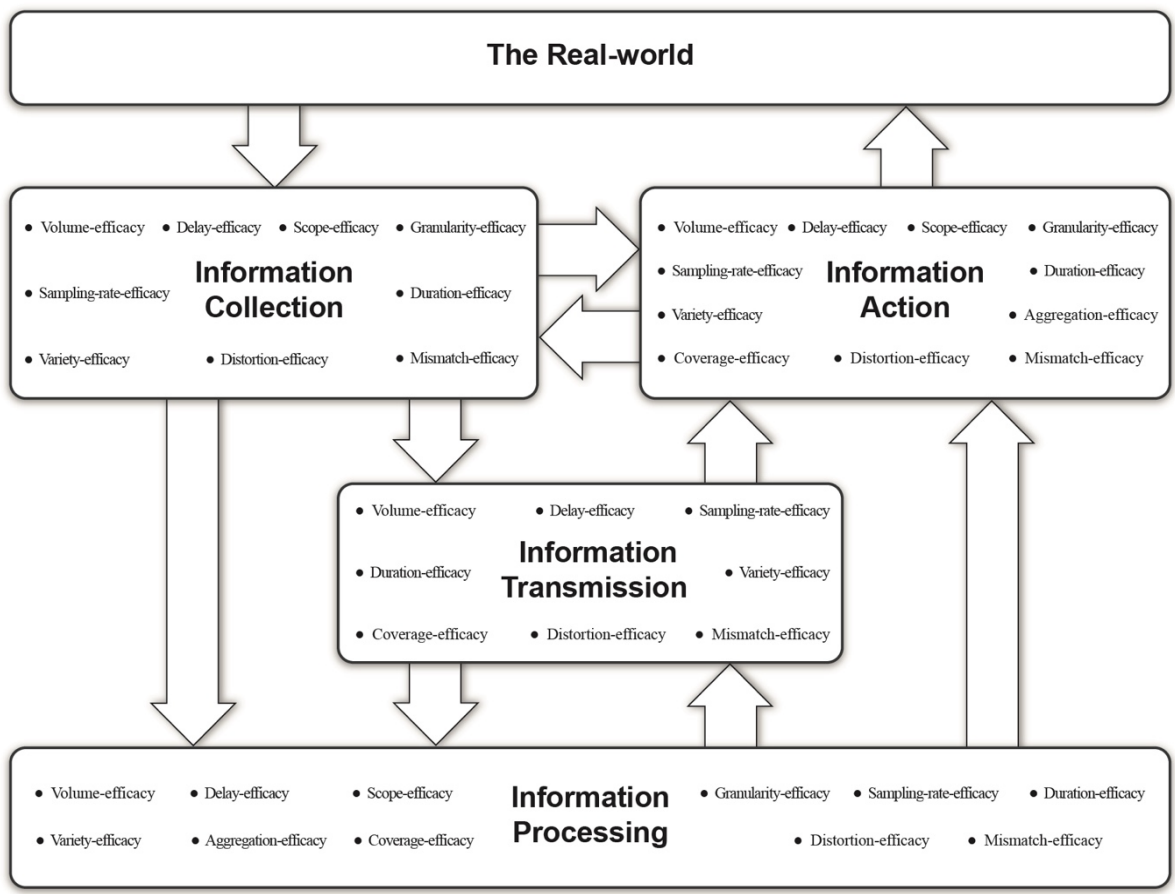

**Figure 7: Triple-ring dynamic configuration of information acquisition-transmission-processing-transmission-action systems**

The typical scenario of the collection-transmission-processing-transmission-action triple-ring configuration is for the remote automatic control systems. After the widely distributed sensors collect all kinds of information, the information is gathered to the control center through the communication network. After the numerical computation, state evaluation and command generation processing of the control center, the control information is automatically distributed to the corresponding control nodes through the communication network to realize the process automatic control of the wide area systems. It should be noted that in this configuration, the information collection link itself affects the nine metrics of the collected information except the degree of aggregation and the degree of coverage. The degree of aggregation of the information can be affected through the information processing link, and the degree of coverage of the information can be affected through the second information transmission link. Therefore, this triple-ring configuration has eleven complete effects.

Fig. 8 presents the acquisition-transmission-data space-transmission-action triple-ring dynamic configuration of the information systems.

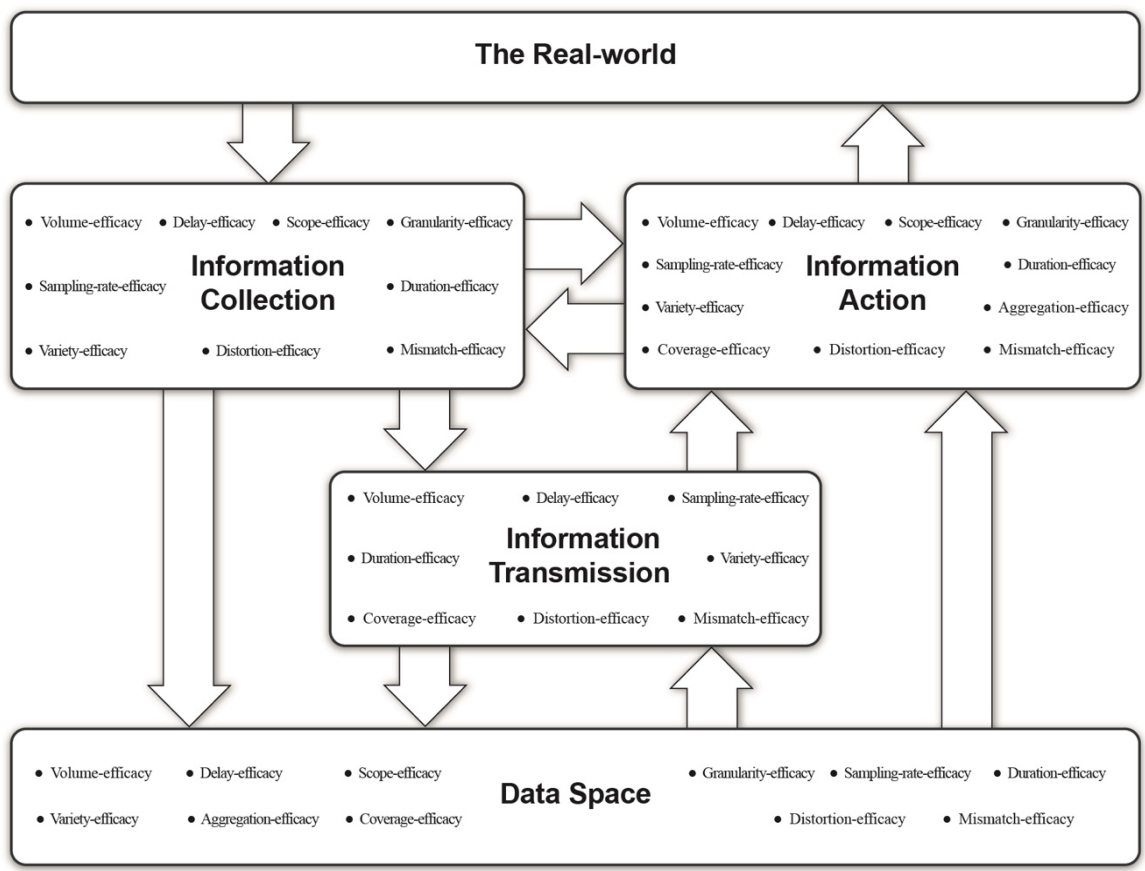

**Figure 8: Triple-ring dynamic configuration of information acquisition-transmission-data space-transmission-action systems**

The typical scenario of the triple-ring configuration of collection-transmission-data space-transmission-action is the wide-area information aggregation and service process of Internet websites. At present, relatively simple Internet websites can rely on all kinds of information publishers, collect information by using Internet terminals, gather the information into the website database through the wide-area distribution of the Internet, form their own data space, and then provide web information services for all kinds of users through the Internet and its terminals. In this configuration, the information collection link itself affects the nine metrics of the collected information except the aggregation degree and the coverage degree. Because the data content of the data space can reflect the aggregation degree and the coverage degree metrics of the information, and can also affect the coverage metric of the information through the second information transmission link, the triple-ring configuration also has eleven complete effects.

Fig. 9 presents the triple-ring dynamic configuration of the information acquisition-processing-data space-processing-action system.

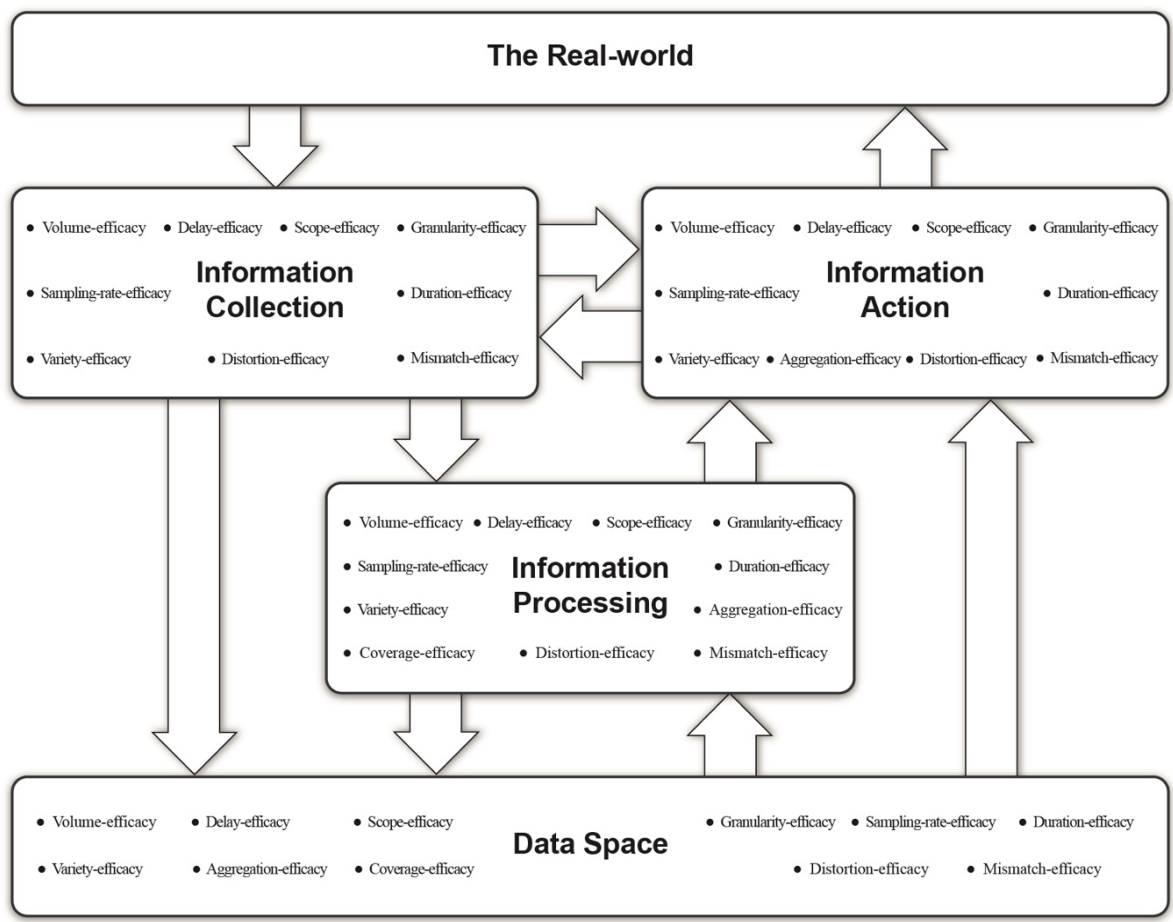

**Figure 9: triple-ring dynamic configuration of information acquisition-processing-data space-processing-action systems**

The typical scenario of the acquisition-processing-data space-processing-action triple-ring configuration is the centralized modeling and simulation systems. Taking the modeling and simulation of electromagnetic radiation characteristics as an example, a specific object is placed in a microwave anechoic chamber environment, a relevant electromagnetic signal is applied to the object, the modeling and simulation system can collect part of the electromagnetic radiation information of the object at a sampling point, and the omnidirectional radiation characteristics of the object in an ideal environment can be obtained through computation and interpolation processing. The electromagnetic radiation information of the object under the simulation condition can be calculated by calling the electromagnetic field environment model information possibly existing in the electromagnetic field data space, so that sufficient simulation information support can be provided for researchers. It can be seen that this is a typical application scenario of information collection-processing-data space-processing-action. Since no information transmission link is involved, the triple-ring configuration only has the other 10 functions except the coverage.

## 6.5 Triple-rings and One-core dynamic configuration of Information Systems

In practice, there is almost no information systems that completely lacks any link in information collection, information transmission, information processing, data space and information action. Therefore, some links are omitted in the previous text to study seven typical configurations, mainly focusing on simplifying the problem and focusing on the key problems. The triple-rings and one-core configuration of information systems shown in Figure 10 is the most complete, most common and most in need of full study of the dynamic configuration of information systems, which can be called the complete configuration.

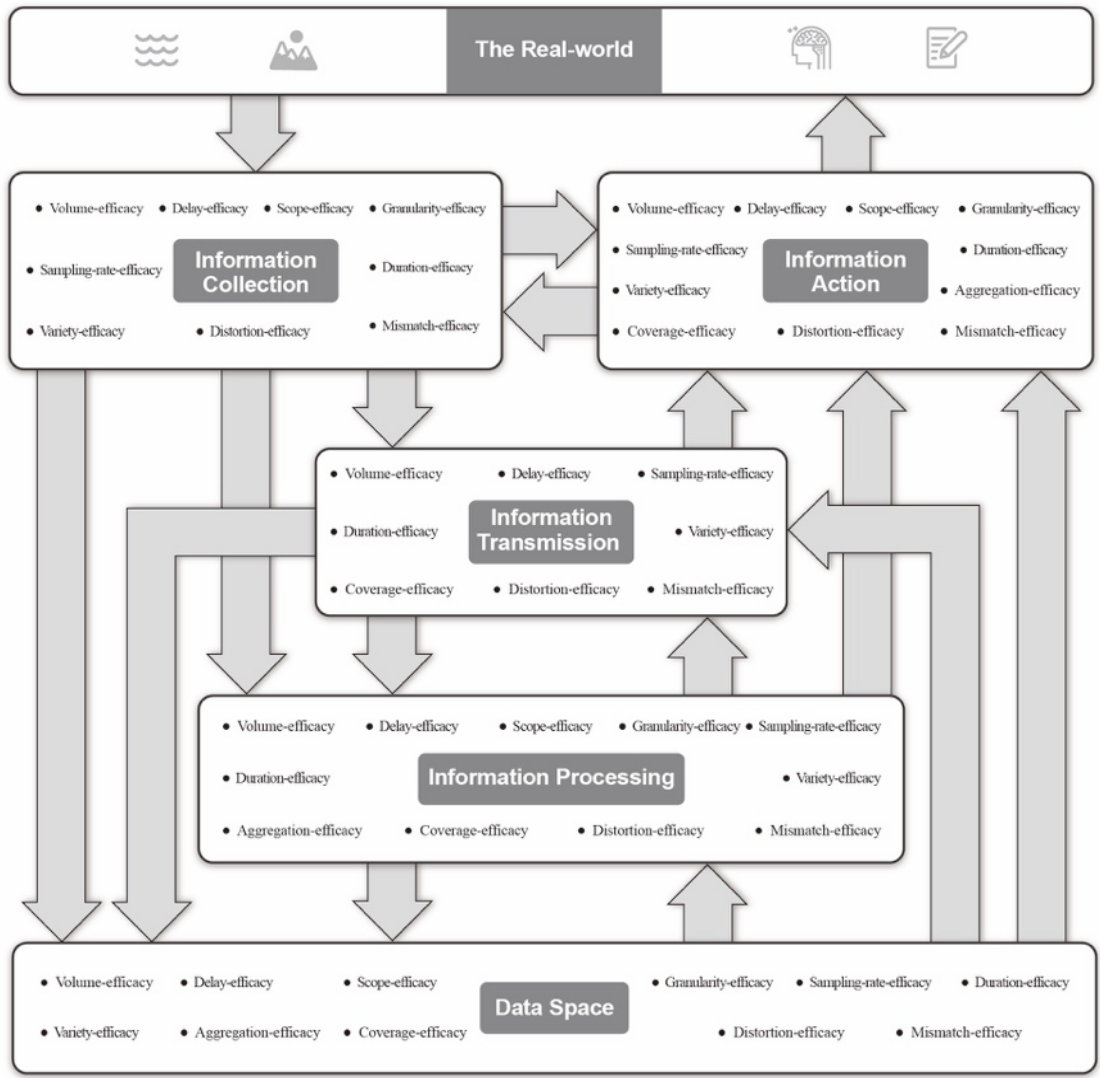

**Figure 10: Triple-ring and one-core dynamic configuration of information systems**

It can be seen from Figure 10 that the complete configuration of the information systems includes not only all the information movement links, but also the possible information flows between the links. At the same time, it should be noted that any part of the complete configuration in engineering practice may be the object of study. Therefore, the design and implementation of information systems need not be limited to the above eight configurations. Generally speaking, as long as the target system maintains the continuity of information flows, we can use the efficacy of various links to analyze the efficacy of the whole system, which is the original intention of putting forward information systems dynamics and guiding the planning, design, development and integration of information systems.

In Figure 10, each aspect of an information systems can affect the metric effects of the entire system of systems. Generally speaking, the same efficacy of each link must have the effect of mutual superposition or mutual restraint. For example, the delay efficacy of each link can obviously be superimposed to form the delay effect of the whole system. The volume metric of the previous link obviously forms the volume requirement for the subsequent link. If the subsequent link cannot meet the requirement, it will inevitably affect the volume efficiency of the whole system of systems.

On the other hand, there are mutual effects among different efficacies. For example, the volume efficiency obviously affects the distortion efficiency of the system. In the case of insufficient volume, the elements of the reflection set will inevitably be abandoned, resulting in an increase in the distortion metric of information. The first ten metrics of efficacies, in order, almost all contribute to the overall system mismatch metric. Because the degree of mismatch reflects the degree to which the output information of the information system deviates from the needs of a specific user, metrics such as volume, delay, scope, granularity, duration, variance, sampling rate and aggregation degree are obviously closely related to the needs of a specific user, not the higher the performance, the better. The coverage metric needs to be adjusted according to the user's wishes. In order to control the scope of information, sometimes many means need to be taken to fully reduce the coverage of information, so the coverage metric is not related to the mismatch metric. The distortion metric is also not positively correlated with the mismatch metric, because in encrypted information systems, a higher distortion is often a lower mismatch for a particular user. Therefore, the study of the mechanism of measured efficacies in various configurations or scenarios of information systems and the revelation of the inherent operation law of information systems provide a broad and practical development prospect for information systems dynamics.

## 6.6 An exemplification: Smart Court SoSs Engineering Project of China

Since 2013, Chinese courts have employed the theories and methodologies of information systems dynamics to promote the construction of Smart Courts nationwide, with remarkable achievements, which brings China as the current leader in legal technology in the world[1].

**(1) Overview of Smart Court SoSs Engineering Project of China**

The construction of Intelligent Courts involves more than 3000 courts, more than 10000 dispatched courts and more than 4000 collaborative departments in China nationwide. The number of information systems, such as infrastructure systems, intelligent applications, data management, network security and operation and maintenance support, has reached more than 13000. Those systems operate relatively independently and simultaneously every day, with a huge scale, a wide spatial distribution and a varying duration. It is an extremely complicated engineering project, which features heterogeneous systems, various functions and tasks, numerous collaborative departments and close sharing and linkage [246].

---

[1] https://www.scl.org/articles/9979-china-as-the-next-leader-in-legal-technology

In the Smart Court information systems, intelligent service, intelligent trial, intelligent execution, intelligent management and judicial openness are information systems that directly face the vast number of users and undertake the task of information collection and information exertion. Wherein, the intelligent service systems include China Mobile Micro-Court, People's Court Mediation Platform, Litigation Service Network, 12368 Litigation Service Hotline, Electronic Service, Online Preservation, Online Identification and other platforms; the intelligent trial systems include trial process management, electronic file transfer application, intelligent trial assistance and other platforms; intelligent execution systems include execution command, execution case flow information management, execution investigation and control, punishment for dishonesty, online judicial auction, "one account a case" management, mobile execution and other platforms; and intelligent management systems include online office, trial supervision and electronic archives; the judicial openness systems include China Trial Process Information Open Network, China Trial Open Network, China Judicial Documents Network and China Execution Information Open Network. Internet, court private networks, mobile private networks and external private networks are information systems that connect users inside and outside the court and undertake the task of information transmission. Automatic cataloging of electronic files, automatic backfilling of case information, legal knowledge service, intelligent recommendation of similar cases, intelligent inspection of court audio and video, intelligent analysis of judgment deviation, and one-key filing of file materials are information systems responsible for information processing tasks. The judicial big data management and service platform gathers the trial and execution data, judicial personnel data, judicial administration data, external data, judicial research data and information operation data of the nationwide courts, which is the core data space reflecting the trial execution and operation management status of the nationwide courts.

**(2) Key efficacies of Intelligent Court information systems**

The effectiveness of the construction and application of the Smart Court information systems depends on the information measure efficacies of various information movement as the consequence of the integration of all the information systems involved. Although almost every system and every information have its own user experience and effects, the key performance indicators of some primary systems can have a more important impact on the eleven types of measure efficacies of the entire systems. Although the common names of these performance indicators may not be exactly the same as the measure efficacies defined in this paper, the substance of the indicators attribute to the measure efficacies indeed. We have formulated the distribution of the key efficacies of Smart Court information systems as shown in Table 4. In practice, we continuously keep adjusting these indicators, so as to continuously improve the operating quality of the Smart Court information systems as a whole.

**Table 4: Distribution of key efficacies of Smart Court information Systems**

| No. | Measure efficacies | Information collection | Information exertion | Information transmission | Information processing | Data space |
|---|---|---|---|---|---|---|
| 1 | Volume-efficacy | • Total input data of application system per unit time | • Total output data of application system per unit time | • Internet access bandwidth<br>• Court private network bandwidth<br>• Mobile private network bandwidth | • Total cloud computing resources<br>• Total amount of on-cloud storage resources<br>• CPU resource utilization on the cloud<br>• Cloud storage resource utilization | • Big data platform aggregates the total amount of judicial data resources<br>• ·Big data platform aggregates the total amount of case data |
| 2 | Delay-efficacy | • Data submission delay of case handling system<br>• Case file information upload delay | • Application system operation response delay | • Video information transmission delay<br>• File information transmission delay | • Judicial big data, judicial artificial intelligence and other computing processing delay | • Judicial big data daily full data aggregation delay |
| 3 | Scope-efficacy | • The number of courts nationwide covered by the case handling system<br>• The number of tribunals nationwide covered by the case handling system<br>• The distribution and number of users covered by the intelligent service system | • The total amount of service data that the legal knowledge service system can provide | | • The total amount of laws, regulations and case information processed by the legal knowledge service system | • The national court coverage of judicial big data |
| 4 | Granularity-efficacy | • Judicial statistical information integrity rate<br>• Integrity ratio of case information items<br>• Video information resolution | • Video information display resolution | | • Legal knowledge decomposition and refinement granularity | • Number of judicial statistical information items on big data platform<br>• National-wide court case coverage integrity on Big Data Platform |
| 5 | Variety-efficacy | • Types and modes of data, text, file, video, audio and other information input by the application system | • Types and modes of data, text, file, video, audio and other information output by the application system | • Internet, court private network, mobile private network, external private network transmission data, texts, files, videos, The number of types of information such as audio | • The number of types of information such as data, text, files, video and audio processed by judicial big data, cloud computing, artificial intelligence, block chain and other systems | • The number of data, text, files, video and audio contained on the Judicial Big Data Platform |
| 6 | Duration-efficacy | • Application MTBF | • Application MTBF | • Network MTBF | • Computing storage facility MTBF<br>• Mean time between failures of information processing system | • Mean time between failures of Big Data Platform |
| 7 | Sampling-rate-efficacy | • Input data sampling rate of application system | • Data output rate of application system per unit time | • Network load utilization rate | • Throughput ratio of computing storage facility<br>• Processing-period of information processing systems | • Data access period of Big Data Platform |
| 8 | Aggregation-efficacy | | • aggregation degree of application system output data | | • The number of case association types<br>• The number of human case association types<br>• The number of related types of the case<br>• The number of associated types of case payment | • Data aggregation degree of big data platform |
| 9 | Coverage-efficacy | | • Application System User Distribution<br>• The number of users of the application system | • The private court network coverage area<br>• Mobile private network coverage area<br>• The number of departments covered by the external private network | • Effectiveness of information encryption<br>• Accuracy of user permission control<br>• Reliability of security isolation between networks | • Storage space and regional distribution of big data platform |

| 10 | Distortion-efficacy | • The input information accuracy of application systems | • The output information accuracy of application systems | • The transmission information distortion of communication system | • The information processing accuracy of processing systems | • Full data confidence of Big Data Platform<br>• Shared data confidence of Big Data Platform |
|---|---|---|---|---|---|---|
| 11 | Mismatch-efficacy | • Adaptability of input data format, type, content and quantity of application system | • Adaptability of output data format, type, content and quantity of application system<br>• User satisfaction rate of information system | • Adaptability of transmission information format and type of communication system | • Accuracy of data-user association computation | • Accuracy of data model of Big Data Platform |

**(3) Growth curve of key measurement efficacy of Intelligent Court information system**

Figure 12 reflects the changes of some key performance indicators of Intelligent Court information systems in recent years. Amongst the indicators, the total amount of data resources of the judicial big data platform reflects the capacity of the Supreme Court to gather the judicial big data of the nationwide courts, and its steady rise shows that the accumulation of judicial big data resources is becoming more and more abundant; The average response delay indicator of court office platform is directly related to the operational experience of almost all staff members. Thanks to the technical improvement, this indicator has dropped to less than 0.8 seconds since November 2020, which has won the unanimous praise from the staff members; The monitoring system of S&T courts is the information system that uses video technology to connect courts around China in real time. Its court coverage rate reflects the scope of video information of courts across the China. Since November 2021, the Supreme Court has steadily connected more than 93% of the S&T courts across the country through video networks; A single case is the smallest granularity of court judicial information. Since August 2015, the coverage rate of nationwide court case information has basically reached and has been stable at 100%, which fully shows that the management of nationwide court judicial big data has reached a very fine level; The types of information gathered by the judicial big data platform reflect the integrity of information management. Since the judicial Big Data Platform was officially launched in December 2013, the types of information have steadily increased, basically realizing the aggregation, management and application of all types of information; The average trouble-free time of the information system reflects the average time of real-time information collection. The figure shows that since March 2018, the average trouble-free time of the court information system has basically stabilized at more than 700 hours. Apart from some significant decline in short time, the duration of real-time information collection in the corresponding period of time is inevitably reduced; The Law-eye platform realizes the monitoring and management of the operation quality and effectiveness of the nationwide court information systems, and its information sampling rate needs to be set reasonably according to the specific characteristics of the monitored objects. The figure shows that 53% of the monitoring information sampling rate is higher than 1 time/hour, and 73% of the monitoring information sampling rate is higher than 1 time/day, which reflects the sampling density of the Law-eye platform; The data aggregation degree of the judicial Big Data Platform reflects the correlation degree of its internal data. The graph shows that since January 2019, the information aggregation degree has been higher than 80%, indicating that the correlation processing and application of information are at a good level; The coverage of information output by the information systems can be characterized by the number of visits. The figure shows that since February 2020, the number of monthly visits of China Mobile Micro Court, the unified online service window for the public, has steadily increased, exceeding 100 million by December 2021, which fully demonstrates its remarkable effect of serve the public; Data confidence is the negative expression of information distortion. The figure shows that since January 2018, the confidence of judicial statistics of judicial Big Data platform has been higher than 97%. At present, it has been stable at more than 99% for a long time, that is, its distortion is less than 1%, which lays a credible foundation for various big data analysis and services; User satisfaction is also a negative expression of the mismatch of information output by the information systems. The graph shows that since January 2020, the user satisfaction of the information system has been higher than 98%, which fully demonstrates the remarkable achievements of the Smart Court SoSs engineering project of China.

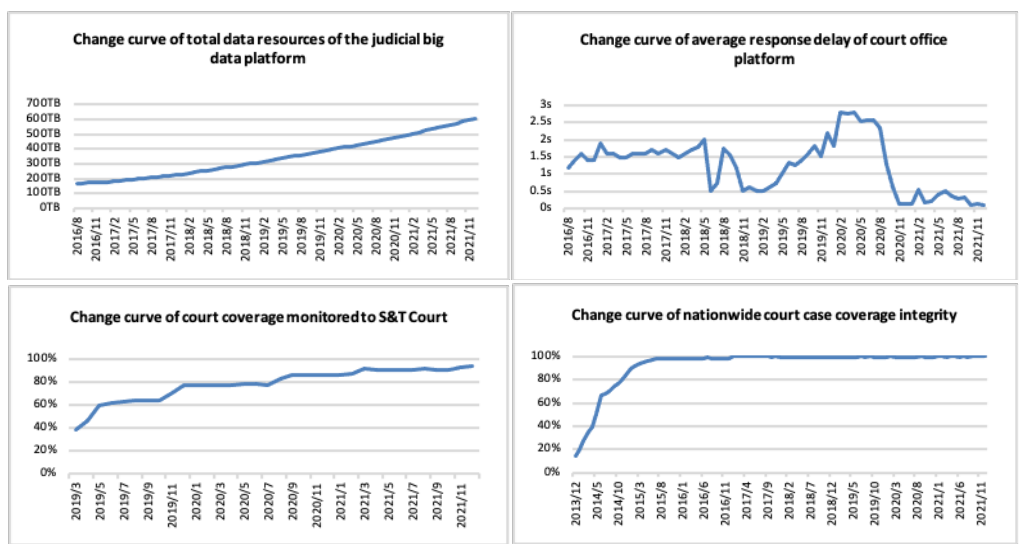

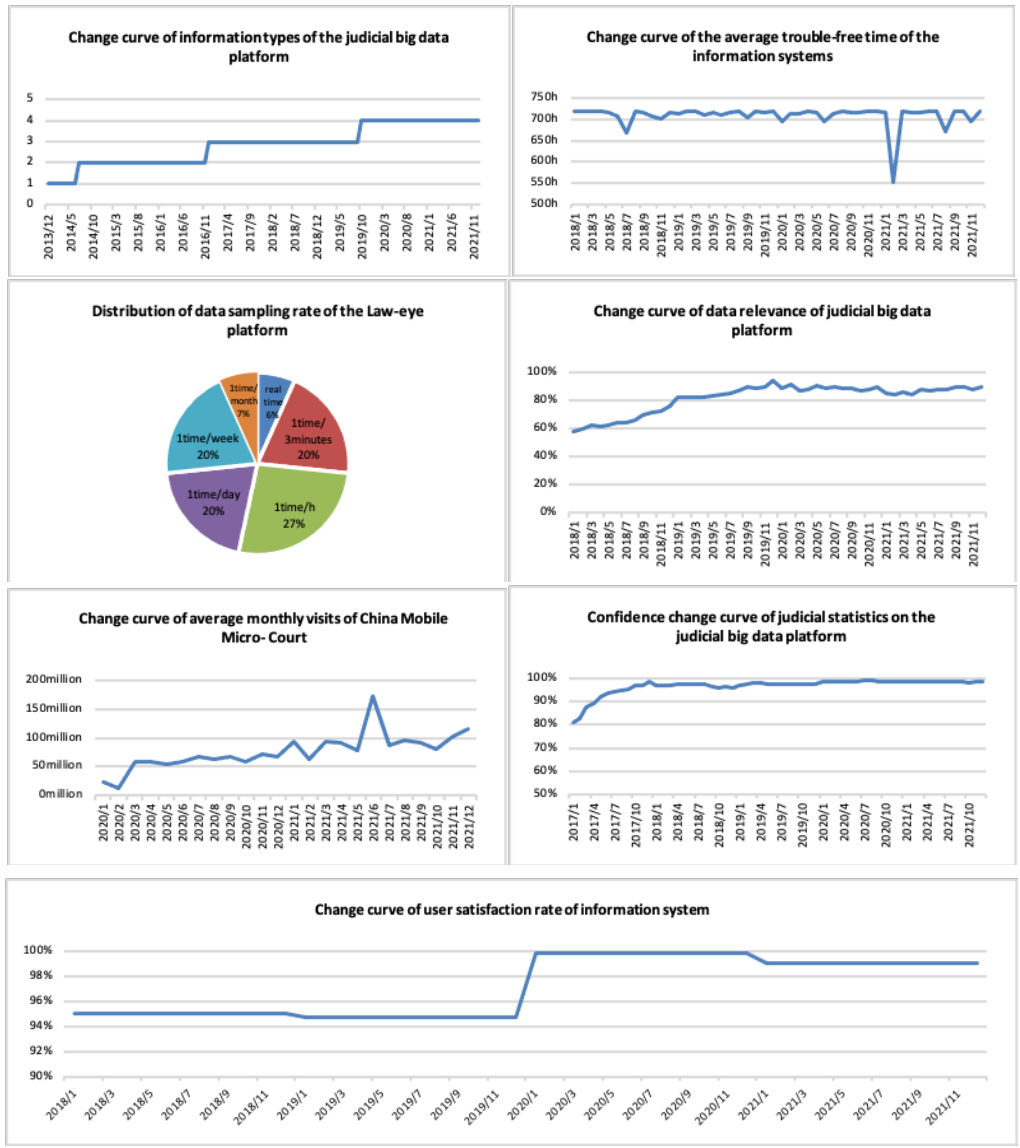

**Fig. 12. Examples of eleven types of metric efficacies of Smart Court information systems**

# 7 Summary

Confronting the grand challenges, such as the lack of universally recognized mathematical basis for the concept of information, the lack of clear and abundant measure system for the value of information, the lack of scientific and rational framework structure for information space, and the absence of well-defined efficacies for analysis of information effects, this paper proposed the structural framework of information space, and the model, characteristics and measure metrics of information. Furthermore, measure efficiencies and dynamics configurations of information systems are proposed, which constitute the technical and mathematical basic theory system of information systems dynamics. These theoretical efforts on information science have been employed and verified in the practice of China Intelligent Court SoSs engineering.

The mathematical foundations for describing information space and the law of information movement in this paper are set theory, measure theory, relational algebra and topology. Although these mathematical subjects are abstract, they all have direct and clear corresponding relationships with the statistical and computational methods widely used in daily life, and are fully compatible with the classical principles of widely adapted information theories such as Shannon information entropy and Nyquist sampling theorem. Therefore, it is fully applicable for the analysis, investigation, design and evaluation of information systems such as big data, streaming media and metaverse, which have attracted much attention nowadays.

In the meanwhile, as any theory can only reflect its value and be constantly improved in practice, the completeness and practicability of information systems dynamics should also be tested and modified in a large number of subsequent applications. Based on the mathematical expressions on information and the measure metrics proposed in this paper, the future investigation on information systems dynamics can be supplemented and enriched in a variety of directions, such as the consistency between information measurement and classical principles of various information technologies, the interaction relationship between information measurement efficacy, and the refinement and decomposition of information system dynamics configuration.

305.

[245]Kalman R . A new approach to linear filtering and predicted problems[J]. J Basic Eng, 1960, 82: 35-45.

[246] Jianfeng Xu, Fuhui Sun, Qiwei Chen. Introduction to smart court system engineering (智慧法院体系工程概论 in Chinese), People's court press, 2021.4.